\documentclass[twocolumn]{aa}
\usepackage{graphicx,times,amssymb,longtable}
\begin{document}
   \title{Rotation and variability of very low mass\\ stars
   and brown dwarfs near $\epsilon$\,Ori\thanks{Based on observations collected at the European Southern 
   Observatory, Chile, observing run 68.C-0213(A)}}

   \author{Alexander Scholz
   \and
   Jochen Eisl{\"o}ffel}

   \offprints{A. Scholz, e-mail: scholz@tls-tautenburg.de}

   \institute{Th{\"u}ringer Landessternwarte Tautenburg,
              Sternwarte 5, D-07778 Tautenburg, Germany}

   \date{Received sooner; accepted later}
   
   \authorrunning{A. Scholz and J. Eisl{\"o}ffel}
   
   \titlerunning{Rotation and variability of VLM objects near $\epsilon$Ori}

   \abstract{We explore the rotation and activity of very low mass (VLM) objects
   by means of a photometric variability study. Our targets in the vicinity of 
   $\epsilon$\,Ori belong to the Ori\,OB1b population in the Orion star-forming complex. 
   In this region we selected 143 VLM stars and brown dwarfs (BDs), whose photometry in 
   RIJHK is consistent with membership of the young population. The variability of these 
   objects was investigated using a densely sampled I-band time series covering four 
   consecutive nights with altogether 129 data points per object. Our targets show three
   types of variability: Thirty objects, including nine BDs, show significant 
   photometric periods, ranging from 4\,h up to 100\,h, which we interpret as 
   the rotation periods. Five objects, including two BDs, exhibit variability with 
   high amplitudes up to 1\,mag which is at least partly irregular. This behaviour 
   is most likely caused by ongoing accretion and confirms that VLM objects 
   undergo a T Tauri phase similar to solar-mass stars. Finally, one VLM star shows 
   a strong flare event of 0.3\,mag amplitude. The rotation periods show dependence on
   mass, i.e. the average period decreases with decreasing object mass, 
   consistent with previously found mass-period relationships in younger and older 
   clusters. The period distribution of BDs extends down to the breakup period,
   where centrifugal and gravitational forces are balanced. Combining our BD periods
   with literature data, we found that the lower period limit for substellar objects
   lies between 2\,h and 4\,h, more or less independent of age. Contrary to stars, these 
   fast rotating BDs seem to evolve at constant rotation period from ages of 3\,Myr 
   to 1\,Gyr, in spite of the contraction process. Thus, they should experience strong 
   rotational braking.\thanks{Table 2 is only available
   in electronic form at the CDS via anonymous ftp to cdsarc.u-strasbg.fr (130.79.128.5)
   or via http://cdsweb.u-strasbg.fr/cgi-bin/qcat?J/A+A/}
   
   }

   \maketitle

.
\section{Introduction}
\label{intro}

Variability studies are a key tool for investigating the physics of stars.
From simple photometric monitoring campaigns alone, it is possible to 
unveil important properties of stars. One example is the rotation
period, which in many cases can be obtained from photometric light curves, 
if the objects exhibit asymmetrically distributed surface features,
e.g. magnetically induced spots (e.g., Bouvier \& Bertout \cite{bb89}, Bouvier
et al. \cite{bcf93}, \cite{bck95}, Herbst et al. \cite{hrw00}, \cite{hbm02}). 
The amplitude of the light curve then contains information 
about the spots, and thus about the magnetic activity of the targets 
(e.g., Krishnamurthi et al. \cite{ktp98}). Other signs of activity can be seen 
in the light curves as well, in particular rapid brightness eruptions like flares 
(Stepanov et al. \cite{sfk95}). Additionally, accretion processes manifest 
themselves in strong variability, because they often produce hot spots where 
matter flows from the accretion disk onto the star's surface. These hot spots 
are again a source of periodic variability, and spot instabilities as well
as accretion rate variations can additionally induce irregular
photometric variability (Fern\'andez \& Eiroa \cite{fe96}, Herbst et al.
\cite{hmw00}). 

This broad output motivated extended photometric monitoring studies of young
stars. The advent of wide-field CCD detectors increases the efficiency of such
projects enormously, since they allow observers to monitor a large number
of objects simultaneously. Many of these studies focus on open 
clusters, because they deliver coeval target samples for which good 
estimates for age and distance are available. That way, one can evaluate 
the evolution and dependence on mass of stellar properties like rotation and 
activity.

Most monitoring studies in young open clusters, however, are concentrated 
on solar-mass stars with masses $>0.5\,M_{\odot}$. The major outcome of
this work is a huge database of more than 1500 rotation periods, mostly 
for T Tauri stars with ages less than 10\,Myr (see the recent reviews 
by Stassun \& Terndrup \cite{st03} and Mathieu \cite{m03}). These 
periods deliver the crucial constraints for any model of rotational 
evolution. The state-of-the-art description of the rotational behaviour 
contains a) magnetic interaction between star and disk, b) angular 
momentum loss through stellar winds, and c) structural effects caused by 
contraction and internal angular momentum transport (see, e.g., 
Barnes \& Sofia \cite{bs96}, Krishnamurthi et al. \cite{kpb97}, 
Bouvier et al. \cite{bfa97}). 

In the last decade, the low-mass end of the known population of open 
clusters was shifted well down into the substellar (or even into the planetary)
mass regime (see B\'ejar et al. \cite{bzr99}, Zapatero Osorio et al. \cite{zbm00}, 
Lucas \& Roche \cite{lr00}, Muench et al. \cite{mll02}, \cite{mll03}). This 
survey work delivered large samples of very low mass (VLM) objects, an object
class herewith loosely defined as 'objects with masses below $0.4\,M_{\odot}$', 
including VLM stars and brown dwarfs. After the detection of these objects, the next
logical step is to explore their properties, e.g. rotation and activity.
From the extended work of Herbst et al. (\cite{hbm01}, \cite{hbm02}) in the
Trapezium cluster and Lamm (\cite{l03}) in NGC2264, first large samples of 
rotation periods for very young VLM stars (ages around 1\,Myr) have become available. 
These samples were complemented by three periods for brown dwarfs in the 
similarly old Chamaeleon I star-forming region (Joergens et al. \cite{jfc03}). 
One important result of these studies is that the mean rotation period 
decreases with decreasing mass. The total range of periods, however, is 
very similar to that of more massive stars; they reach from several hours up to two 
weeks. The fast rotation of young VLM objects has been interpreted mainly as 
a consequence of imperfect disk-locking (Lamm \cite{l03}, Lamm et al. 
\cite{lmb04}). 

For more evolved VLM objects, however, the rotation period database is
very sparse. As of the end of 2003, six periods have been published for
VLM objects in open clusters with ages between 3\,Myr and 125\,Myr 
(Mart\'{\i}n \& Zapatero Osorio \cite{mz97}, Terndrup et al. \cite{tkp99},
Bailer-Jones \& Mundt \cite{bm01}, Zapatero Osorio et al. \cite{zcb03}),
complemented by a few periods for ultracool dwarfs in the field (e.g., 
Bailer-Jones \& Mundt \cite{bm01}, Clarke et al. \cite{ctc02}). All these
periods are shorter than one day, and thus give tentative evidence for a 
lack of slow rotators among VLM objects, which is confirmed by rotational
velocity studies (e.g., Terndrup et al. \cite{tsp00}). 

\begin{figure}[h]
\centering
\resizebox{\hsize}{!}{\includegraphics[angle=-90,width=6.5cm]{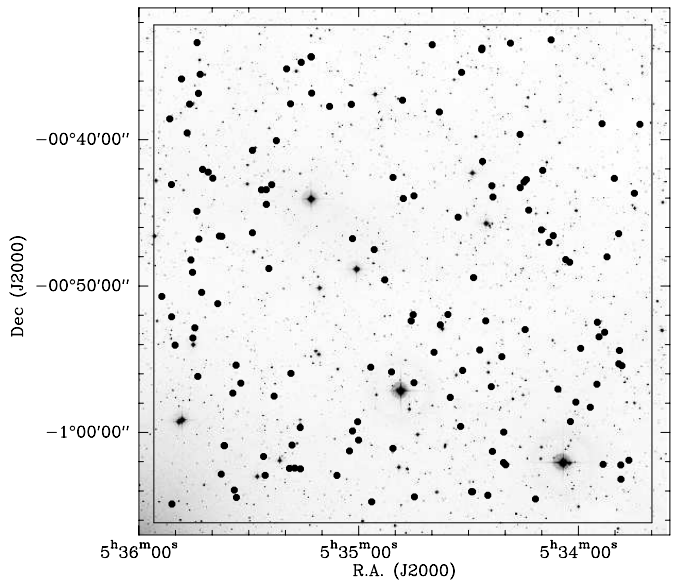}}
\caption{Target field of our monitoring campaign. The borders of our field of
view are shown as solid lines. All VLM candidates in this region, selected
by multi-filter photometry, are shown with large dots. The bright star 
$\epsilon$\,Ori is located close to the lower left corner of the field.}
\label{field}
\end{figure}

The described lack of rotation periods for VLM objects motivated a long-term
project to study their rotational evolution. In the first two papers of
this project, we published 23 rotation periods for objects in the $\sigma$\,Ori
cluster (Scholz \& Eisl{\"o}ffel \cite{se04a}, hereafter SE1) and nine for
VLM Pleiades members (Scholz \& Eisl{\"o}ffel \cite{se04b}, hereafter SE2).
In this paper, we report a variability study of VLM objects near
the star $\epsilon$\,Ori. This region (and also the $\sigma$\,Ori cluster)
belongs to the OB1b association of the Orion star forming complex, which 
harbours a large population of young stars and brown dwarfs, as recently
shown by Sherry (\cite{s03}). The age of the young objects near $\epsilon$\,Ori 
lies between 2 and 10\,Myr (see Wolk \cite{w96}), thus the objects are on 
average probably somewhat older than those in the $\sigma$\,Ori cluster (age 3\,Myr,
Zapatero Osorio \cite{zbp02}). The object density around $\epsilon$\,Ori is 
similarly high as in the $\sigma$\,Ori cluster. Therefore, we decided
to use a field near $\epsilon$\,Ori as target for a monitoring campaign
(see Fig. \ref{field}). The particular aim of the $\epsilon$\,Ori project 
was to enlarge the rotation period database for very young brown dwarfs, and
to improve the statistical significance of our previous $\sigma$\,Ori 
results. 

The structure of this paper is as follows. We first describe the monitoring 
campaign and the reduction of the images (Sect. \ref{obs}), the photometry
process and the calibration (Sect. \ref{photo}). The targets of the
variability study were selected with colour-magnitude diagrams, whose construction
and evaluation is explained in Sect. \ref{tar}. The subsequent section concentrates
on the time series analysis with the main focus on period search (Sect. \ref{tsa}).
Then, we discuss the origin of the variability and show that the light curve 
variations are caused by cool, magnetically induced spots, hot spots from 
accretion flow, and flare activity (Sect. \ref{ori}).
Finally, we concentrate on the periods and discuss the mass-period relationship
as well as the rotational evolution of VLM objects (Sect. \ref{rot}). We give
our conclusions in Sect. \ref{conc}. 

\section{Observations and image reduction}
\label{obs}

\begin{figure}[h]
\centering
\resizebox{\hsize}{!}{\includegraphics[angle=-90,width=6.5cm]{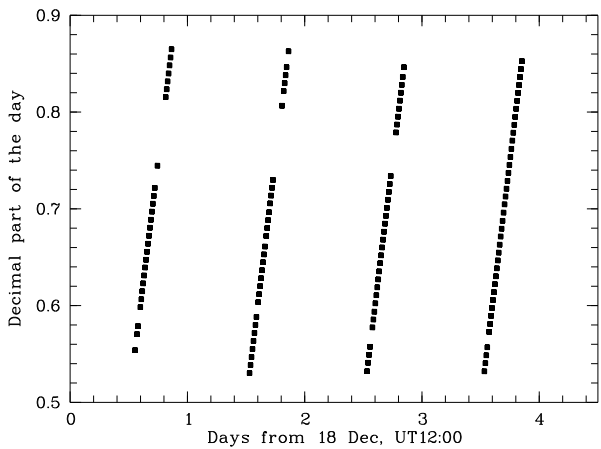}}
\caption{Sampling of the time series in the $\epsilon$\,Ori field. Plotted is the
non-integer fraction of the observing time vs. the observing time.}
\label{sampling}
\end{figure}

All images were obtained during a four-night observing run from 18 to
22 December 2001 with the ESO/MPG Wide Field Imager (WFI) at the 2.2-m telescope 
on La Silla. In these four nights, we stayed on our time series field
over at least seven hours. We obtained at least 27$\times$500\,sec time series 
exposures per night, altogether 129 images. The distribution 
of the time series images, i.e. our sampling, is nearly regular during the nights 
(see Fig. \ref{sampling}). In the first night, we took an additional deep R-band 
image of the field. For photometric calibration, we observed photometric 
standard stars from the catalogue of Landolt (\cite{l92}). All four nights were 
photometric; the seeing varied between 0\farcs7 and 1\farcs7. 

The WFI consists of eight $2K \times 4K$ CCDs in a $4\times 2$ mosaic, so that
the whole detector has $8192 \times 8192$\,pixels. The pixel size is 0\farcs25,
giving a field-of-view of $34\times 33$\,arcmin$^2$. To account for the differences
between the CCDs, we did each step of reduction, photometry, and calibration on every CCD
on its own. The only exception is the absolute photometric calibration (see Sect. 
\ref{photo}).

The reduction of the WFI I-band images follows the strategy used by, e.g., Alcal\'a et al.
(\cite{ars02}) and L\'opez Mart\'{\i} et al. (\cite{les04}). After bias subtraction 
and flatfield correction, we constructed a so-called superflat from ten images
of dark sky regions. This image contains a large-scale structure (called illumination),
which is probably caused by scattered light in the twilight flats, and small-scale
fringe structures from night sky emission. By smoothing the superflat and subtracting 
the result from the original superflat, we obtained separate images for illumination 
and fringes. Illumination was corrected by dividing all images by the illumination
mask. The fringe pattern, however, has to be subtracted, because it is due to additional light
on the detector. Since the strength of the night sky emission varies with airmass and
weather, the fringe mask has to be scaled before subtracting it from the images.
The appropriate scaling factors were determined with an automatic procedure: 
First, we subtracted the median from the respective image, so that its background
is zero. Then we divided this image by the fringe mask. The average of the ratio
image is the required scaling factor. Finally, the scaled fringe mask was subtracted
from the time series images. Visual inspection of the images reduced as described here
showed that the procedure removes at least 99\% of the fringes.

The R-band image was reduced in the same way, only the fringe correction was not 
necessary, because no fringes are visible in the R-band. This is also valid for
the standard star frames, even in the I-band, because of their short exposure times.
Accordingly, the fringe correction was skipped for these images.

\section{Photometry and astrometry}
\label{photo}

An object catalogue of the time series field was obtained by running SExtractor
(Bertin \& Arnouts \cite{ba96}) on the I-band reference image, which was
the image with the best seeing. The pixel positions in the object catalogue
were transformed to sky coordinates by fitting the known sky coordinates of 
HST guide stars (Morrison et al. \cite{mrm01}) in the field to their
measured pixel coordinates. The coordinate precision is $\pm1\farcs0$.

Following our target field over more than 7\,h per night led to small positional
offsets between the reference image and the other images. These offsets are typically 
below $5\farcs0$ and were measured by determining the positions of several bright 
stars in all images and comparing them with the positions in the
reference image. By applying these offsets to the object catalogue we obtained
the object positions for each image. Instrumental magnitudes were then measured 
for all registered objects by fitting their Point Spread Function (PSF) with the 
{\it daophot} routines (Stetson \cite{s87}). 

The instrumental magnitudes of the Landolt standard stars were determined
by aperture photometry, and scaled to consistent exposure times.
The zero-points, extinction coefficients, and colour 
coefficients for the R- and I-band were derived with a multi-linear fit:
(I,R -- Landolt magnitudes; i,r -- instrumental magnitudes, X -- airmass):
\begin{eqnarray}
I = i - 1.746 - 0.077 X + 0.192 (r-i) \\
R = r - 0.623 - 0.090 X + 0.000 (r-i)
\end{eqnarray}
The uncertainty of this transformation is dominated by the error of the zero-point, 
which is 0.06\,mag in the I- and 0.04\,mag in the R-band. Note the considerable 
colour term in the I-band, which indicates significant differences between the 
Landolt I-band and the WFI I-band. Since most of the Landolt stars have moderate 
colours (R-I $<1.0$), this may cause systematic calibration errors for very red 
targets. For these objects, we expect to overestimate the I-band flux. 

As noted above, the absolute calibration was performed only once for the
whole mosaic. For an optimal calibration, it would have been desirable to
calculate the transformation of Eqs. (1) and (2) for every CCD on its own. This
was, however, not possible, because of a lack of standard stars in the
observed Landolt fields. Recent tests showed that there are indeed 
systematic zero-point offsets between the WFI CCDs, which cannot be
corrected with our calibration procedure. We expect that these systematic
offsets account for a large part of the uncertainties in the derived
zero-points.

The transformations in Eqs. (1) and (2) were applied to the instrumental magnitudes 
obtained from the reference image of the $\epsilon$\,Ori field, after shifting them 
to the same exposure time as the standard stars. We obtained a deep catalogue of 
R- and I-band photometry of our time series field.

\begin{figure}[h]
\centering
\resizebox{\hsize}{!}{\includegraphics[angle=-90,width=6.5cm]{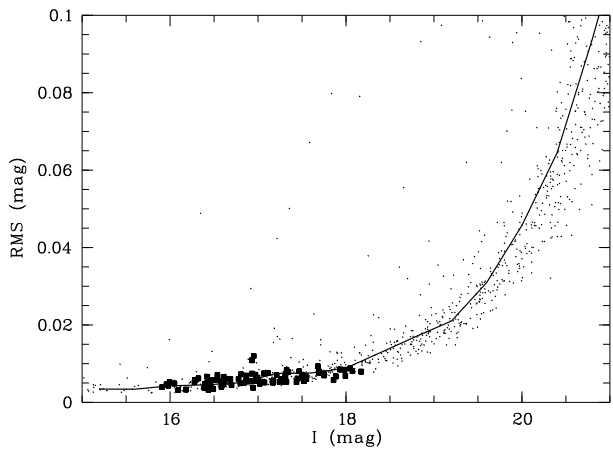}}
\caption{Root mean square of the light curves in the $\epsilon$\,Ori field vs.
I-band magnitude for one CCD. The solid line is the median rms. The values for
the reference stars in the relative calibration process are marked with filled
squares.}
\label{rms}
\end{figure}

For the time series images of the $\epsilon$\,Ori field, we are only interested in
differential photometry, i.e. the task is to obtain photometry relative to non-variable
reference stars in the field. These reference stars were selected with the algorithm
presented in SE1. The basic idea of this process is to select an initial sample
of stars with reliable photometry and to compare the light curve rms of each of these
preliminary reference stars to the average light curve of all other
stars in the initial sample. That way, we are able to identify variables and to reject
them. For details of the procedure we refer to SE1.

The typical number of stars in the initial sample of reference stars was 180 per CCD.
Typically half of them were marked as possible variable stars and rejected. The average
light curve of the remaining reference stars was then subtracted from all light curves. 
We obtained differential light curves for all objects in the $\epsilon$\,Ori field.
The rms of these light curves is shown in Fig. \ref{rms} as a function of I-band
magnitude for one CCD. The median of these values is over-plotted as a solid line. For 
the brightest stars, we reached a median rms of 5\,mmag. Fig. \ref{rms} shows also the 
rms of the reference stars (filled squares). The mean rms of the reference star light curves 
is 7\,mmag. 

The described procedure for the relative calibration neglects the colour dependence
of the atmospheric extinction, since it is done with stars of arbitrary spectral
type. Since our targets are redder than most other stars in the field, it could
be that our 'white' relative calibration is inadequate. To investigate this in more
detail, we examined if the colour dependence of the extinction is measurable
in our light curves. For about 200 non-variable objects, we fitted the relation between
instrumental magnitudes and airmass linearly. The slope of the fit is the
average extinction for the respective target. These values were then plotted against
the R-I colour. For $0.3< R-I <2.0$\,mag, there is no measurable colour dependence,
the extinction is on average 0.005\,mag/airmass. If we make the same test with the
relative magnitudes, there is also no colour dependence. The extinction 
becomes zero, as we should expect. Thus, our 'white' relative calibration
is valid.

For the reddest targets with $R-I>2.0$, we also see no significant excess in
extinction. However, in this colour range the error bars of the photometry
are large and thus the extinction values are not very reliable. In addition,
there are only very few objects in this colour range. Thus, we cannot definitely
exclude that the light curves of the reddest of our objects are affected
by colour effects. 

%9.0 for journal version, 7.0 for referee version
\begin{figure*}[ht]
\resizebox{9.0cm}{!}{\includegraphics[angle=-90,width=6.5cm]{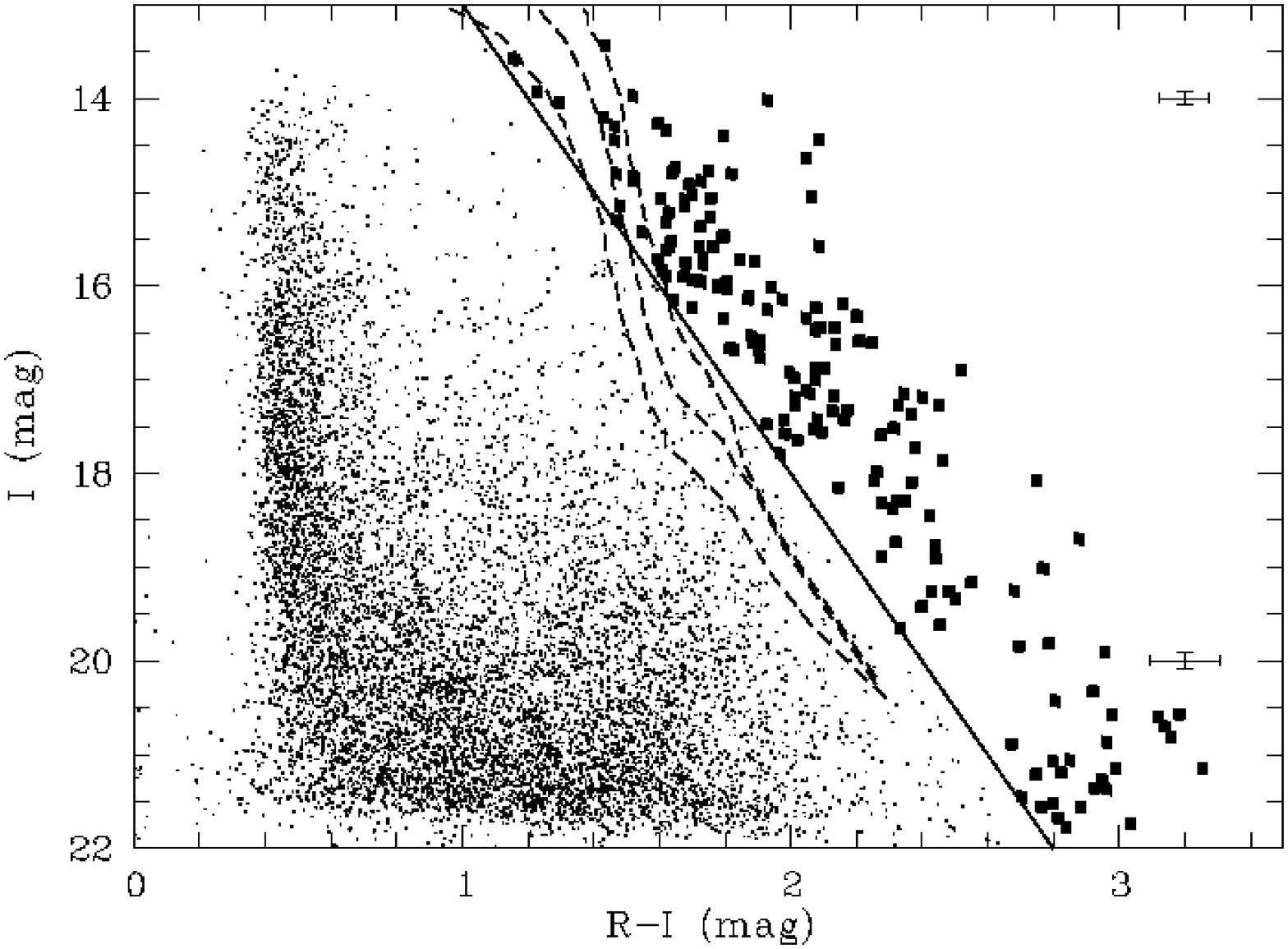}} \hfill
\resizebox{9.0cm}{!}{\includegraphics[angle=-90,width=6.5cm]{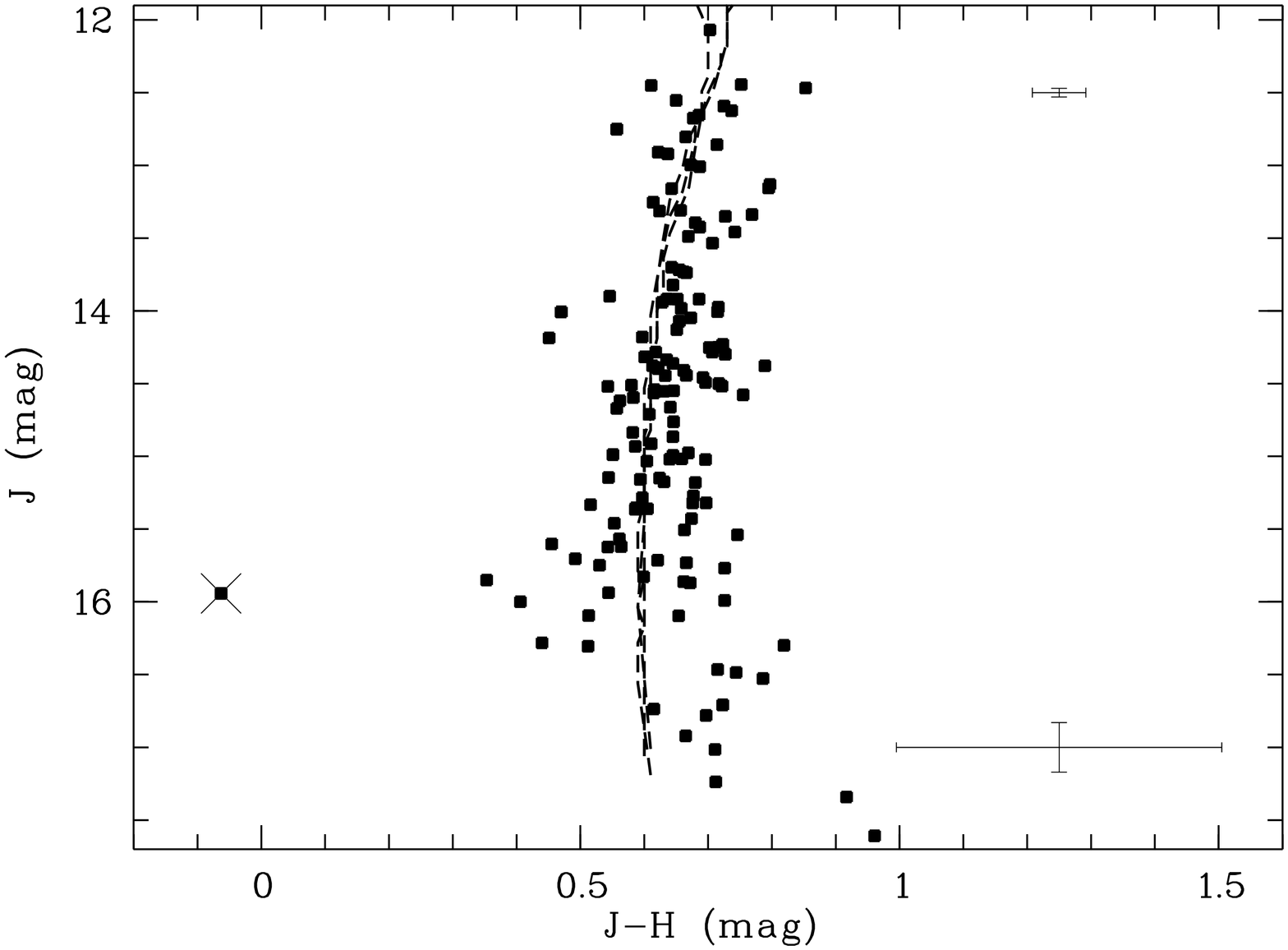}}
\caption{Colour-magnitude diagrams (CMD) for objects in our $\epsilon$\,Ori 
field. Probable cluster members are marked with filled squares. Error bars indicate 
typical photometry errors for the candidates. Left panel: (I,R-I) CMD with the 
separation line between candidates and field stars shown as straight line. The 
dashed lines indicate the positions of isochrones of Baraffe et al. (\cite{bca98})
for 3\,Myr (right), 5\,Myr (middle), 10\,Myr (left). Right panel: (J,J-H) CMD for 
the candidates constructed from 2MASS data. The marked outlier was excluded from 
the candidate list. The three isochrones are nearly indistinguishable.}
\label{cmd}
\end{figure*}

\section{Target selection}
\label{tar}

In this work, we are only interested in the photometric variability of young VLM
members of the Ori\,OB1b association. For convenience, we call these objects
$\epsilon$\,Ori cluster members, although they may not form a cluster similar to the
one around $\sigma$\,Ori, as found by Sherry (\cite{s03}). We used multi-filter
photometry to identify probable cluster members by means of colour magnitude
diagrams (CMD). Young VLM objects are redder than most field stars and should
therefore populate a distinct area in the red part of the CMD.

The cluster member selection is based on the (I,R-I) CMD constructed from our
WFI photometry (Fig. \ref{cmd}, left panel). The majority of data points in this
diagram forms the broad main sequence of fore- and background stars. On the red
side of this sequence, there is a clear accumulation of objects, which cannot 
be caused by field stars alone (see below), and should therefore represent the
$\epsilon$\,Ori cluster isochrone. To define the position of this isochrone 
exactly, the CMD was divided in 1\,mag wide horizontal bands. The histogram of the 
R-I colours of each band shows a broad maximum at $R-I<1.0$ for the field stars 
and an additional distinct peak on the red side, indicating the position of the 
isochrone for the $\epsilon$\,Ori cluster. This position is a roughly linear 
function of the I-band magnitude. 

Having thus defined an empirical cluster isochrone, we should now estimate the 
maximum distance of a VLM cluster member from this isochrone. The errors of the
photometric calibration are $<0.25$\,mag down to the detection limit. Additionally,
we should account for the uncertainty of the isochrone definition, which we
estimate to be about 0.05\,mag. Hence, the maximum distance between
isochrone and cluster member is 0.3\,mag. Consequently, we shifted the isochrone 
0.3\,mag to the left and selected all objects to the red side of this line
as potential cluster members. This preliminary candidate list comprises
175 objects with I-band magnitudes fainter than 14\,mag and R-I colours between 1.3 and 
3.2. We did not constrain the selection on the red side, since we expect 
intrinsic reddening for at least some of our candidates, as can be concluded 
from surveys in similarly old clusters (e.g., Oliveira et al. \cite{ojk02}). 

The position of our empirical isochrone was compared with the isochrone
from the evolutionary models of Baraffe et al. (\cite{bca98}). The age of the
$\epsilon$\,Ori cluster is probably between 2 and 10\,Myr. In Fig. 
\ref{cmd} (left panel), we therefore show the Baraffe et al. isochrones
for 3, 5 and 10\,Myr (dashed lines, from right to left). At the bright end
of our selection band, the isochrones agree well with the position of our
candidates. Comparing the photometry of the brightest candidates with the models, 
we estimate an upper mass limit of roughly 0.6$\,M_{\odot}$ for our objects.
For targets fainter than $I=15$\,mag, the theoretical isochrones are clearly 
offset compared with the empirical isochrone, indicating that the measured 
R-I colours are larger than in the models. This can probably
be explained with the above noted calibration problems for red objects,
which result in too high I-band fluxes and, thus, too high R-I colours. 
Another probable reason for the discrepancy is model uncertainties, which
are considerable for optical wavelengths (Delfosse et al. \cite{dfs00}).

The next step of the candidate selection was to examine if their 
near-infrared magnitudes are in agreement with the predictions of
the Baraffe et al. models, which appear to be much more precise in the
near-infrared than in the optical regime. A large part of our
candidates could be identified in the 2MASS database\footnote{Catalogue 
available under {\it http://www.ipac.caltech.edu/2mass}},
which delivers magnitudes in the J-, H-, and K-band. We did not use
the K-band magnitudes for candidate selection, because for young
objects they could be influenced by the radiation from a circumstellar
disk, as found by Muench et al. (\cite{mal01}) and Oliveira et al. (\cite{ojk02}).

Fig. \ref{cmd} (right panel) shows the (J,J-H) CMD for all candidates
for which 2MASS photometry is available. The isochrones of Baraffe et al.
(\cite{bca98}) are again plotted as dashed lines. Our potential
cluster members clearly accumulate around all three isochrones, which are
nearly indistinguishable in this wavelength regime. Thus, the theoretical
tracks can be compared with our photometry, and they are more or less
insensitive to the exact age of the objects. There is only one clear outlier, 
which was rejected from the candidate list. Our cluster member list thus comprises 
143 candidates with photometry in five bands. We note, however, that the (J,J-H) 
CMD is not a particularly good tool to discriminate between 
cluster members and field stars, since both types of objects show no clear 
separation in this diagram. Coordinates and photometry of the 143 probable
young VLM objects are listed in Table 2.

The masses of all candidates were estimated by comparing their near-infrared
colours with the 5\,Myr isochrone from Baraffe et al. (\cite{bca98}). The
mass-magnitude relation from the theoretical isochrone was fitted with
a low degree polynomial for the J- and H-band. The resulting function was then
applied to the J- and H-band magnitudes of the candidates. We obtain two
masses, one based on J-band and the other on H-band photometry. The two
values are very similar; in Table 2 we give their average 
as our mass estimate. These masses could be systematically too high or too 
low due to model shortcomings or wrong values for distance and age
of the $\epsilon$\,Ori cluster. Recent studies indeed find evidence 
that the Baraffe et al. masses are systematically too high for low 
mass brown dwarfs, but they are in good agreement with observed values for 
VLM stars (Mohanty et al. \cite{mjb04}). Nevertheless, for relative
comparisons these mass estimates are still useful.

\begin{figure}[h]
\centering
\resizebox{\hsize}{!}{\includegraphics[angle=-90,width=6.5cm]{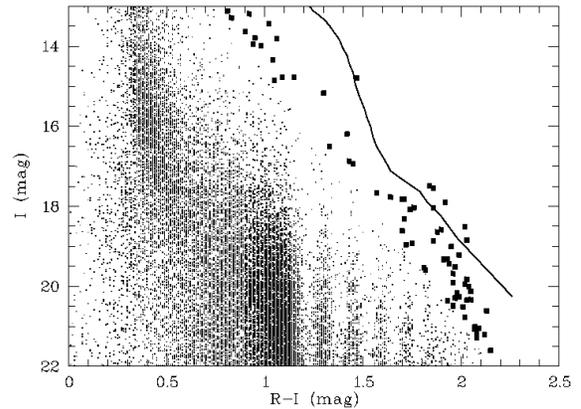}}
\caption{Simulated (I,R-I) CMD for a 1\,sq. deg. field centred on $\epsilon$\,Ori.
The solid lines shows the 5\,Myr Baraffe et al. (\cite{bca98}) isochrone. All
objects in a 0.3\,mag wide band around this line are marked as larger dots.}
\label{contam}
\end{figure}

The rate of contaminating objects can be estimated by comparing our 
(I,R-I) CMD with a diagram which contains {\it only} field stars. Such
a diagram can be simulated using the Besan\c{c}on Galaxy model (Robin \& Cr\'ez\'e 
\cite{rc86}, Robin et al. \cite{rrd03}). With this model we generated an artificial 
(I,R-I) CMD of a 1\,sq. deg. field towards $\epsilon$\,Ori.\footnote{This simulation 
was done with the 1994 version of the models, available under 
{\it http://www.obs-besancon.fr/www/model1994/}}
The result is shown in Fig. \ref{contam} together with the 5\,Myr isochrone
of Baraffe et al. (\cite{bca98}). There are 72 objects in a 0.3\,mag wide
band around this isochrone, which resembles the selection band we used
in the real CMD. An equivalent number of field stars should have been picked 
up by our candidate selection. Scaling to the WFI field, the number of 
contaminating stars is 22, giving a contamination rate of 16\%. A definite 
confirmation of cluster membership of our objects needs follow-up spectroscopy,
which is not yet available. Thus, we leave the detailed discussion of this 
object sample for future work. In this paper, we will use these candidates as
target sample for the variability study.

%5.9 vor journal version, 4.4 for referee version
\begin{figure*}[t]
\resizebox{5.9cm}{!}{\includegraphics[angle=-90,width=6.5cm]{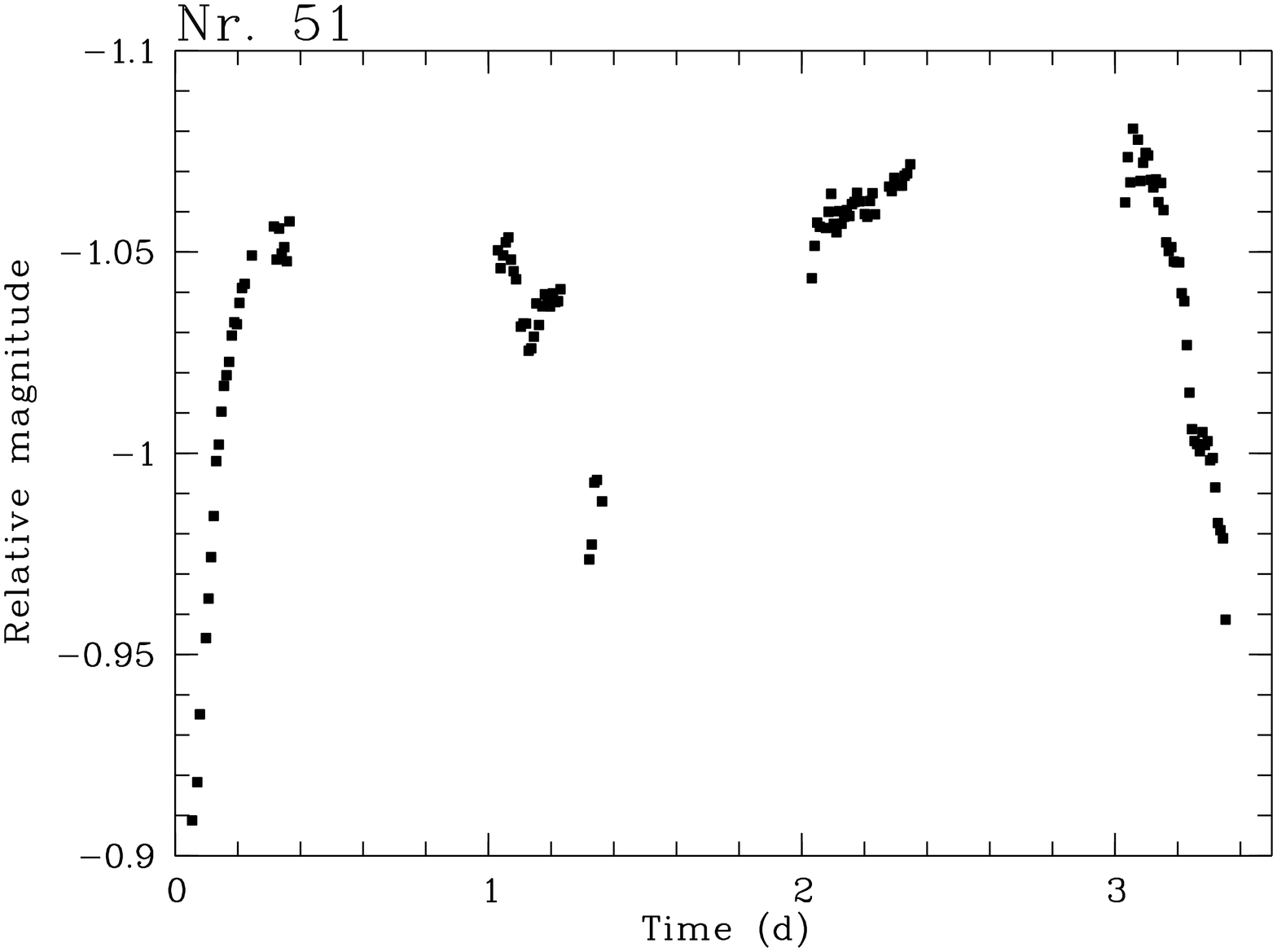}} \hfill %no 51
\resizebox{5.9cm}{!}{\includegraphics[angle=-90,width=6.5cm]{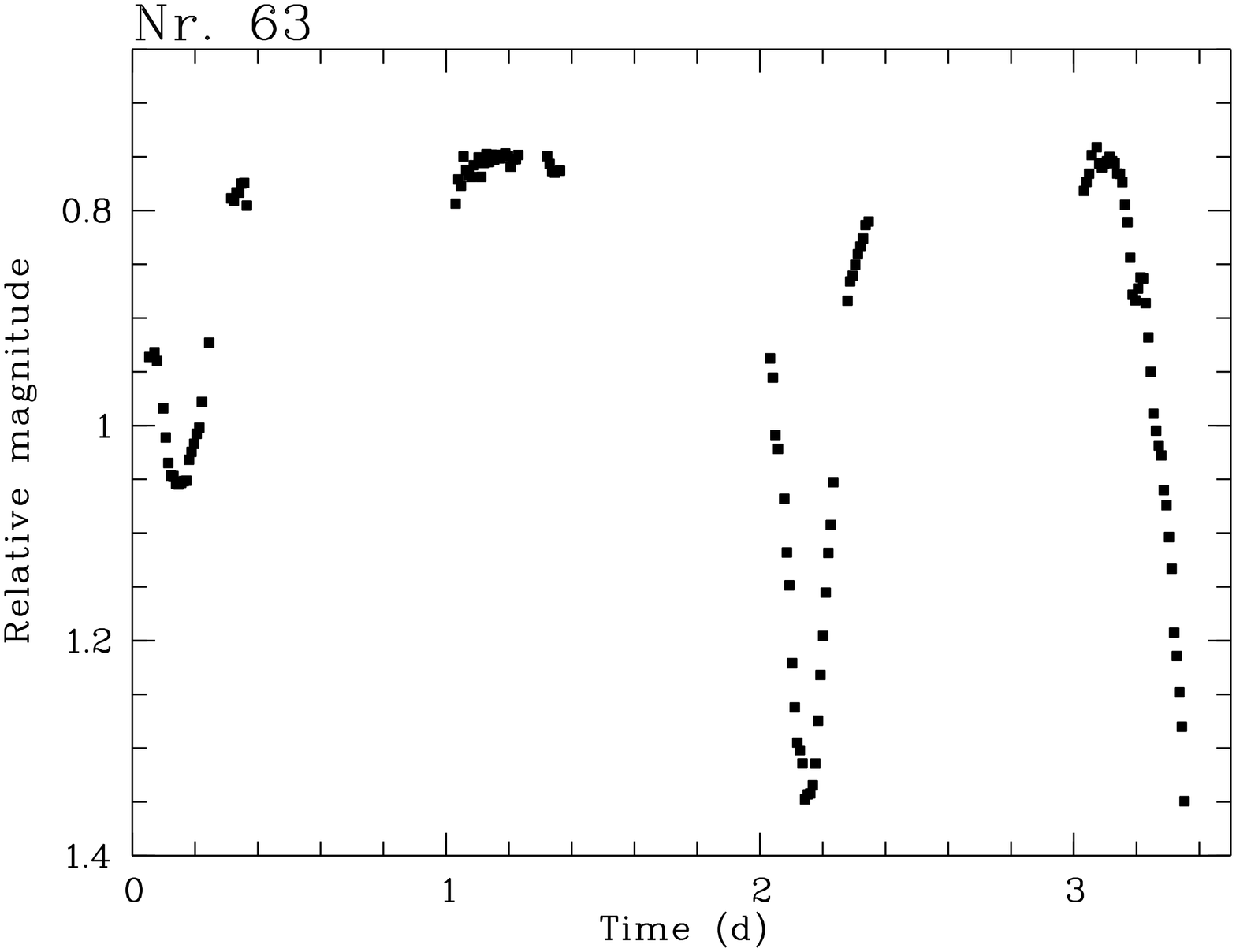}} \hfill %no 63
\resizebox{5.9cm}{!}{\includegraphics[angle=-90,width=6.5cm]{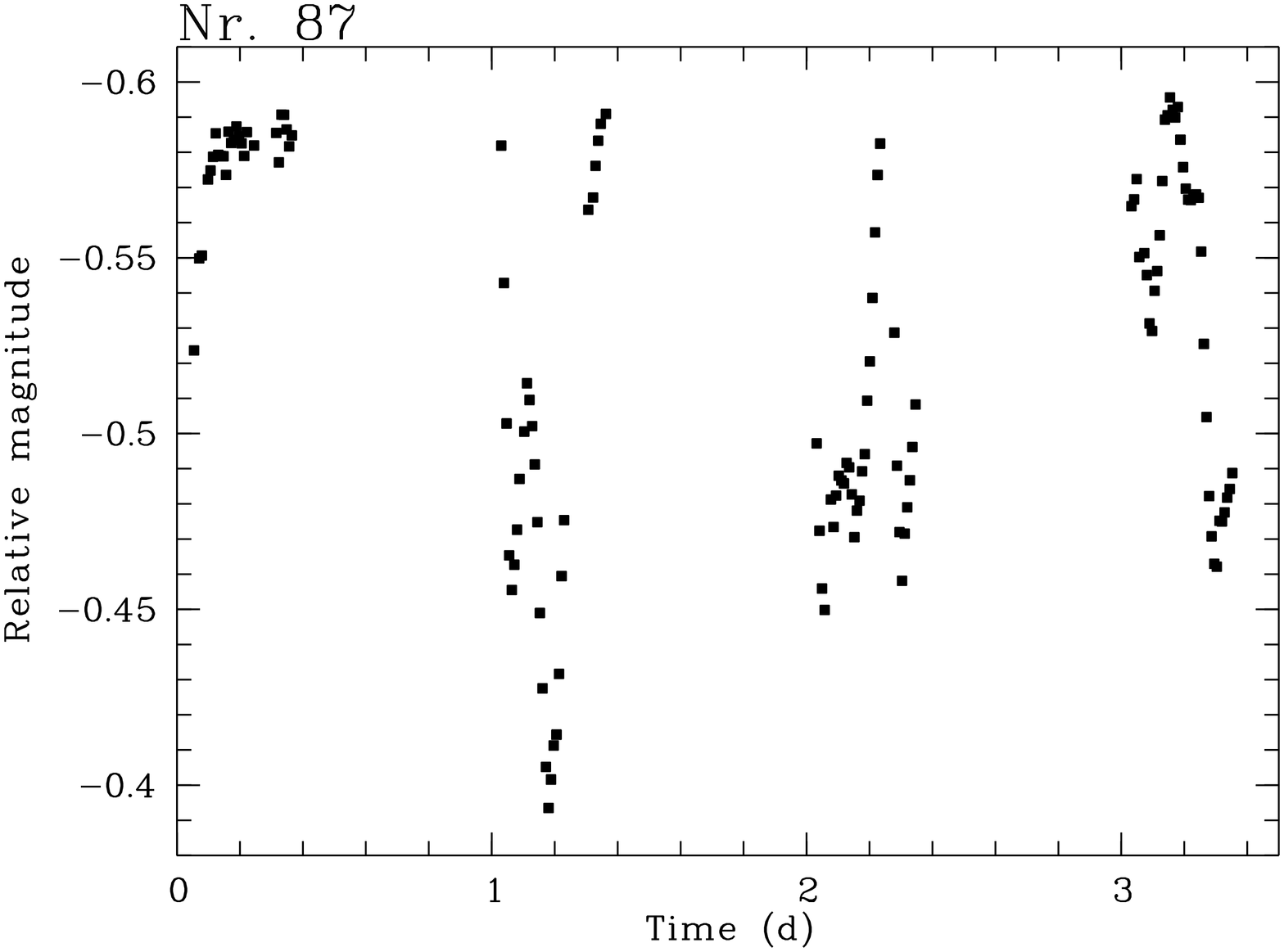}} \\ %no 87
\resizebox{5.9cm}{!}{\includegraphics[angle=-90,width=6.5cm]{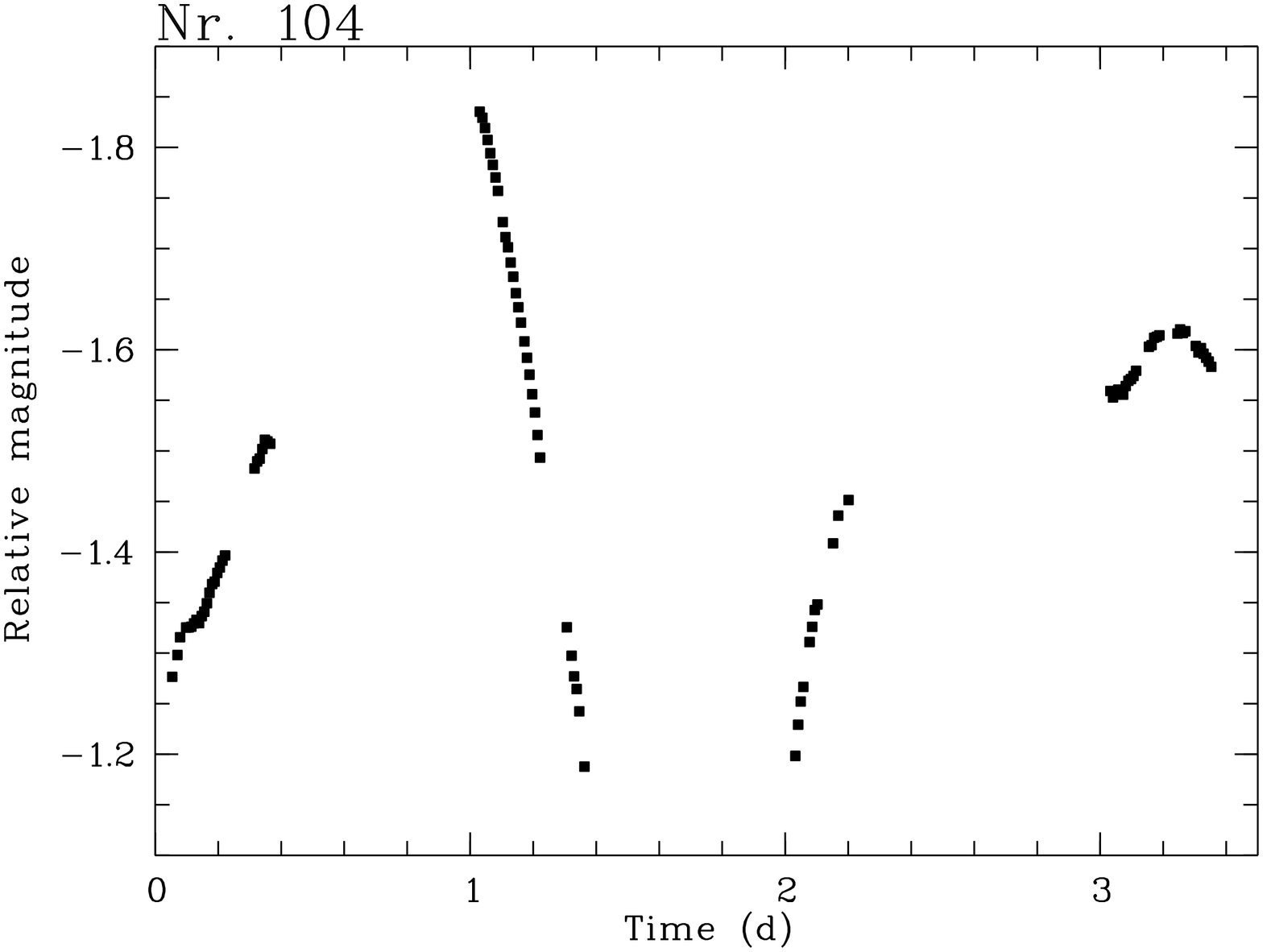}} \hfill %no 104
\resizebox{5.9cm}{!}{\includegraphics[angle=-90,width=6.5cm]{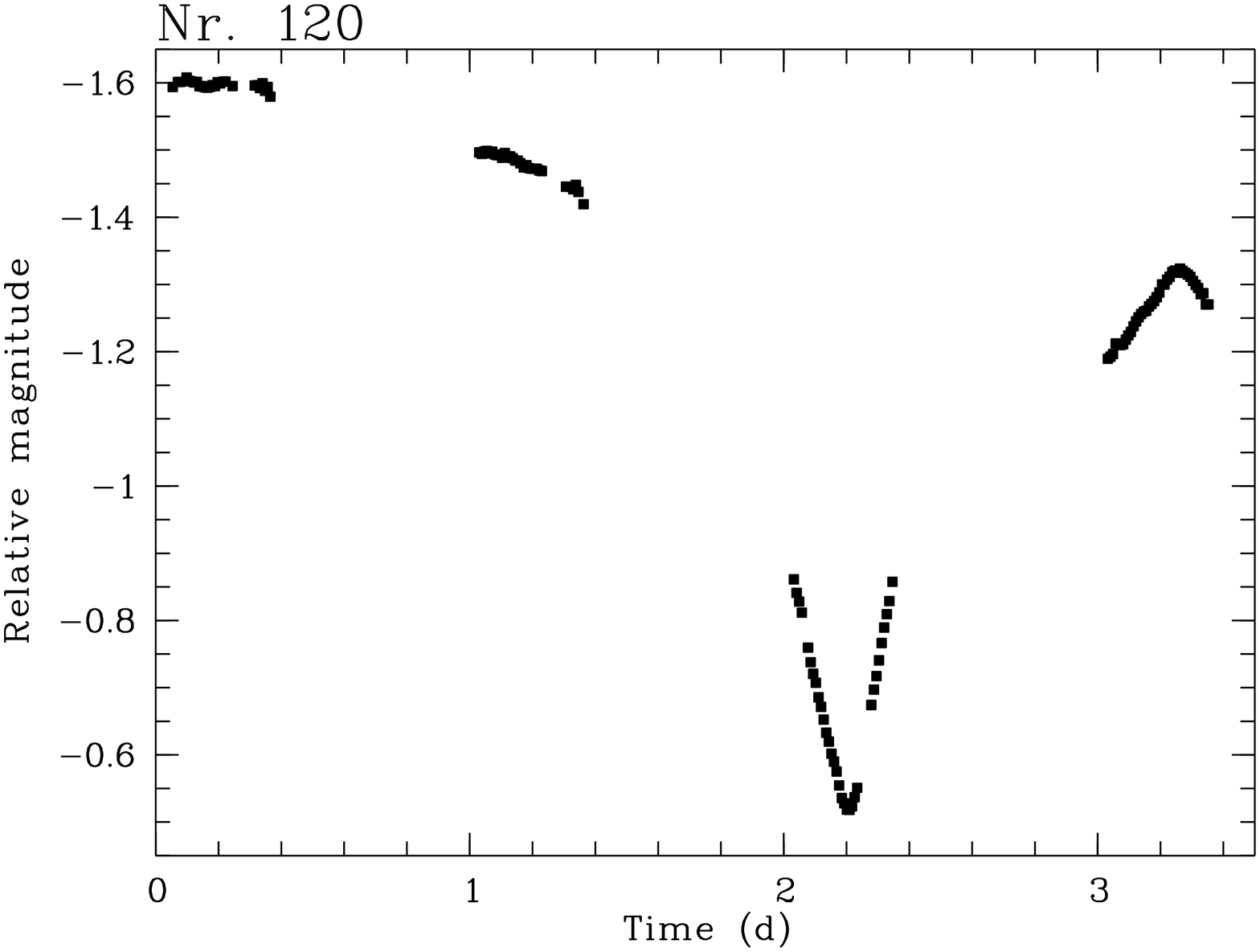}} \hfill %no 120
\resizebox{5.9cm}{!}{\includegraphics[angle=-90,width=6.5cm]{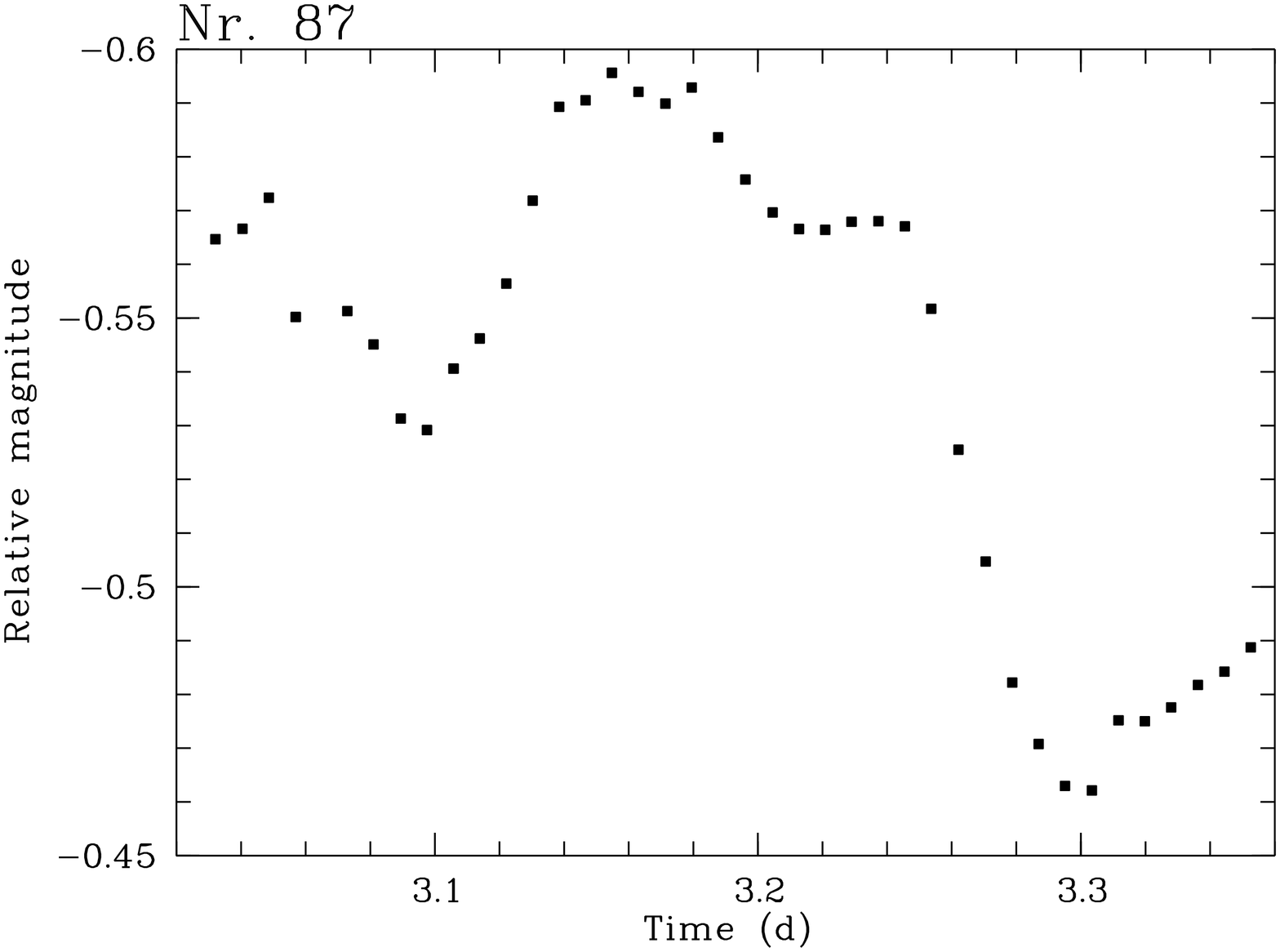}} \\ %no 87
\caption{Light curves of objects with high-amplitude, partly irregular variations. 
Five panels show the complete light curves for objects no. 51, 63, 87, 104, 120, which 
show strong variations on timescales of a few hours. The lower right panel 
shows a part of the light curve of object no. 87, whose brightness varies very rapidly 
on timescales below one hour. Plotted are only the data points for the third observing
night.}
\label{strange}
\end{figure*}

\section{Time series analysis}
\label{tsa}

Our time series analysis was focused on the search for photometric
periods. Prior to the period search, all light curves were inspected
visually to detect obvious signs of variability. In particular, we 
looked for rapid and/or irregular variations, which would not have
been detected by the period search.

\begin{figure}[t]
\centering
\resizebox{\hsize}{!}{\includegraphics[angle=-90,width=6.5cm]{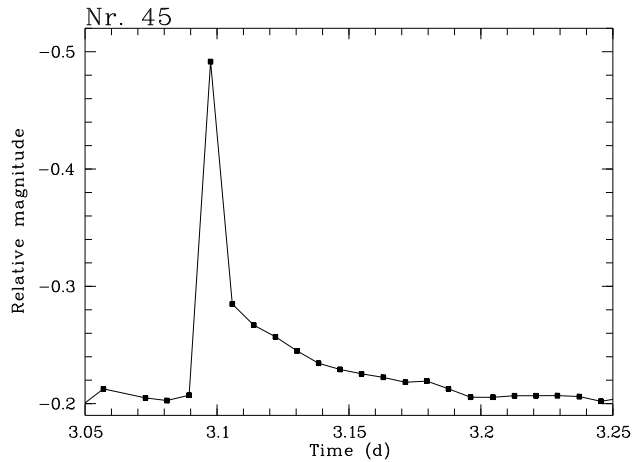}}
\caption{Flare event in the light curve of object no. 45. Shown are only
the data points from the 20th December, the complete light curve is plotted
in Fig. \ref{phase1}.}
\label{flareev}
\end{figure}

A large fraction of our candidates shows clear signs of variability, 
which confirms that we have identified a pre-main sequence population.
In many cases, periodic behaviour is apparent. One target (no.
45) shows a huge eruptive event with a rapid 0.3\,mag brightness
increase followed by an exponential decline to the normal level,
the typical characteristics of a flare. Its light curve is plotted
in Fig. \ref{flareev}. The data points affected by this flare were
excluded for the remaining analysis. Four other targets show 
variability with high amplitudes, which could contain a periodic 
component, but additionally there are irregular fluctuations on timescales
of roughly one day. Finally, there is one suspicious target (no. 87) whose
light curve shows rapid, irregular variations with an amplitude
of 0.2\,mag. All these unusual light curves are shown in Fig. 
\ref{strange}. A discussion of the origin of this behaviour follows 
in Sect. \ref{ori}.

We did not attempt to examine formally the generic
variability by analysing the rms of the light curves with
respect to their photometric errors, mainly because of the difficulties
to estimate the latter. As noted in the previous sections,
the time series images still contain remnants of the fringe
structures with amplitudes up to 1\% (see Sect. \ref{obs}). As a
consequence, the background level and the background slope varies 
around many objects, depending on the quality of our fringe 
correction. In regions with strong fringing, the distances between 
fringe maxima and fringe minima are rather small, so that the small
positional offsets between our time series images are sufficient 
to allow the objects to move significantly with respect to the
fringe pattern. Moreover, it is not guaranteed that the background
around the object is exactly the same as the background at the 
position of the object. It needs to be stressed that that both effects
are not constant for all objects, they will be enhanced in regions
with strong fringing and negligible in regions without fringing.
Therefore, additional noise of up to 1\% may be introduced for objects 
which lie in regions of strong fringing. A precise determination 
of the photometric error for each individual object, which is 
indispensable for a generic variability analysis, is thus 
not possible. Since the {\it mean} error of our relative photometry
is below 1\% for many objects, the spatially variable 'fringe' noise 
probably causes a considerable fraction of the total noise.
For the period analysis, which will be described in the 
following section, the fringe remnants are not a problem, since they 
could cause only small-scale stochastic brightness variations.

\subsection{Period search}
\label{per}

Our period analysis is based on periodogram techniques, which are
to be preferred over the string length method and other phase-space
based routines, as we have shown in SE2. We start with the Scargle 
periodogram (Scargle \cite{s82}). If a light curve produces a significant 
peak in the Scargle periodogram, this suspected periodicity is examined 
with several independent procedures, including the application of the 
CLEAN algorithm (Roberts et al. \cite{rld87}) and comparison with 
the light curve of nearby reference stars. These tests should ensure that 
the periodicity is intrinsic to the target and that it is not an artefact of 
the window function of the data. Prior to the period analysis, all light curves 
were filtered, so that 3$\sigma$ outliers were excluded. A period was only 
accepted if the following criteria were fulfilled:

\begin{itemize}
\item{The Scargle periodogram shows a significant peak, where
significant means that the False Alarm Probability (FAP) calculated
following Horne \& Baliunas (\cite{hb86}) is below 1\% (see below
for a discussion of FAP).}
\item{The periodograms of at least ten nearby reference stars do not
contain a significant peak at the same position.}
\item{The phased light curve of the candidate shows the period
clearly. (This subjective criterion has to be used with caution,
since the high number of data points allows a reliable period
detection even when the noise level is relatively high (see Sect. 
\ref{sens}).}
\item{The scatter in the light curve is significantly reduced
after subtraction of the sine-wave-approximated period. (For this
test, we used the statistical F-test, which is well-suited for the
comparison of scatter in data.)}
\item{The light curve of nearby reference stars phased to the
period of the target object do not show the same periodicity.}
\item{The CLEANed periodogram confirms the peak, i.e. it is not
caused by the window function.}
\item{The empirical FAP, determined with the bootstrap algorithm
(see below), is below 1\%.}
\end{itemize}
For more details about these criteria, we refer to SE1.

Reliable False Alarm Probabilities for unevenly sampled light curves 
can only be determined with simulations. However, the peak height
in the Scargle periodogram can be translated into an estimate of the
FAP, as shown by Horne \& Baliunas (\cite{hb86}). The crucial point then is to
choose an appropriate value for the number of independent frequencies
$N_i$. For regularly sampled data, $N_i$ is usually somewhat larger
than the number of data points $N$. With clumped data points, as in our case,
$N_i$ decreases considerably (see Horne \& Baliunas \cite{hb86}). 
For a first estimate of the FAP from the Scargle periodogram alone, 
we used $N_i = N/2$.

\begin{table}[h]
\caption{Candidates with significant periodic variability (see text for
explanations).}
\begin{tabular}{cclcccc} \hline
No. & M($M_{\odot}$) & P(h) & $\Delta$P(h) & A(mag) & FAP$\mathrm{_{E}}$(\%) & N\\
\hline
~~~~4 &   	 0.17 &  ~~56.1 &  2.33 &  0.027  & $<0.01$  & 128 \\ % 6503
 ~~17 &   	 0.05 &  ~~9.46 &  0.14 &  0.027  & $<0.01$  & 128 \\ % 6601
 ~~20 &   	 0.08 &  ~~35.2 &  0.98 &  0.034  & $<0.01$  & 129 \\ % 6705
 ~~24 &   	 0.22 &   100.0 &  10.5 &  0.023  & $<0.01$  & 124 \\ % 6622
 ~~26 &   	 0.27 &  ~~18.0 &  0.47 &  0.037  & $<0.01$  &  97 \\ % 4622
 ~~28 &   	 0.21 &  ~~96.8 &  8.50 &  0.042  & $<0.01$  & 117 \\ % 5394
 ~~29 &   	 0.27 &   100.4 &  11.3 &  0.027  & $<0.01$  &  98 \\ % 5762
 ~~39 &   	 0.06 &  ~~19.2 &  0.34 &  0.038  & $<0.01$  & 123 \\ % 8548
 ~~44 &   	 0.12 &  ~~31.0 &  1.78 &  0.027  & $<0.01$  & 128 \\ % 4021
 ~~45 &   	 0.09 &  ~~14.6 &  0.42 &  0.022  & $<0.01$  & 118 \\ % 8185
 ~~51 &   	 0.16 &  ~~15.7 &  0.51 &  0.073  & $<0.01$  & 127 \\ % 7895
 ~~56 &   	 0.12 &  ~~33.7 &  1.09 &  0.030  & $<0.01$  & 129 \\ % 3361
 ~~60 &   	 0.12 &   102.1 &  11.6 &  0.017  & $<0.01$  & 128 \\ % 7663
 ~~61 &   	 0.18 &  ~~40.1 &  0.72 &  0.069  & $<0.01$  & 118 \\ % 7723
 ~~63 &   	 0.04 &  ~~15.5 &  0.31 &  0.426  & $<0.01$  & 128 \\ % 7912
 ~~65 &   	 0.02 &  ~~4.70 &  0.08 &  0.049  & $<0.01$  & 126 \\ % 9111
 ~~69 &   	 0.12 &  ~~3.80 &  0.05 &  0.020  & ~~0.01   & 112 \\ % 7694
 ~~70 &   	 0.02 &  ~~5.79 &  0.10 &  0.024  & $<0.01$  & 127 \\ % 2591
 ~~80 &   	 0.22 &  ~~45.1 &  2.70 &  0.023  & $<0.01$  & 127 \\ %10184
 ~~81 &   	 0.19 &  ~~30.0 &  0.94 &  0.029  & $<0.01$  & 125 \\ % 2857
 ~~83 &   	 0.18 &  ~~38.5 &  1.75 &  0.027  & $<0.01$  & 115 \\ % 1850
 ~~84 &   	 0.14 &  ~~11.4 &  0.23 &  0.020  & $<0.01$  & 127 \\ %10369
 ~~93 &   	 0.05 &  ~~87.6 &  9.58 &  0.130  & $<0.01$  & 129 \\ %10466
  110 &   	 0.11 &  ~~35.9 &  2.31 &  0.025  & $<0.01$  & 129 \\ %11329
  113 &   	 0.06 &  ~~16.9 &  0.52 &  0.093  & $<0.01$  & 125 \\ %12287
  118 &   	 0.19 &  ~~98.0 &  4.85 &  0.089  & $<0.01$  & 129 \\ %12068
  120 &   	 0.29 &  ~~82.5 &  2.64 &  0.952  & $<0.01$  & 129 \\ %10860
  123 &   	 0.05 &  ~~34.3 &  1.32 &  0.037  & $<0.01$  & 129 \\ %  845
  126 &   	 0.04 &  ~~4.06 &  0.05 &  0.016  & $<0.01$  & 128 \\ % 1212
  132 &   	 0.11 &  ~~32.3 &  1.84 &  0.022  & $<0.01$  & 128 \\ %11111
\hline				     	      
\end{tabular}	
\label{periods}		     		     
\end{table}		

The final FAP were then determined using the bootstrap approach 
(see, e.g., Kuerster et al. \cite{ksc97}). The basic idea is to 
produce 10000 randomised data sets by retaining the observing times 
and shuffling the original data values. The resulting randomised 
light curves have exactly the same sampling and the same noise level 
as the original time series. For all these random light curves, we
computed the Scargle periodogram and recorded the power of the 
highest peak. The fraction of data sets for which this value
exceeds the peak height in the original light curve is the empirical
FAP for the detected period. This time-intensive simulation has to
be done for each target individually to account for the slightly
different sampling caused by our filtering. Therefore, a first
quick estimate of the FAP was obtained only from the peak height
in the Scargle periodogram, as described above. It turned out that
this Scargle FAP is comparable to the empirical FAP from the bootstrap
algorithm, if we choose $N_i= N/2$, as given above. These bootstrap
results might underestimate the FAP in a few cases, in particular
when additional short-term fluctuations are present in the light curves,
but this effect is probably not significant (see SE1 for a detailed
discussion).

\begin{figure*}[htbd]
\resizebox{5.9cm}{!}{\includegraphics[angle=-90]{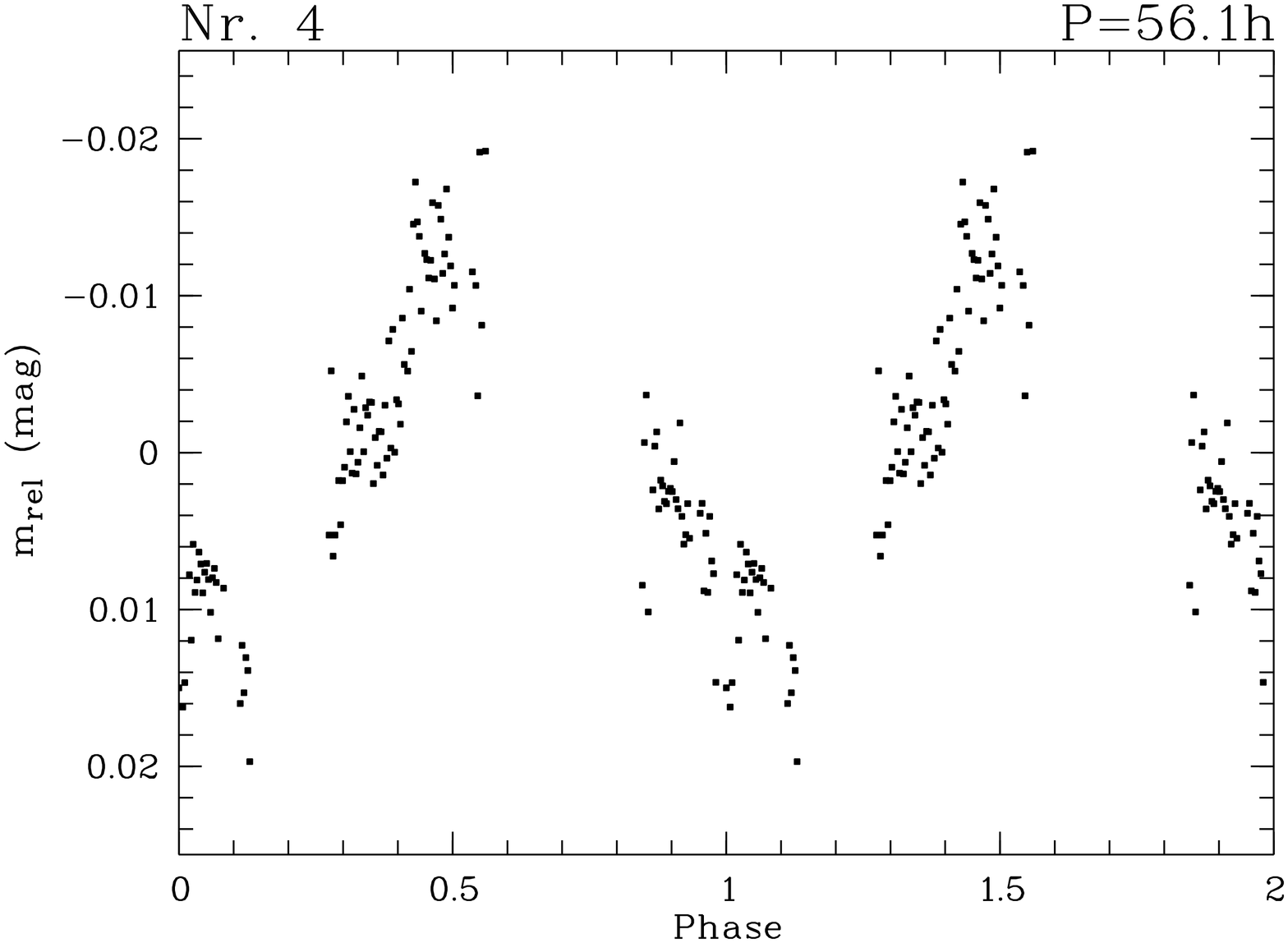}} \hfill %pp6503
\resizebox{5.9cm}{!}{\includegraphics[angle=-90]{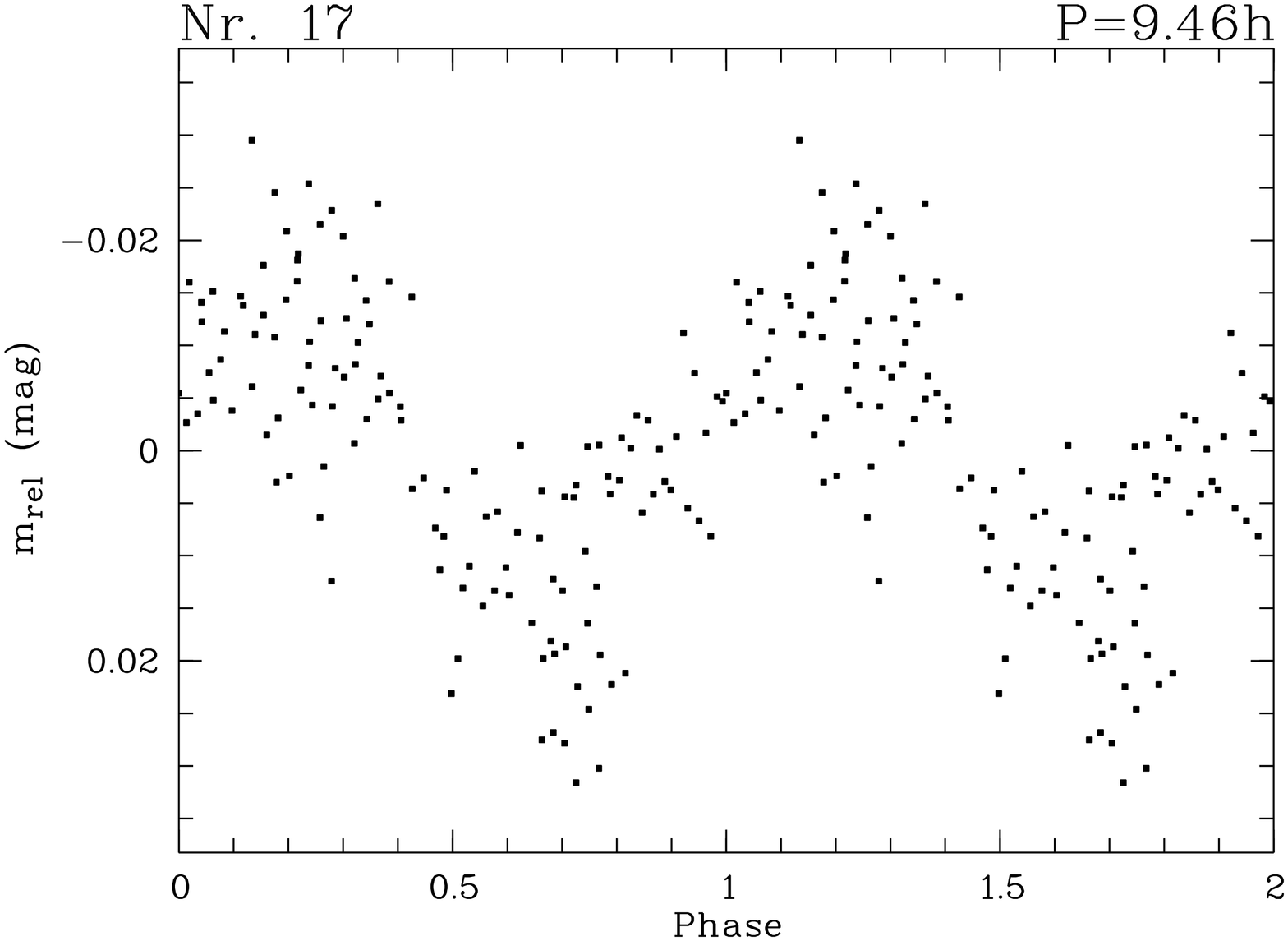}} \hfill %pp6601
\resizebox{5.9cm}{!}{\includegraphics[angle=-90]{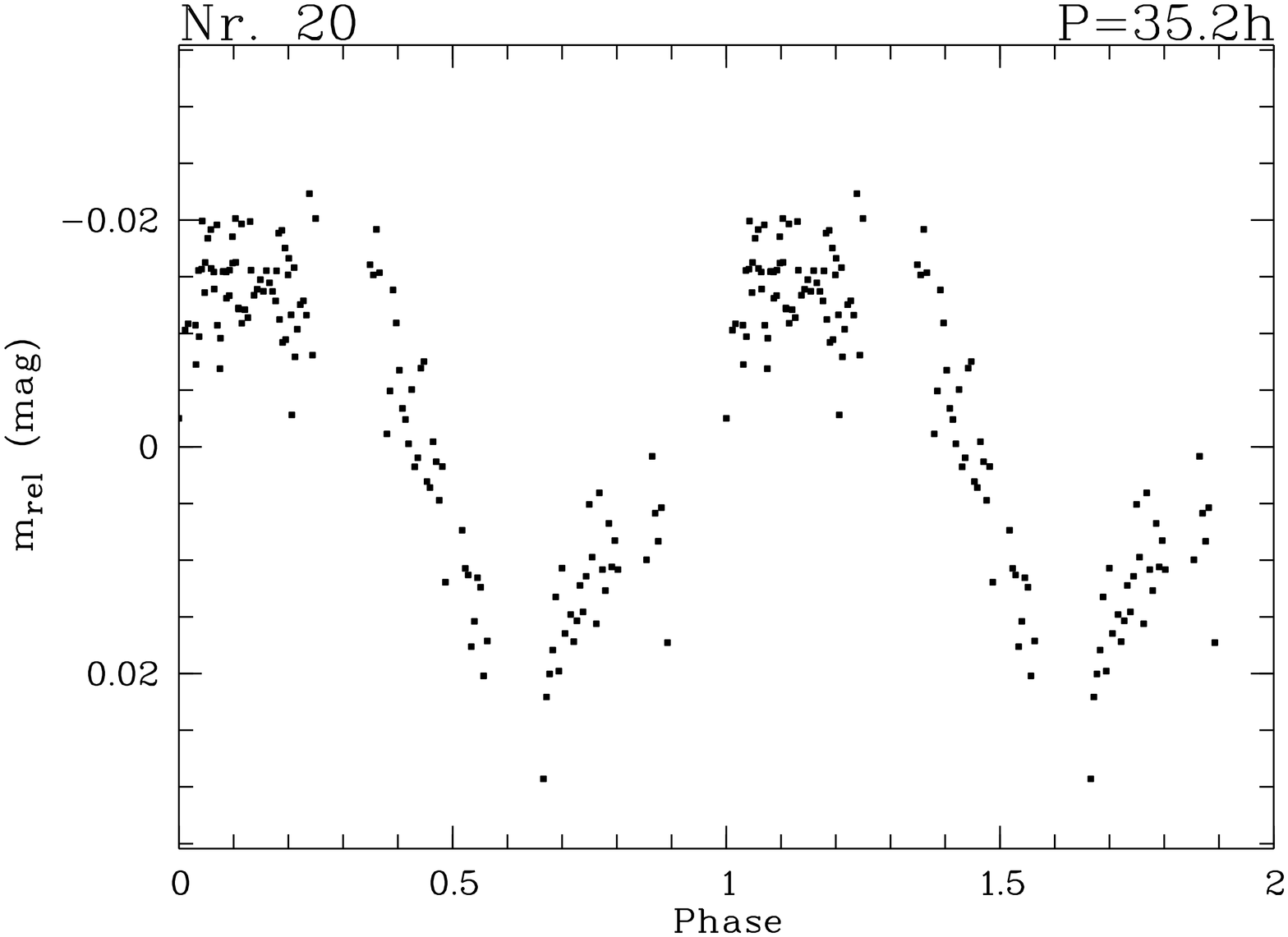}} \\     %pp6705
\resizebox{5.9cm}{!}{\includegraphics[angle=-90]{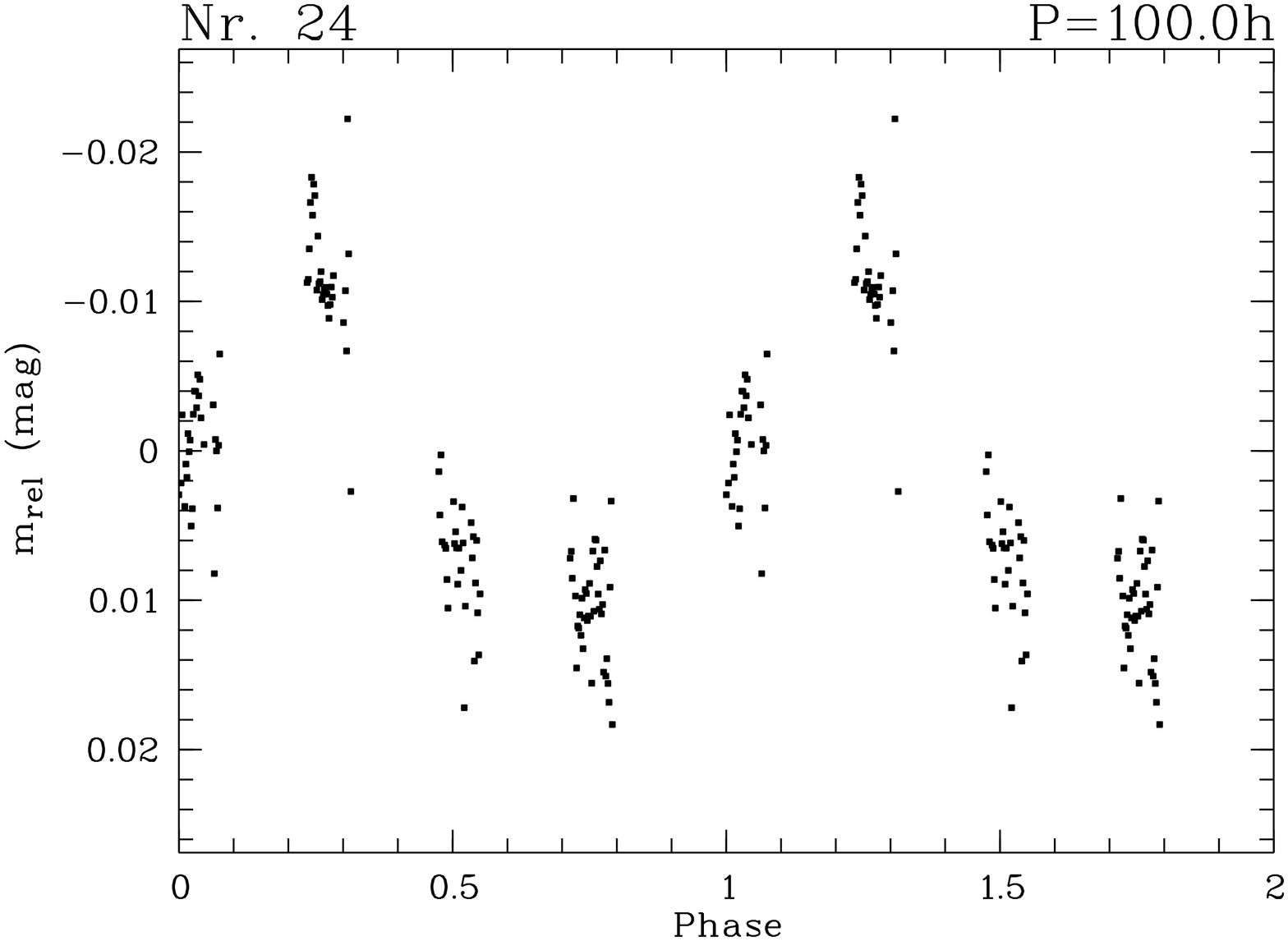}} \hfill %pp6622
\resizebox{5.9cm}{!}{\includegraphics[angle=-90]{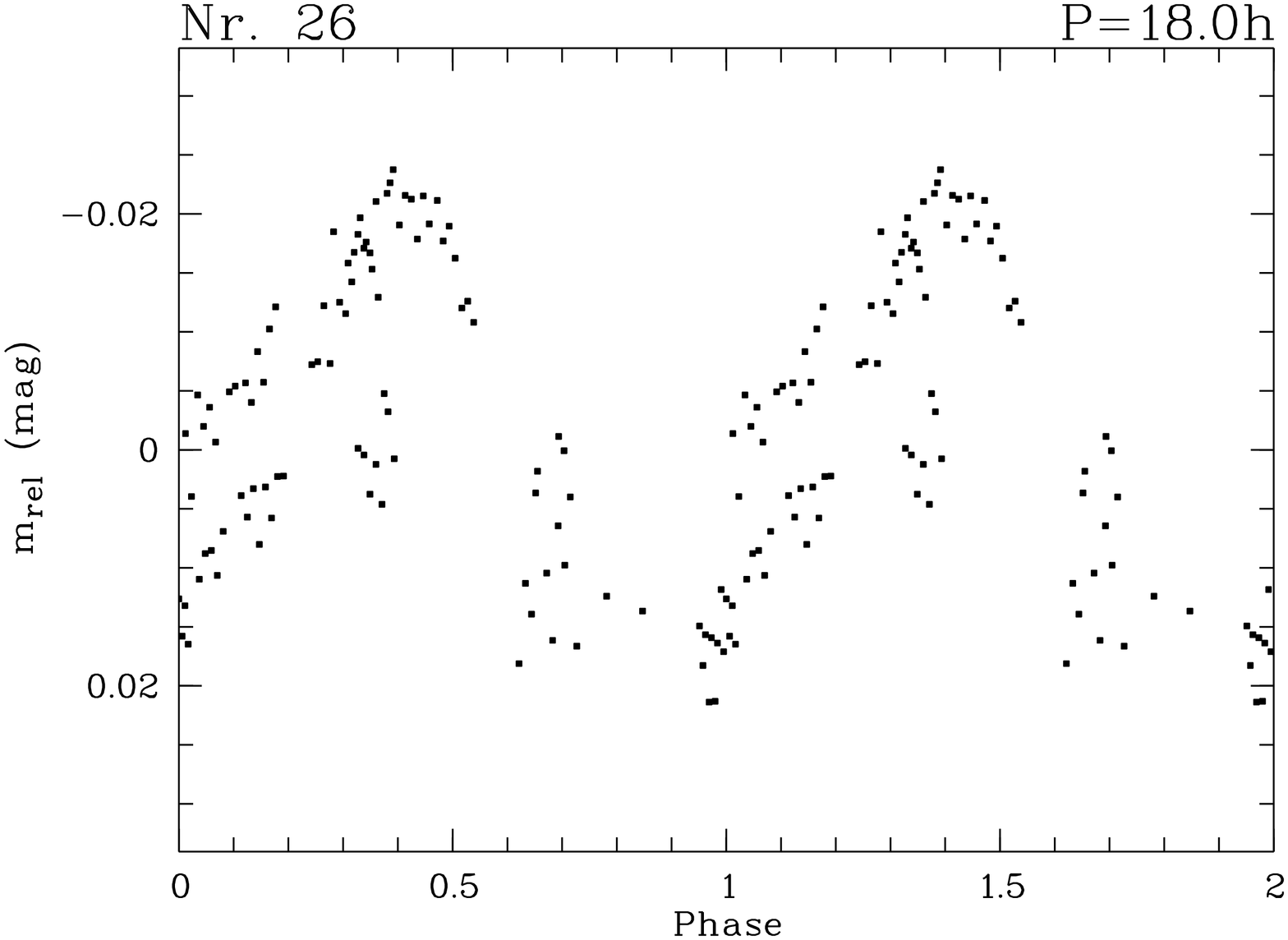}} \hfill %pp4622
\resizebox{5.9cm}{!}{\includegraphics[angle=-90]{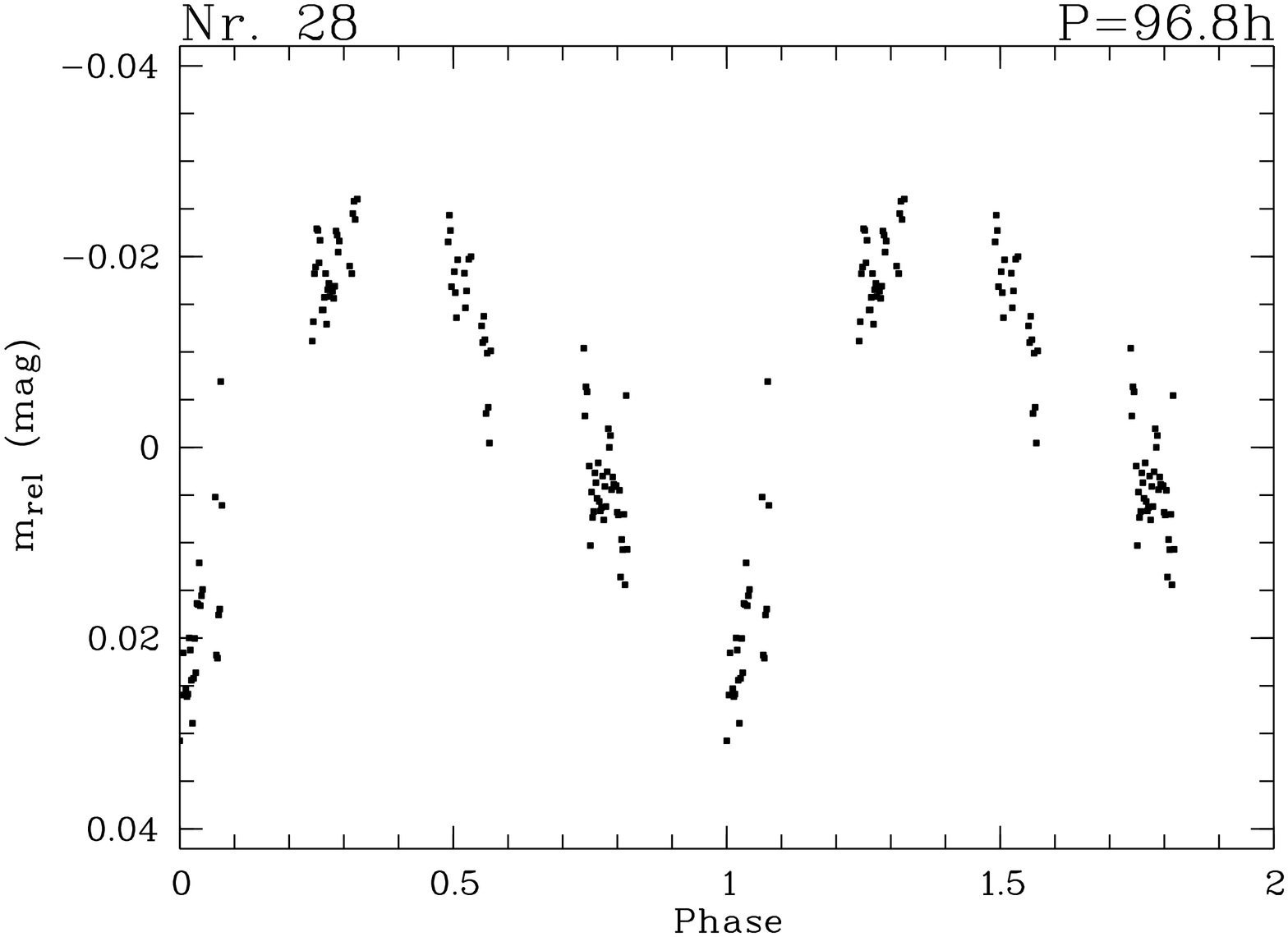}} \\     %pp5394
\resizebox{5.9cm}{!}{\includegraphics[angle=-90]{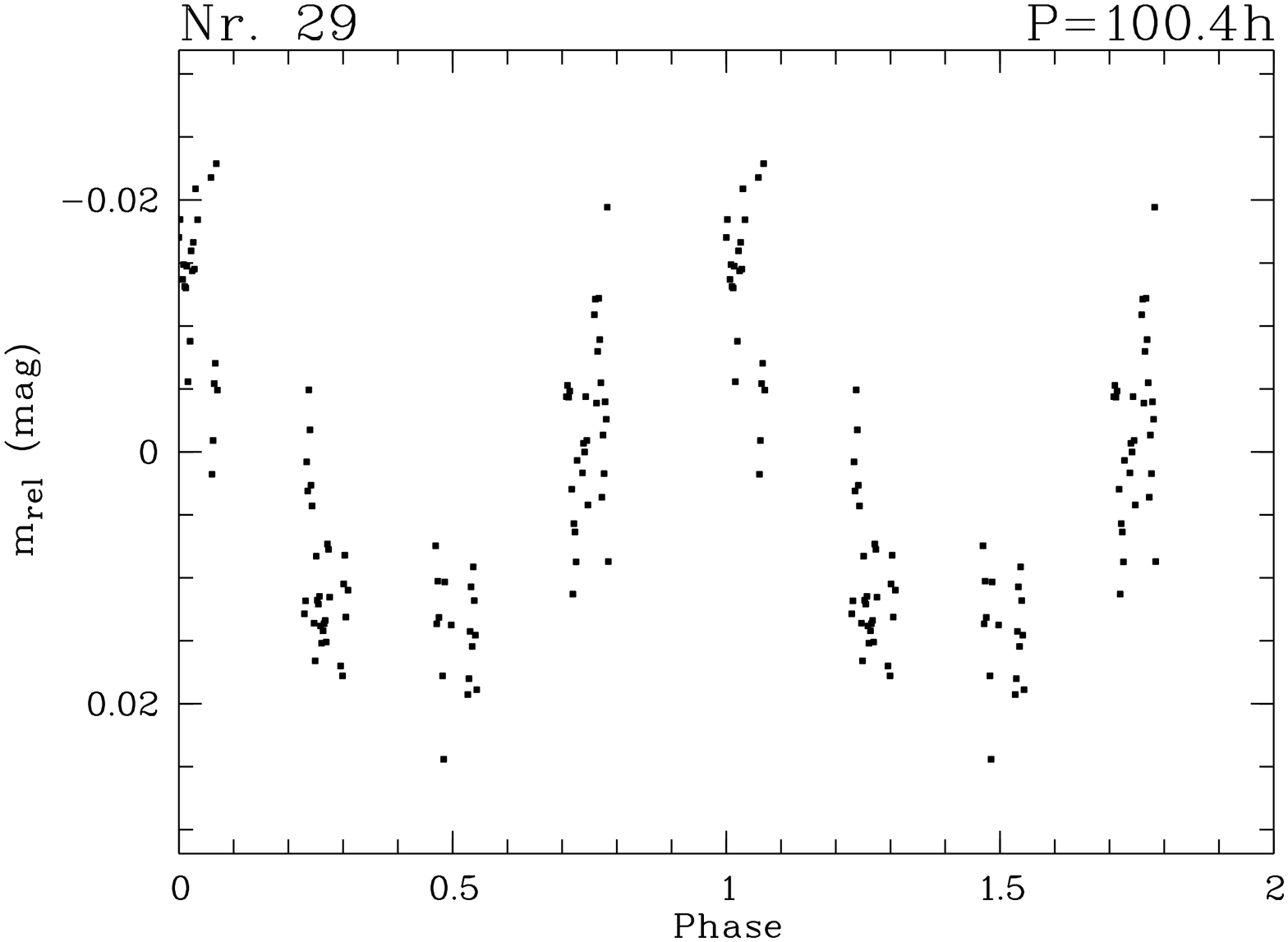}} \hfill %pp5762
\resizebox{5.9cm}{!}{\includegraphics[angle=-90]{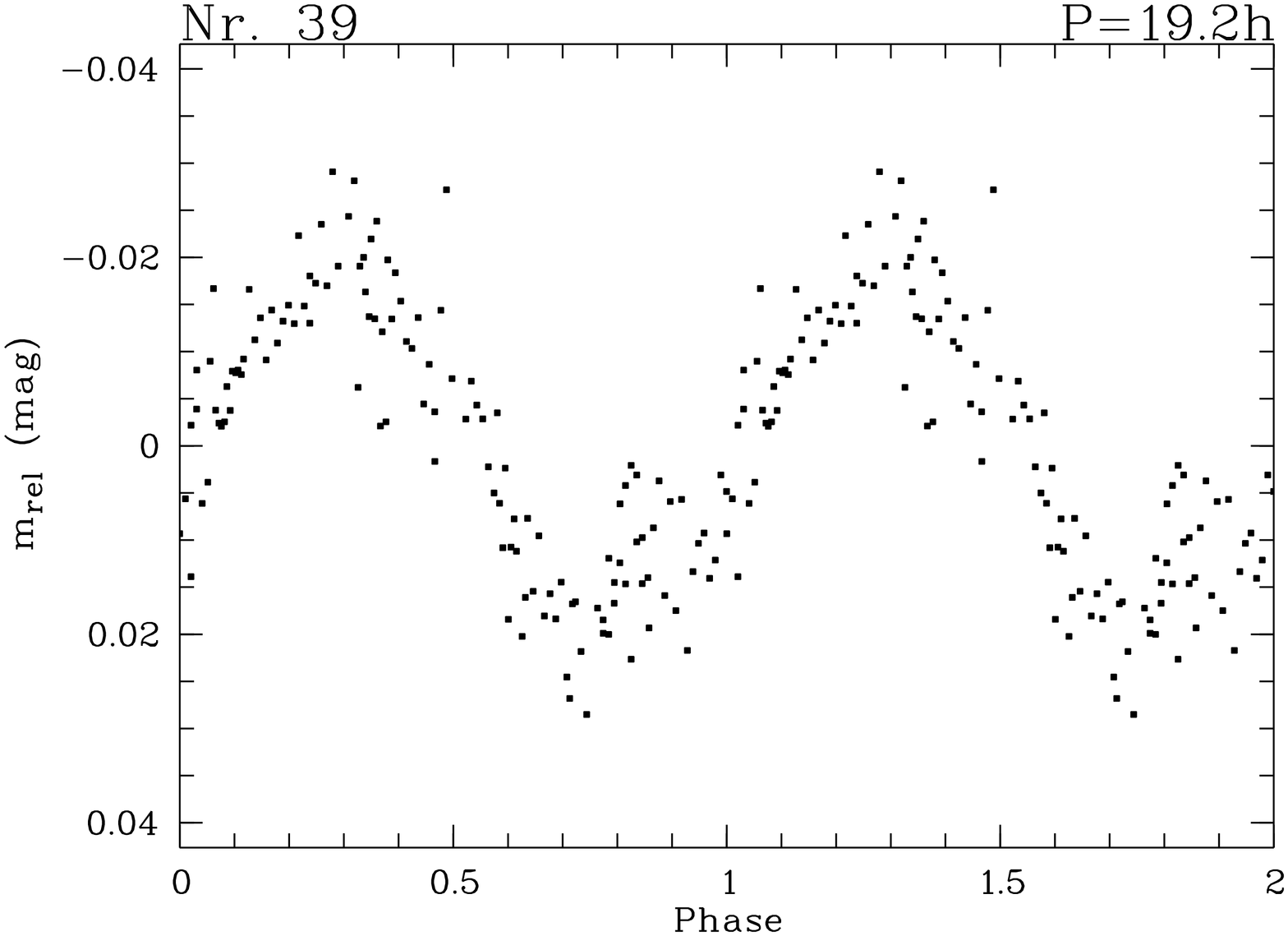}} \hfill %pp8548
\resizebox{5.9cm}{!}{\includegraphics[angle=-90]{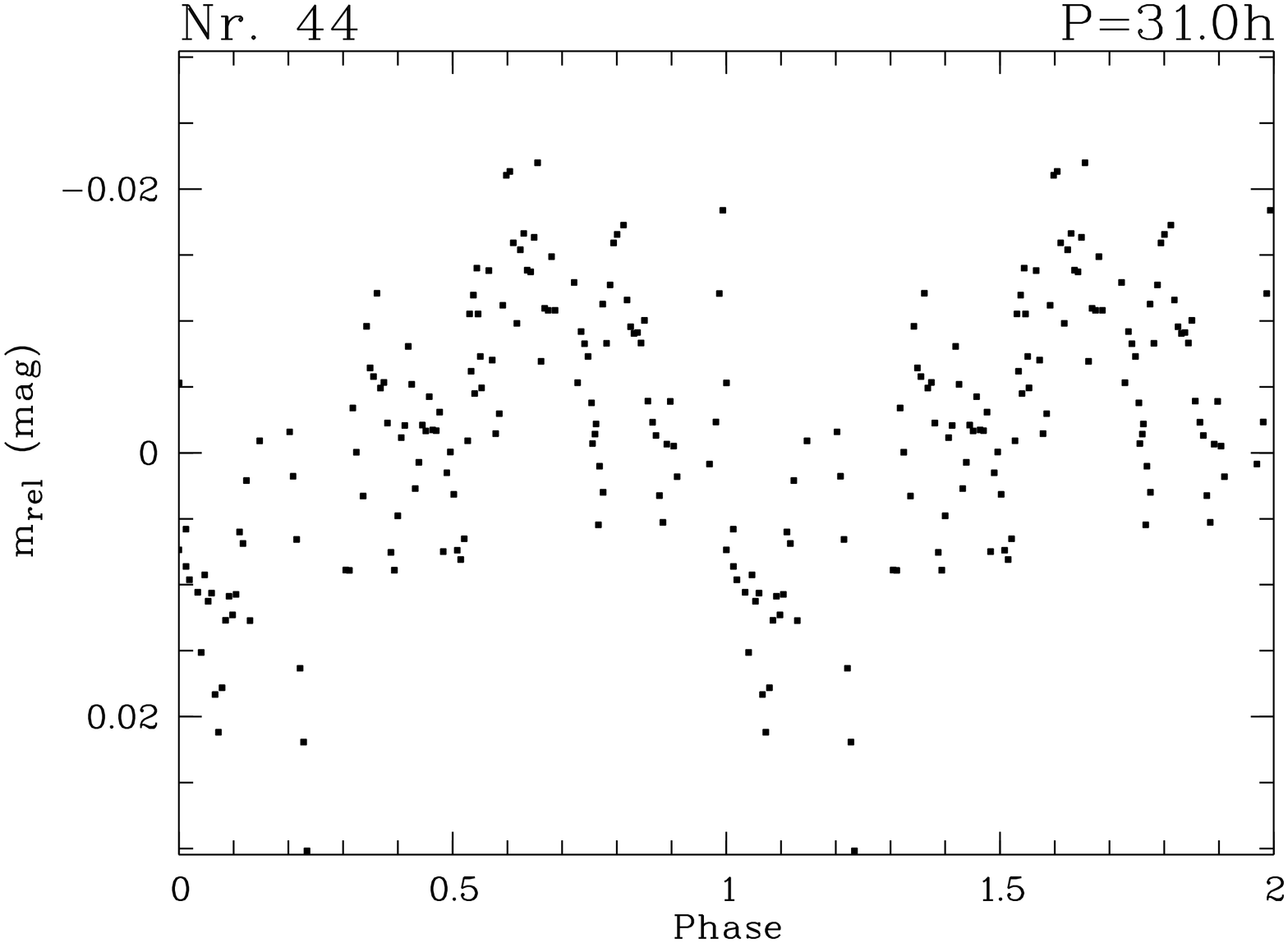}} \\     %pp4021
\resizebox{5.9cm}{!}{\includegraphics[angle=-90]{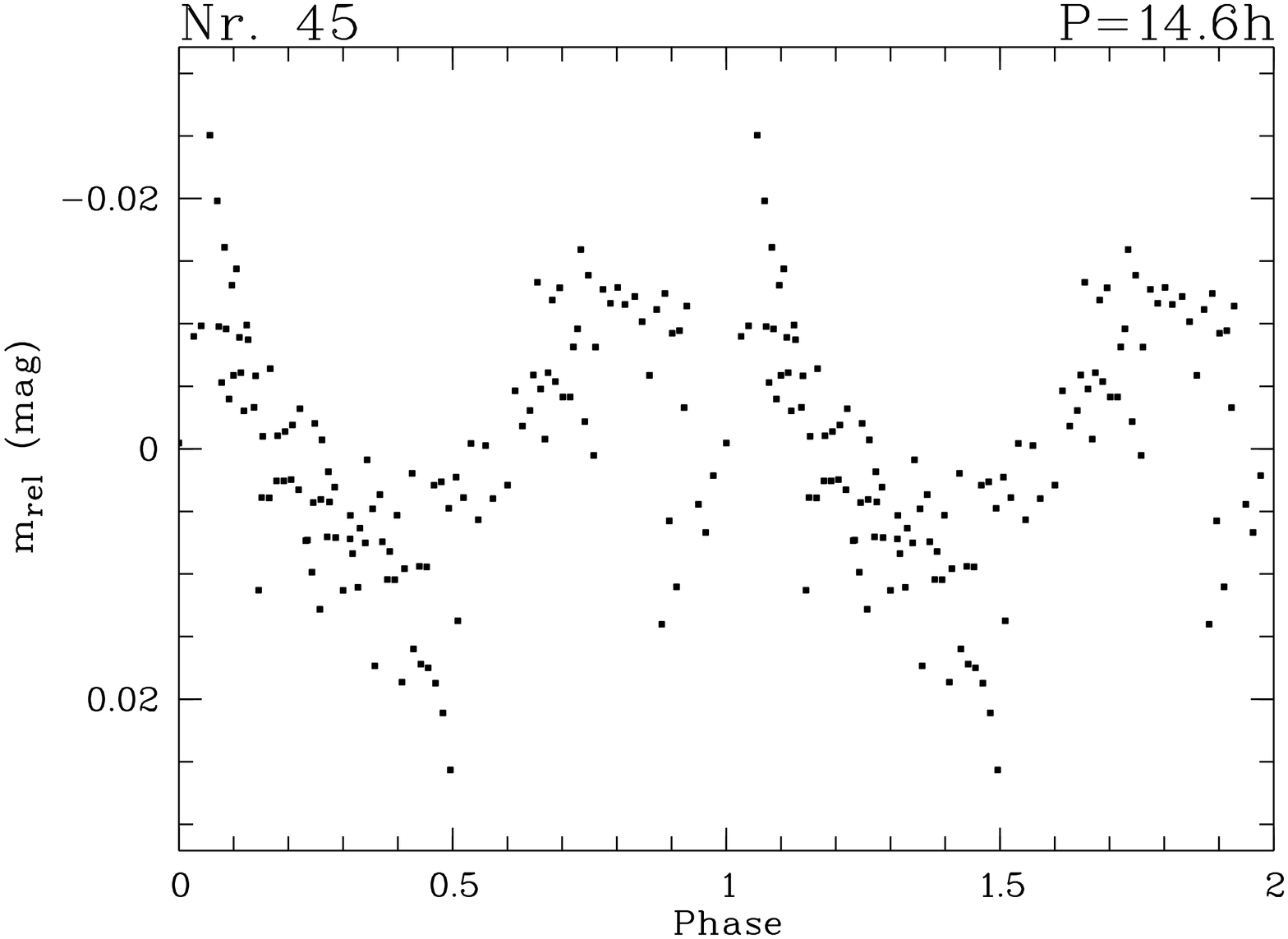}} \hfill %pp8185
\resizebox{5.9cm}{!}{\includegraphics[angle=-90]{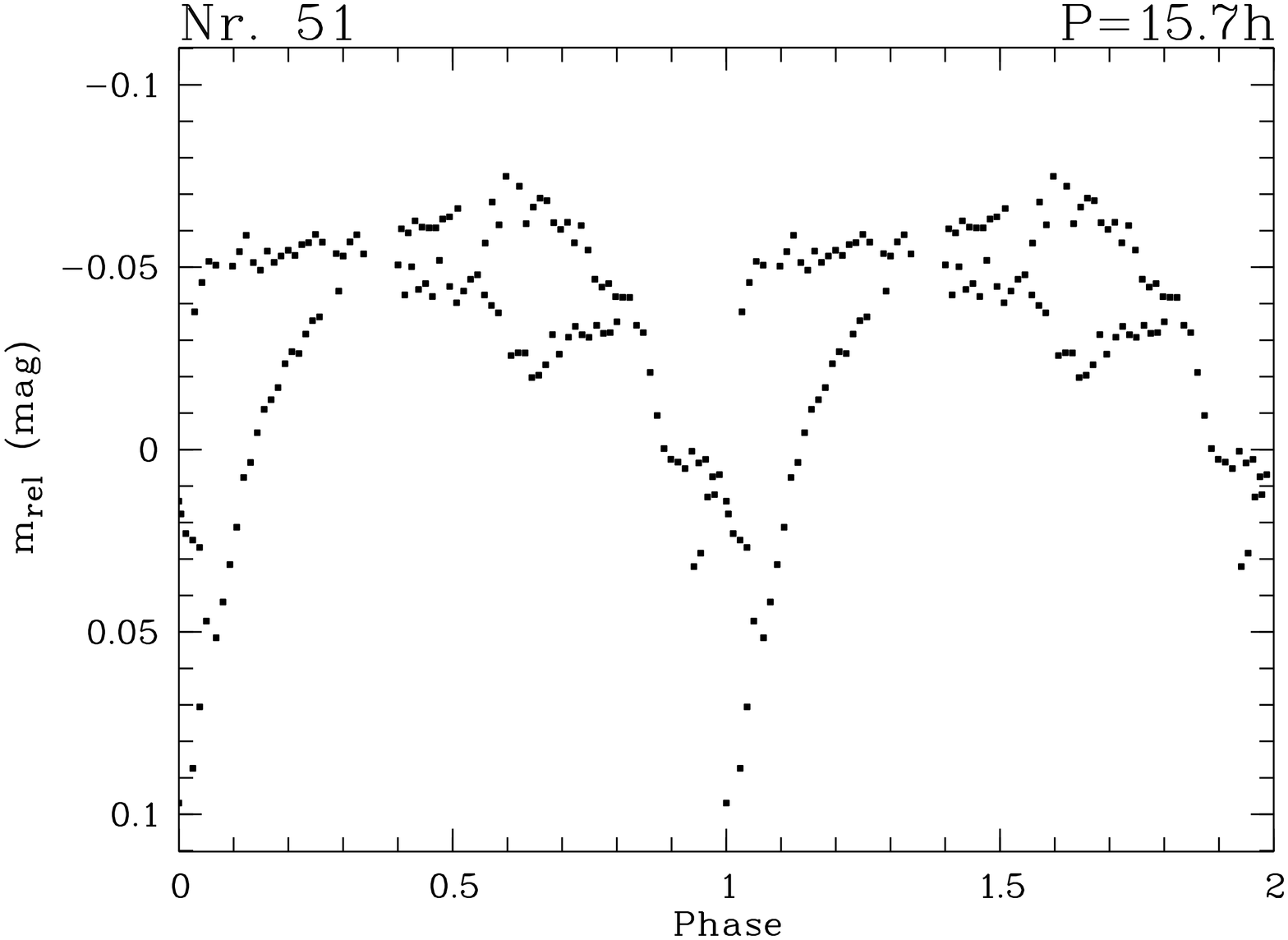}} \hfill %pp7895
\resizebox{5.9cm}{!}{\includegraphics[angle=-90]{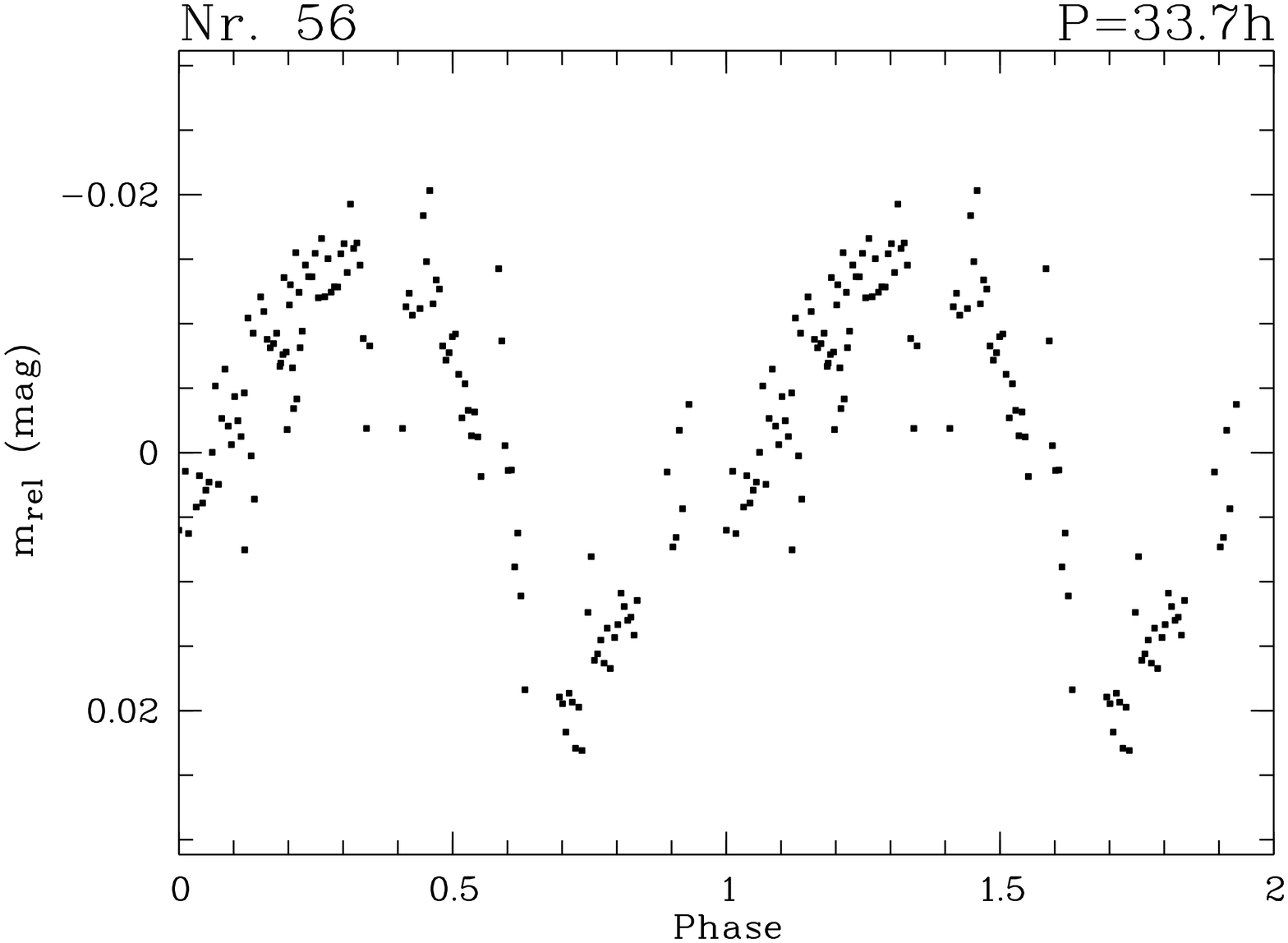}} \\     %pp3361
\resizebox{5.9cm}{!}{\includegraphics[angle=-90]{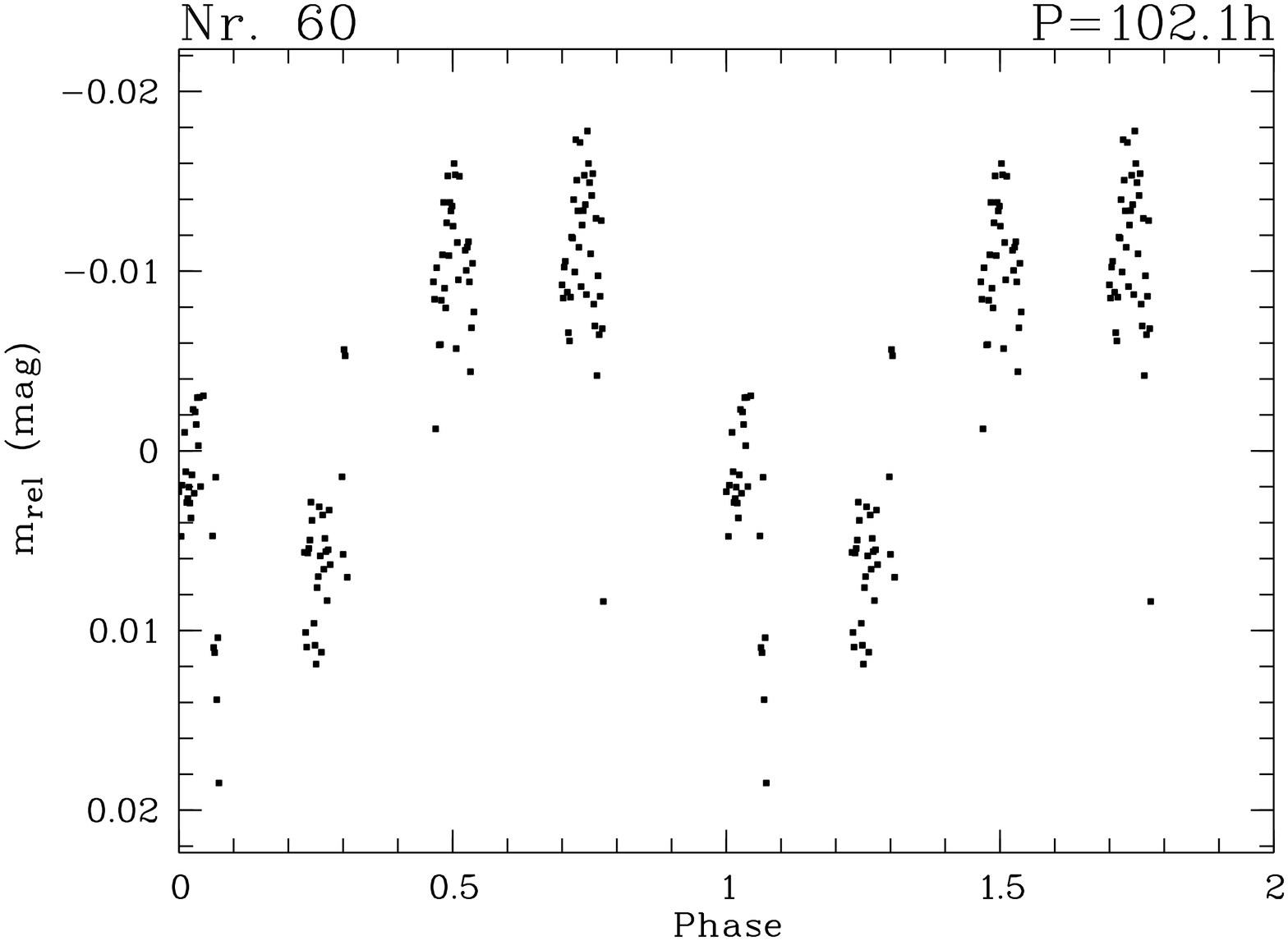}} \hfill %pp7663
\resizebox{5.9cm}{!}{\includegraphics[angle=-90]{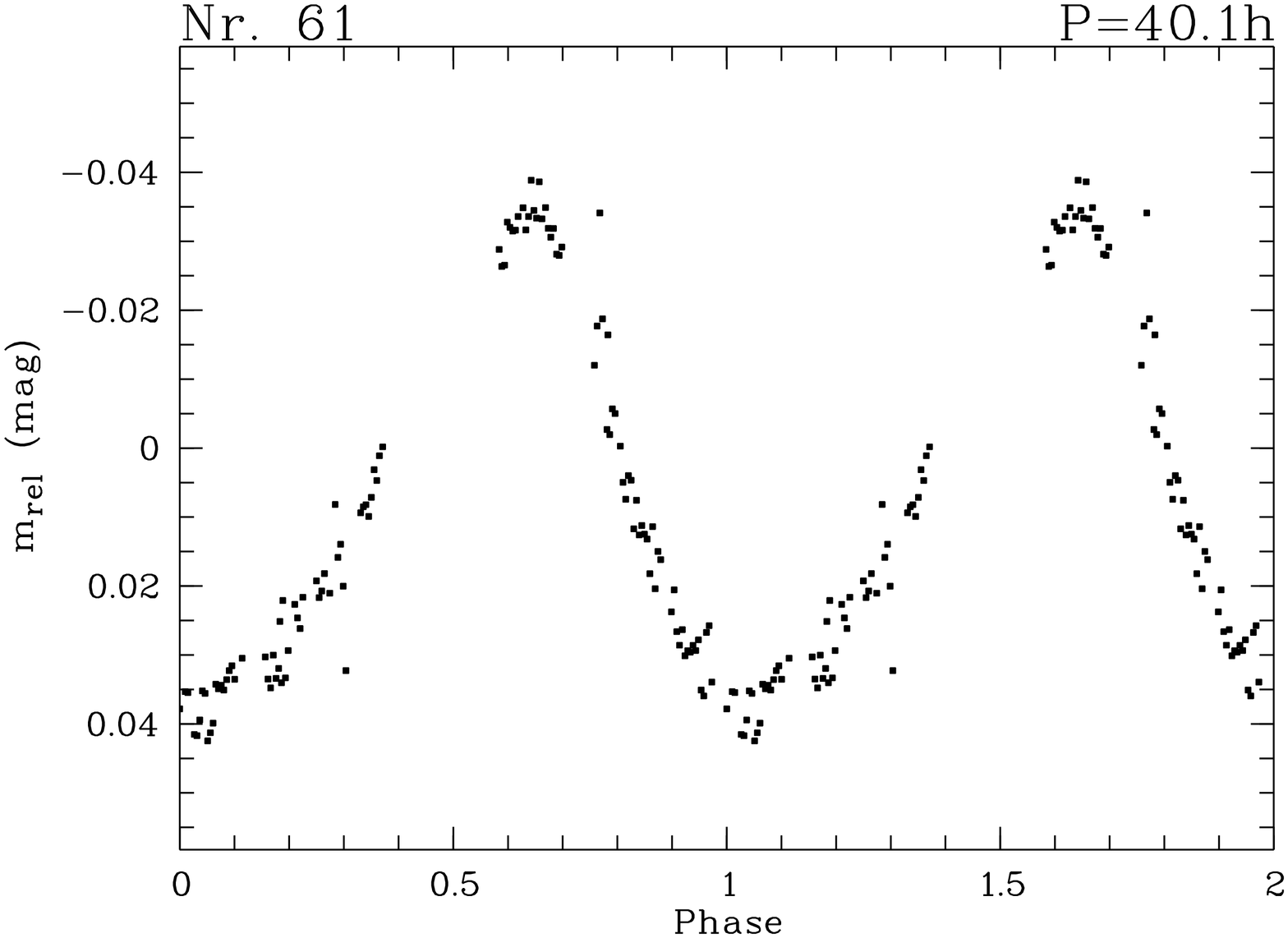}} \hfill %pp7723
\resizebox{5.9cm}{!}{\includegraphics[angle=-90]{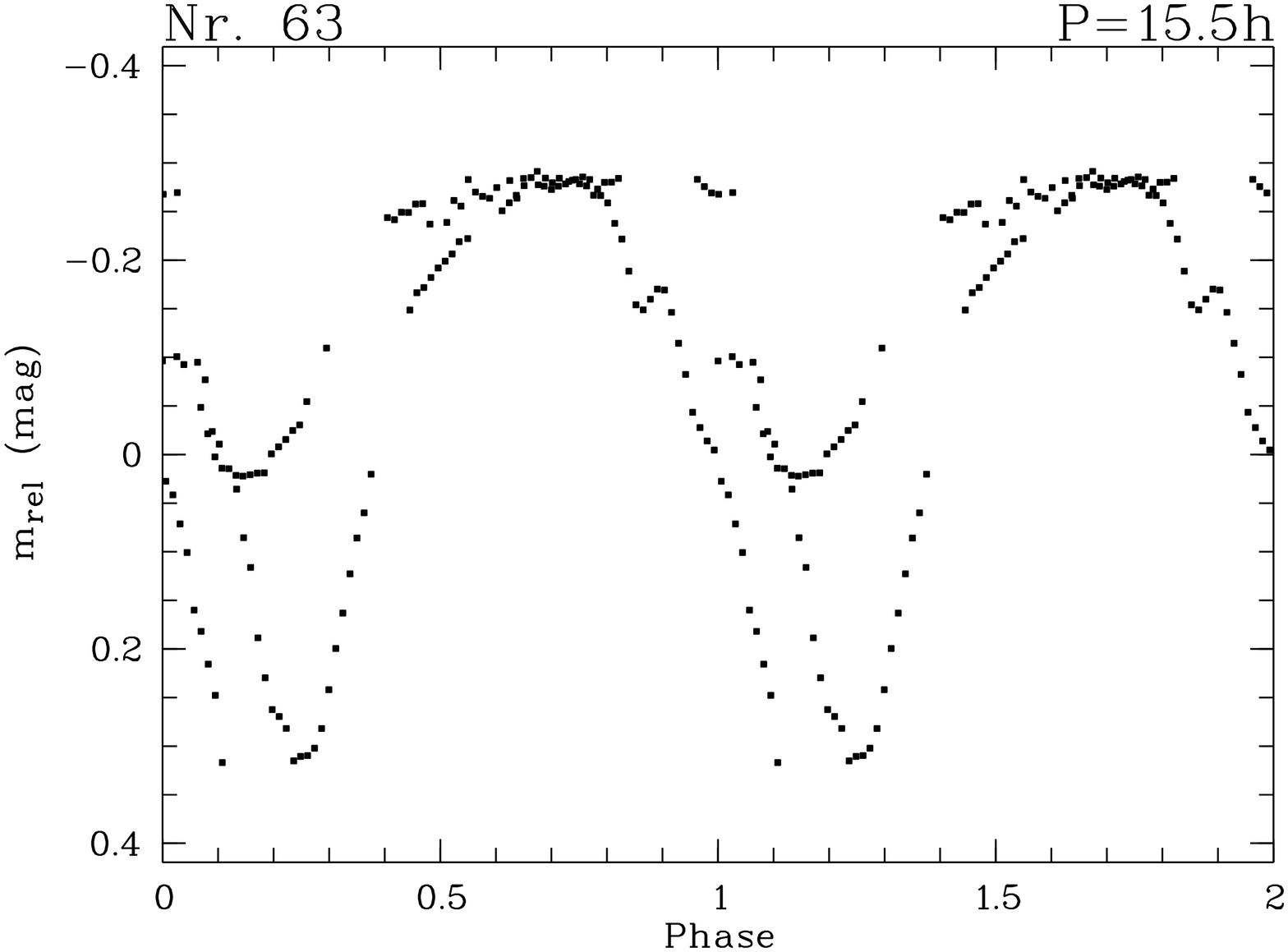}} \\     %pp7912
\caption{Phased light curves for the detected periodicities (part 1). At the top
of the diagrams we give the object number and the period from Table \ref{periods}.}
\label{phase1}
\end{figure*}

\begin{figure*}[htbd]
\resizebox{5.9cm}{!}{\includegraphics[angle=-90]{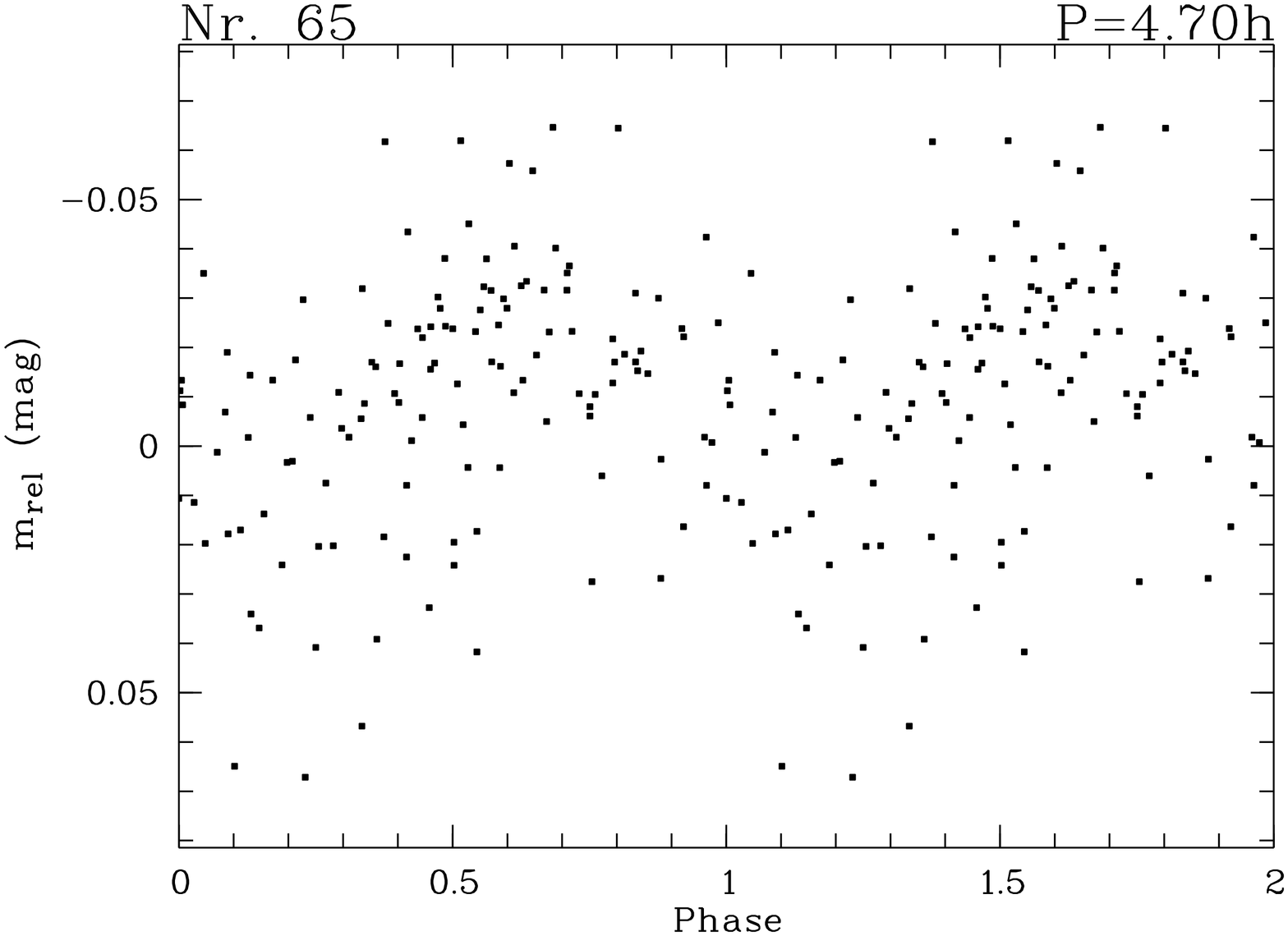}} \hfill  %pp9111.ps
\resizebox{5.9cm}{!}{\includegraphics[angle=-90]{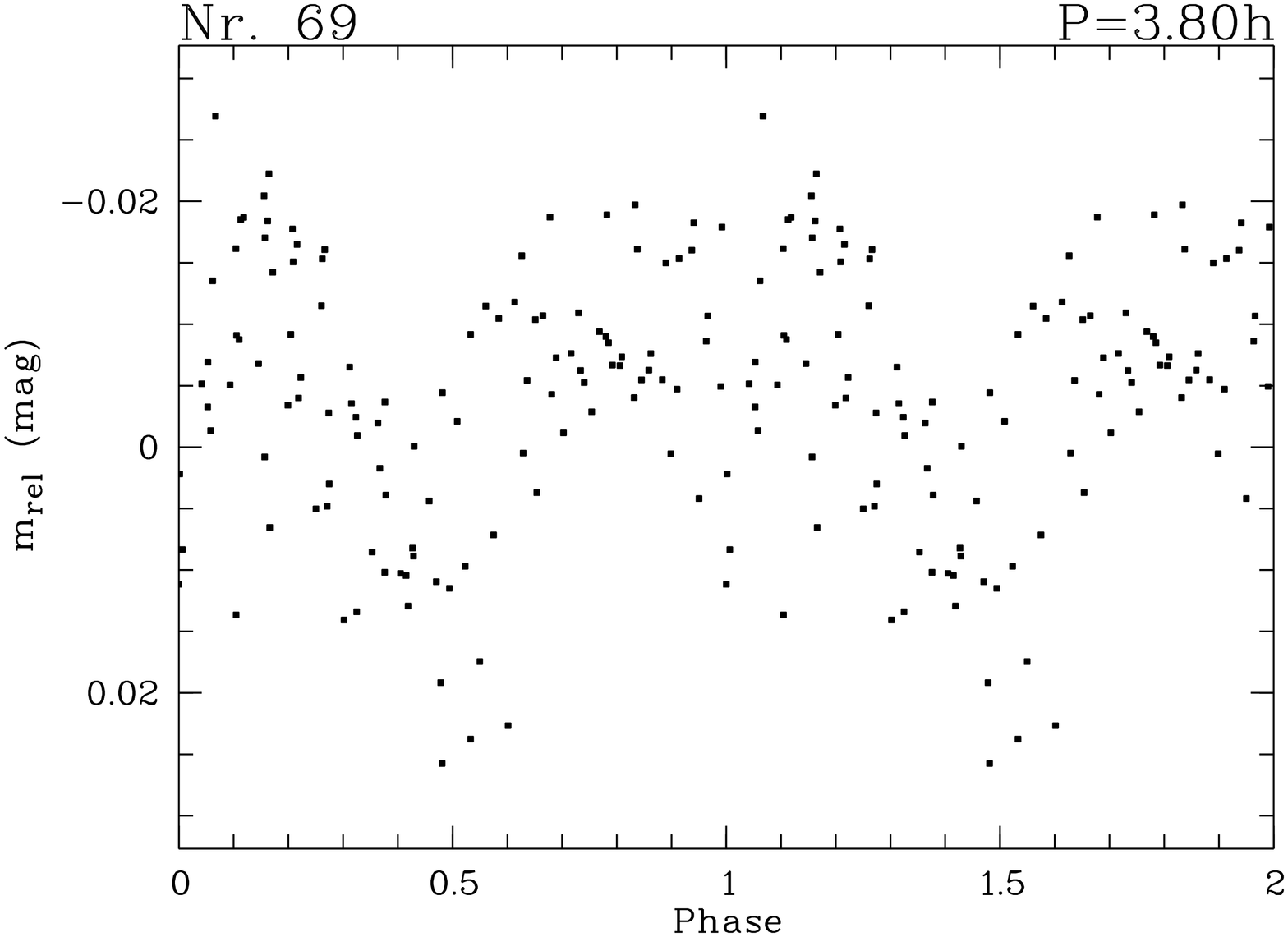}} \hfill  %pp7694.ps
\resizebox{5.9cm}{!}{\includegraphics[angle=-90]{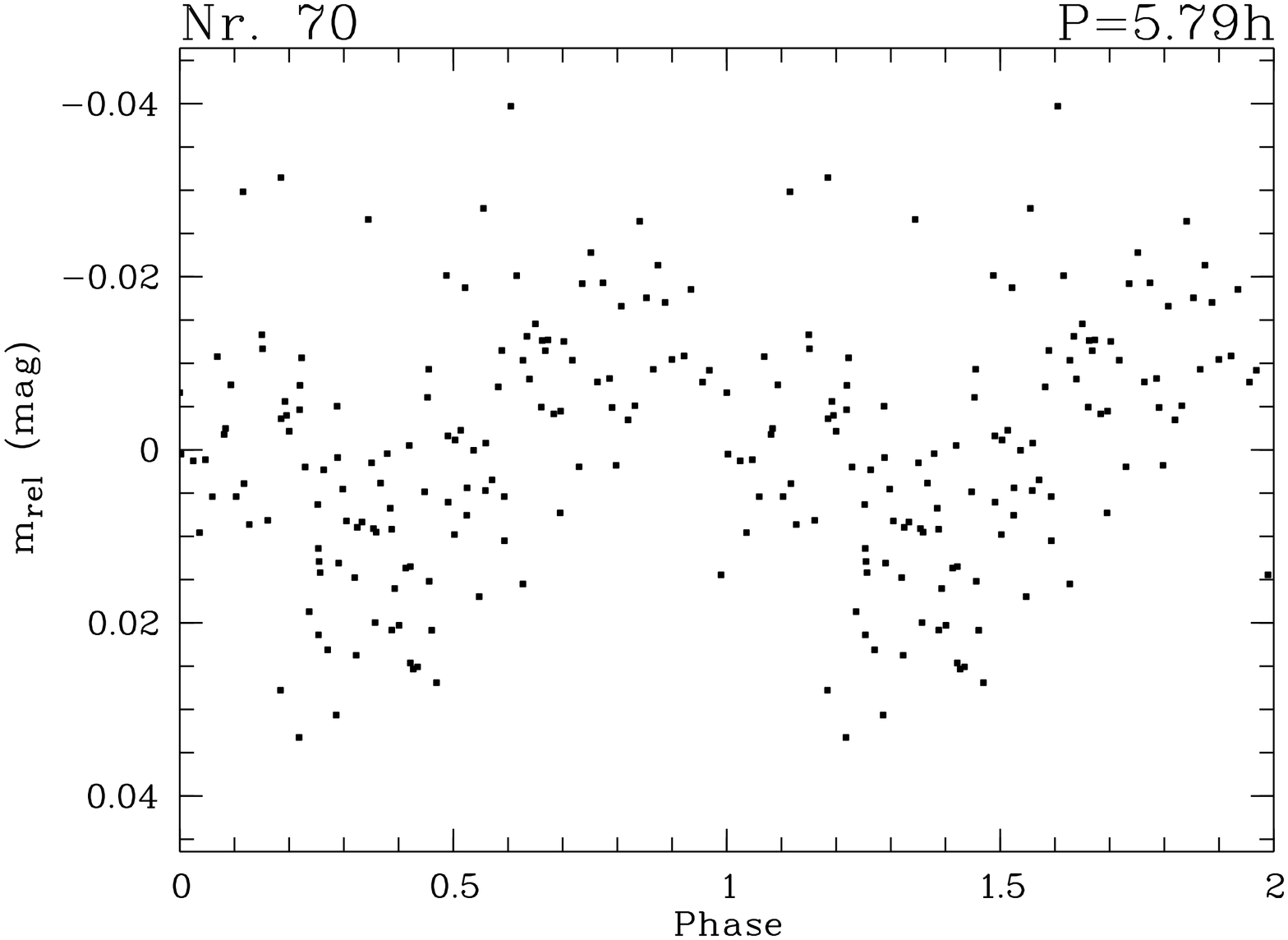}} \\      %pp2591.ps
\resizebox{5.9cm}{!}{\includegraphics[angle=-90]{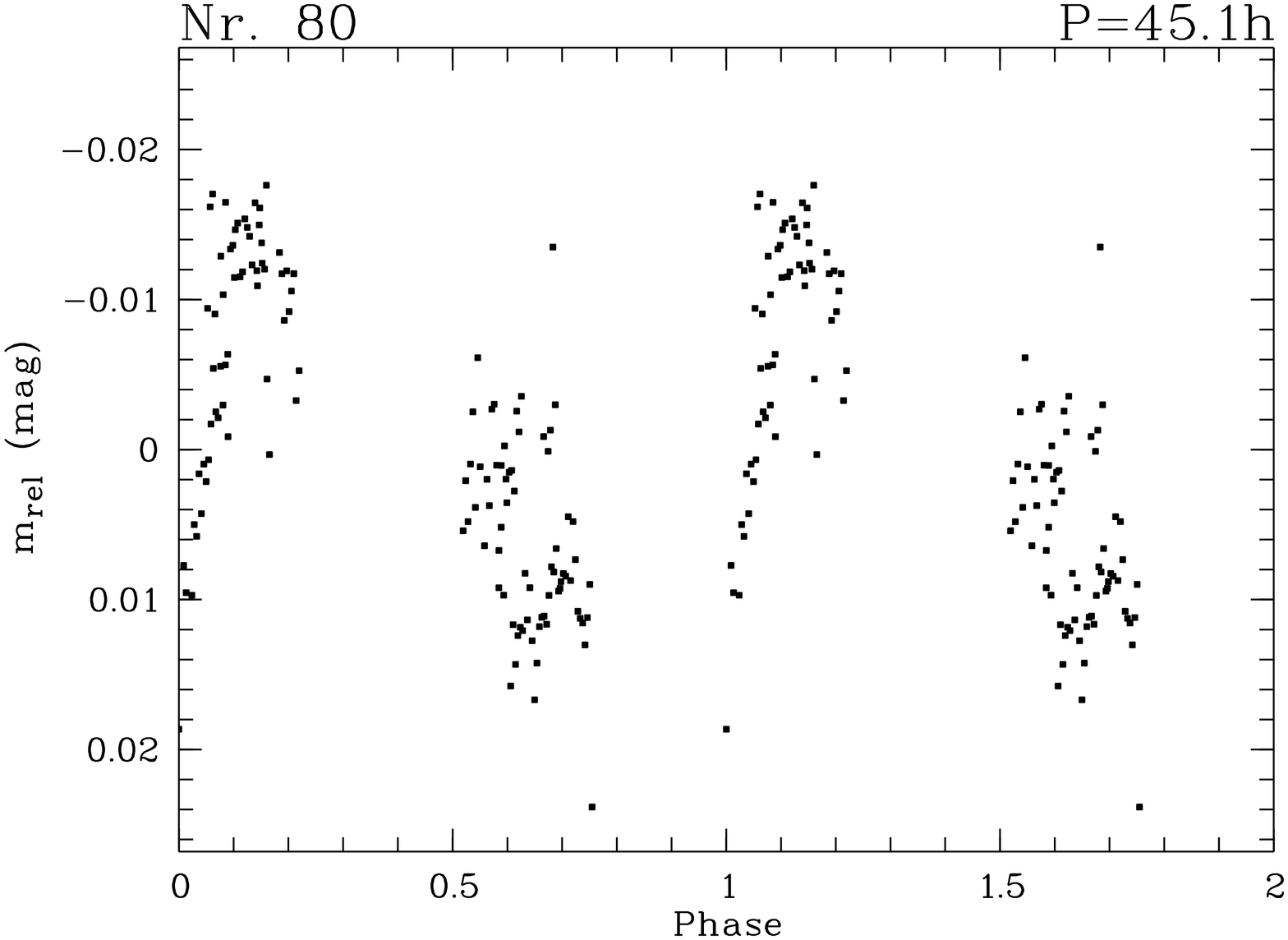}} \hfill %pp10184.ps
\resizebox{5.9cm}{!}{\includegraphics[angle=-90]{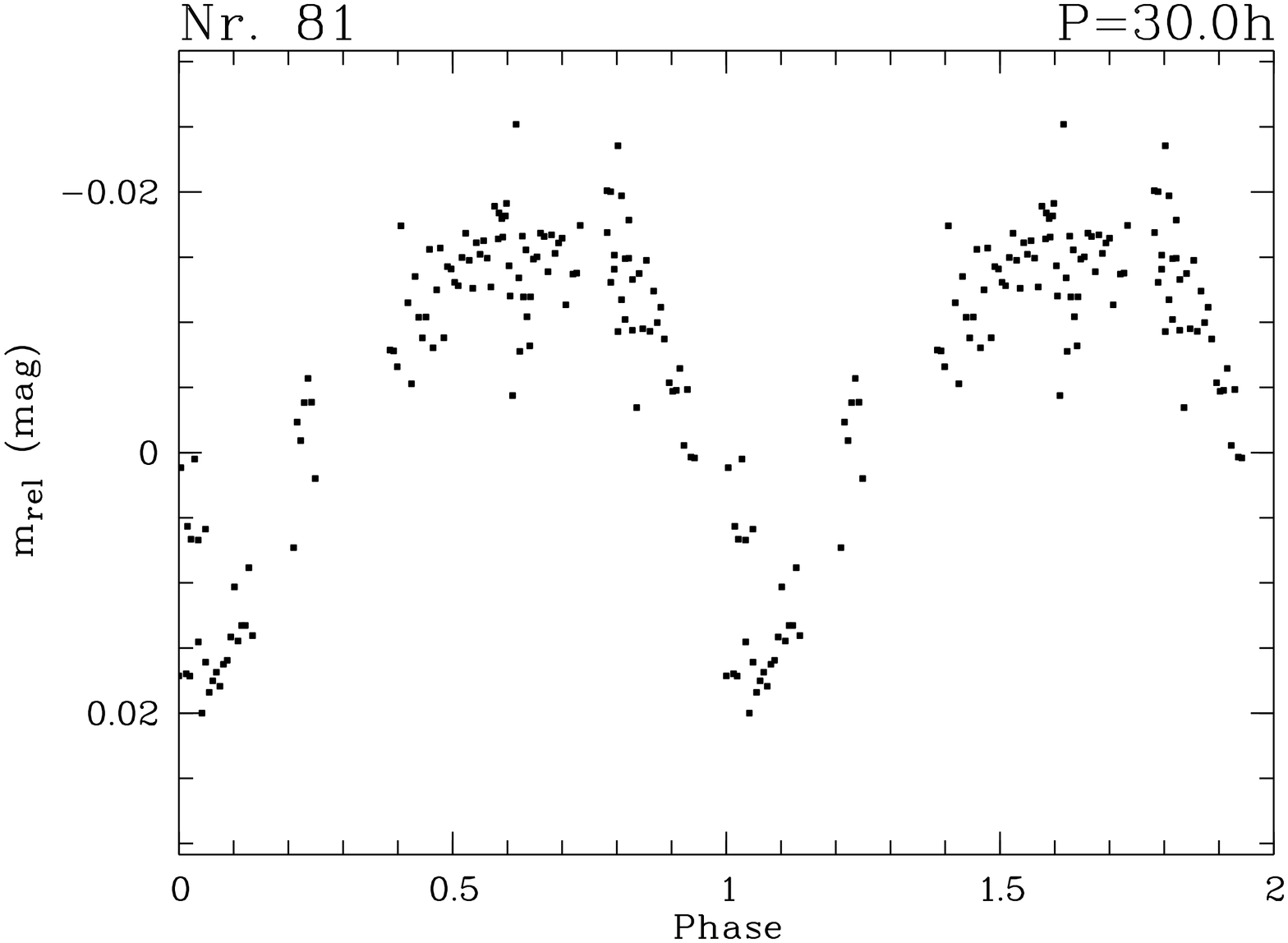}} \hfill  %pp2857.ps
\resizebox{5.9cm}{!}{\includegraphics[angle=-90]{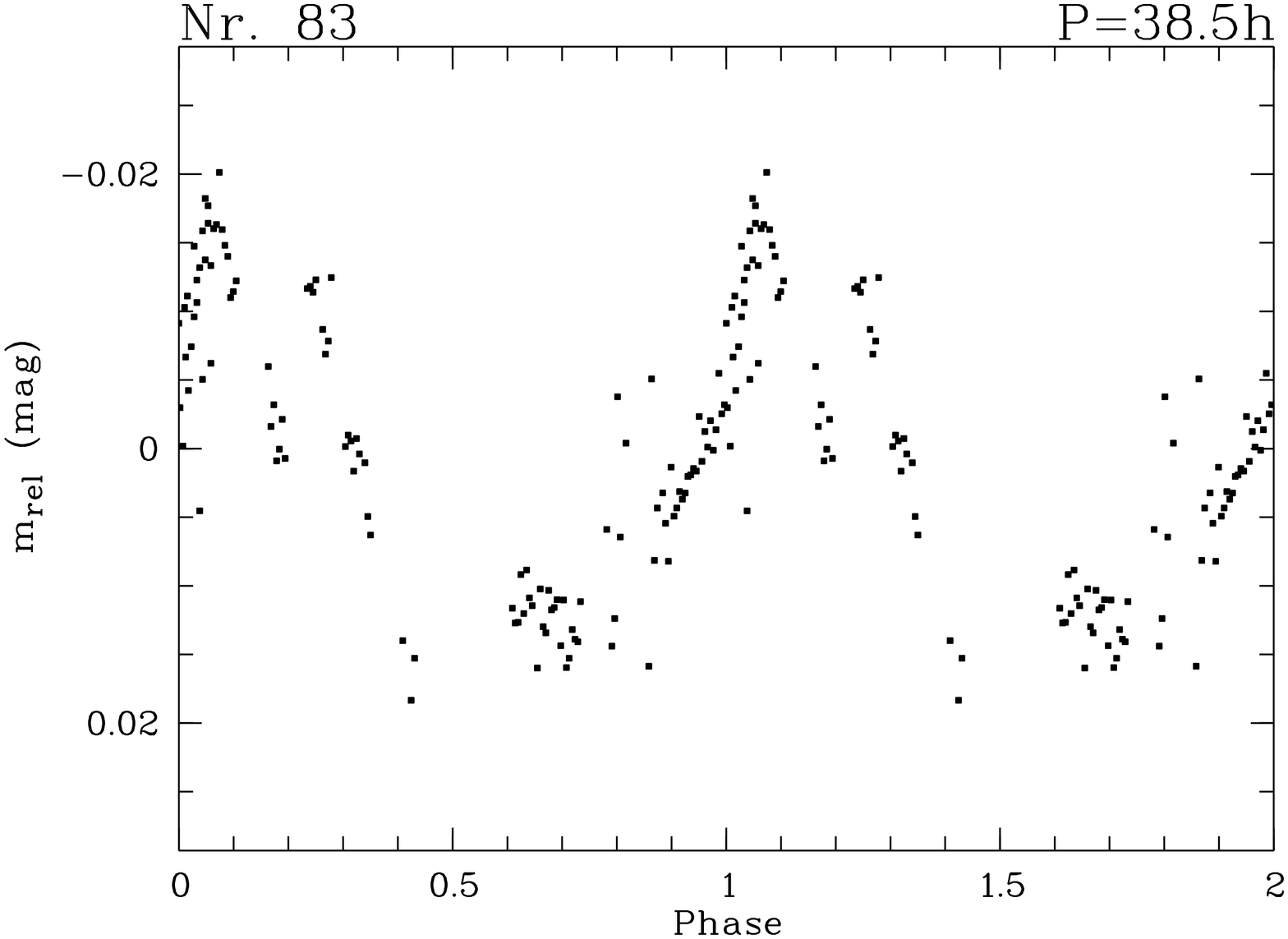}} \\      %pp1850.ps
\resizebox{5.9cm}{!}{\includegraphics[angle=-90]{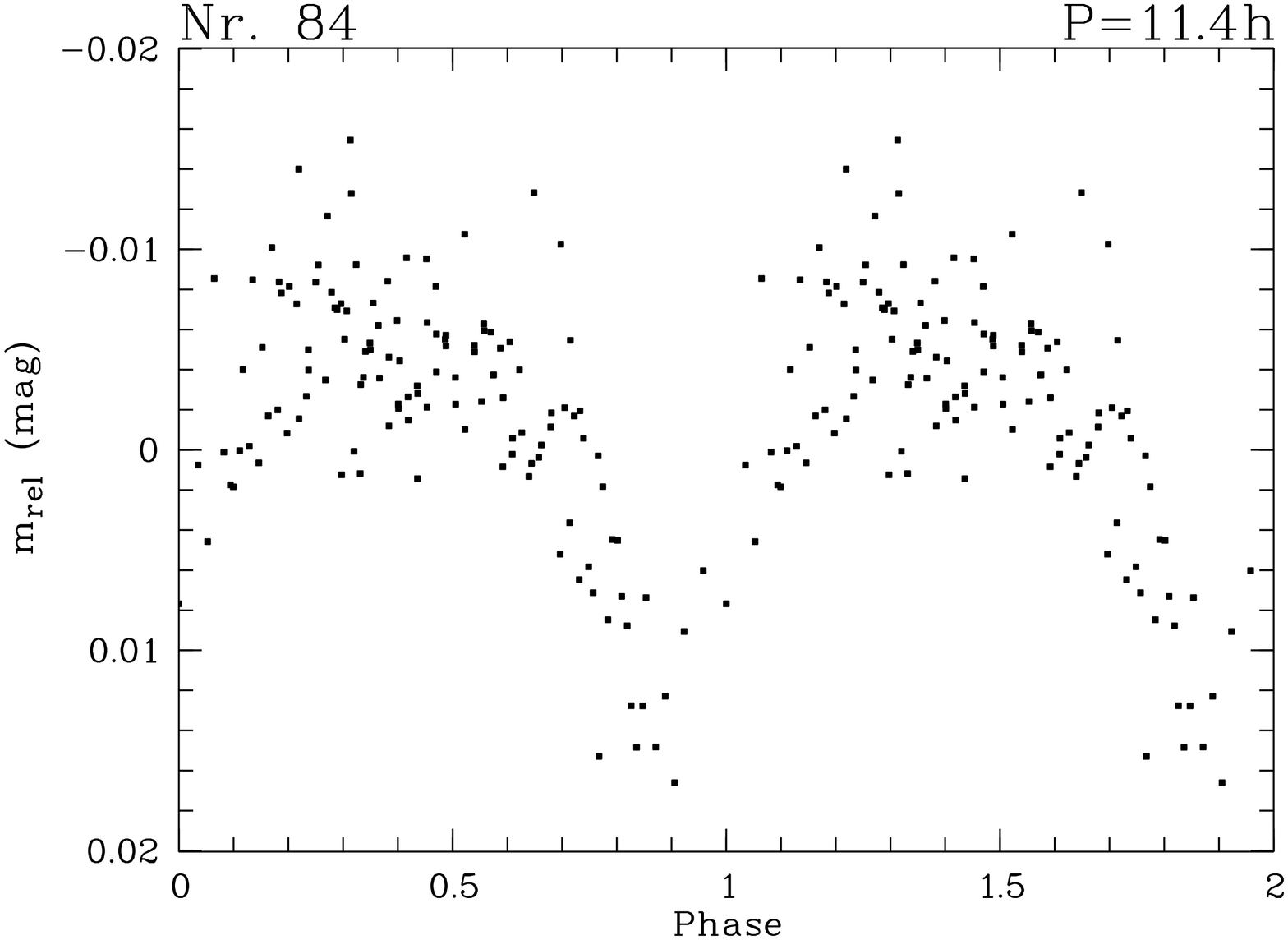}} \hfill %pp10369.ps
\resizebox{5.9cm}{!}{\includegraphics[angle=-90]{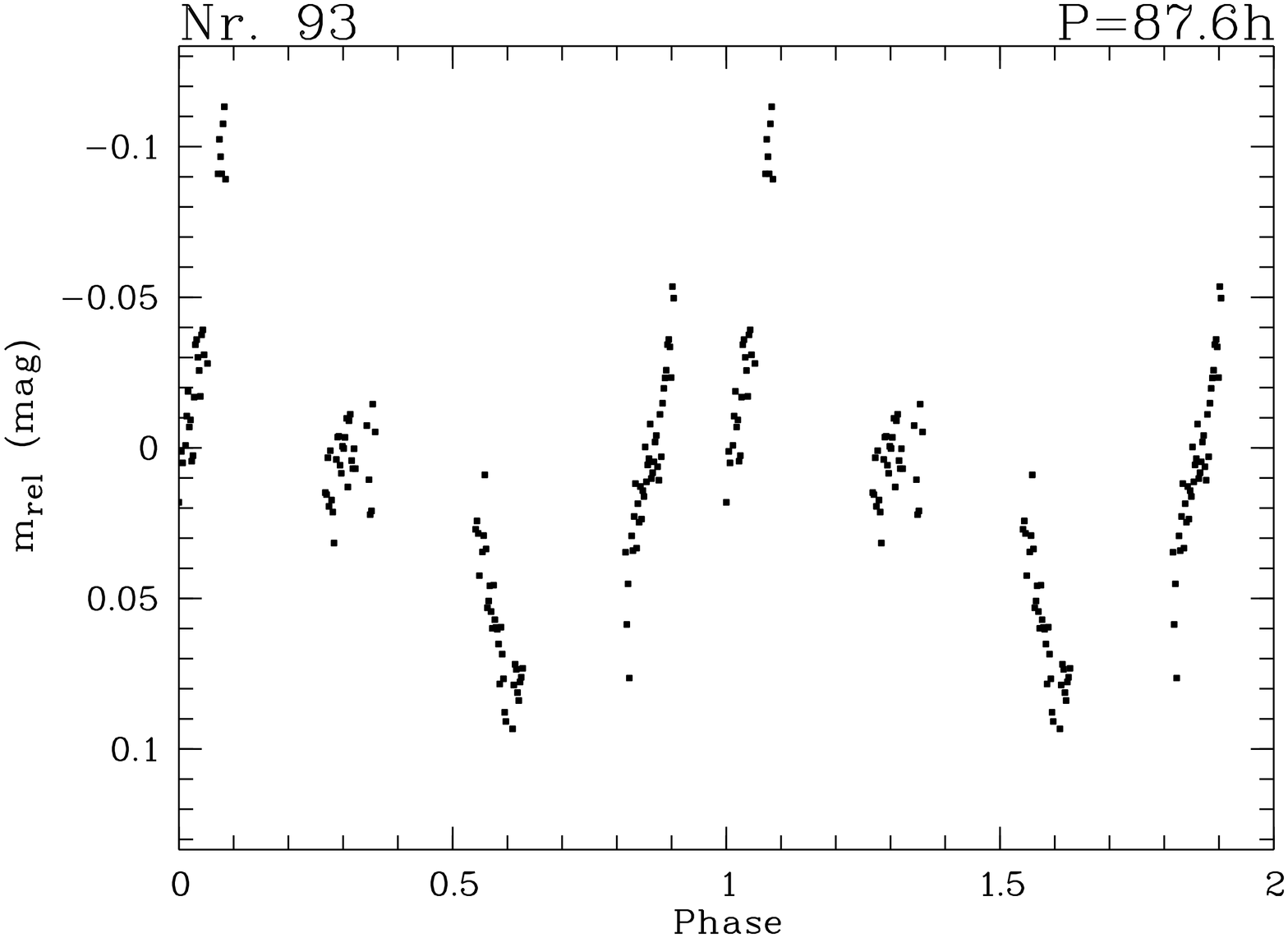}} \hfill %pp10466.ps
\resizebox{5.9cm}{!}{\includegraphics[angle=-90]{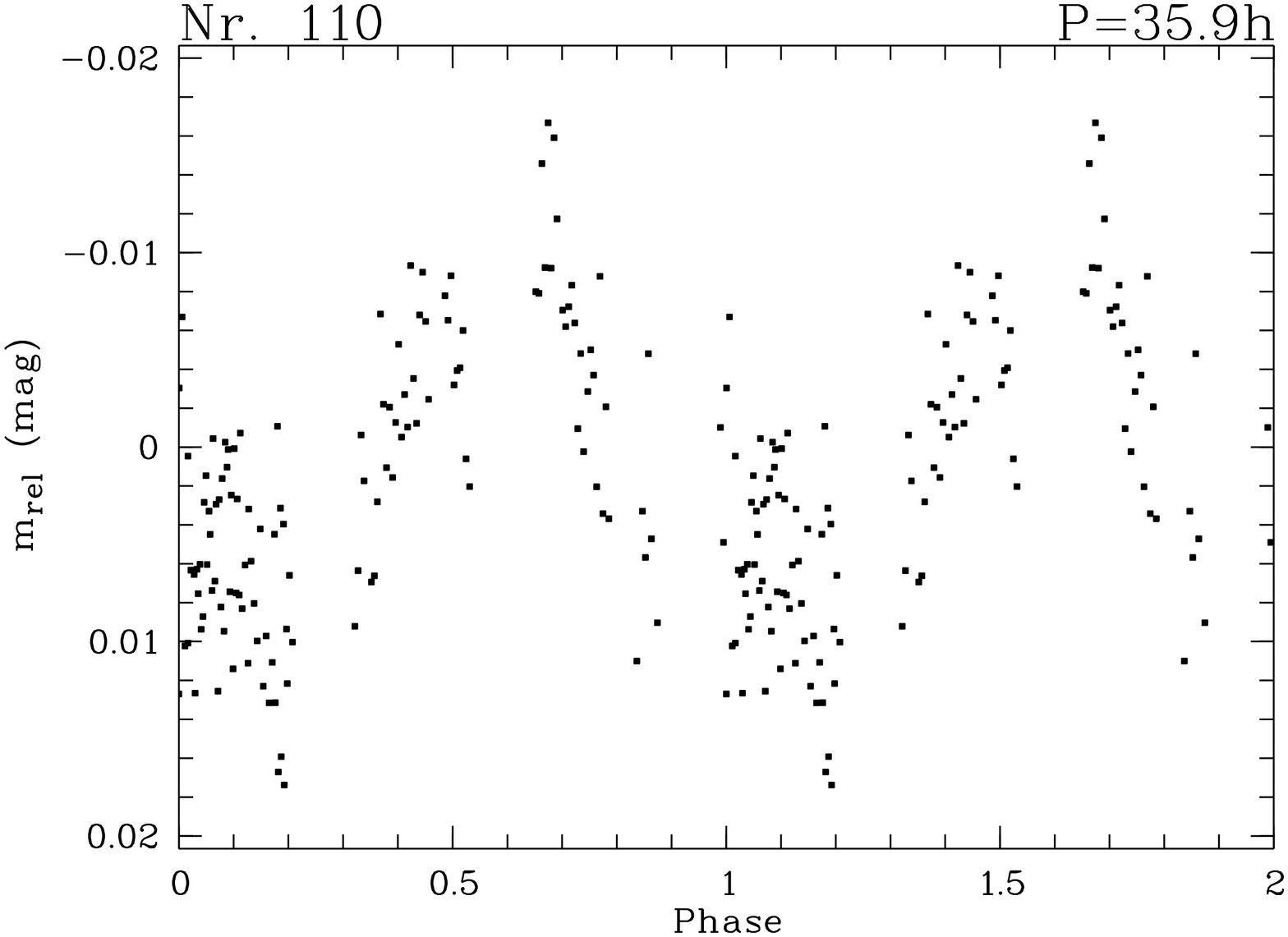}} \\     %pp11329.ps
\resizebox{5.9cm}{!}{\includegraphics[angle=-90]{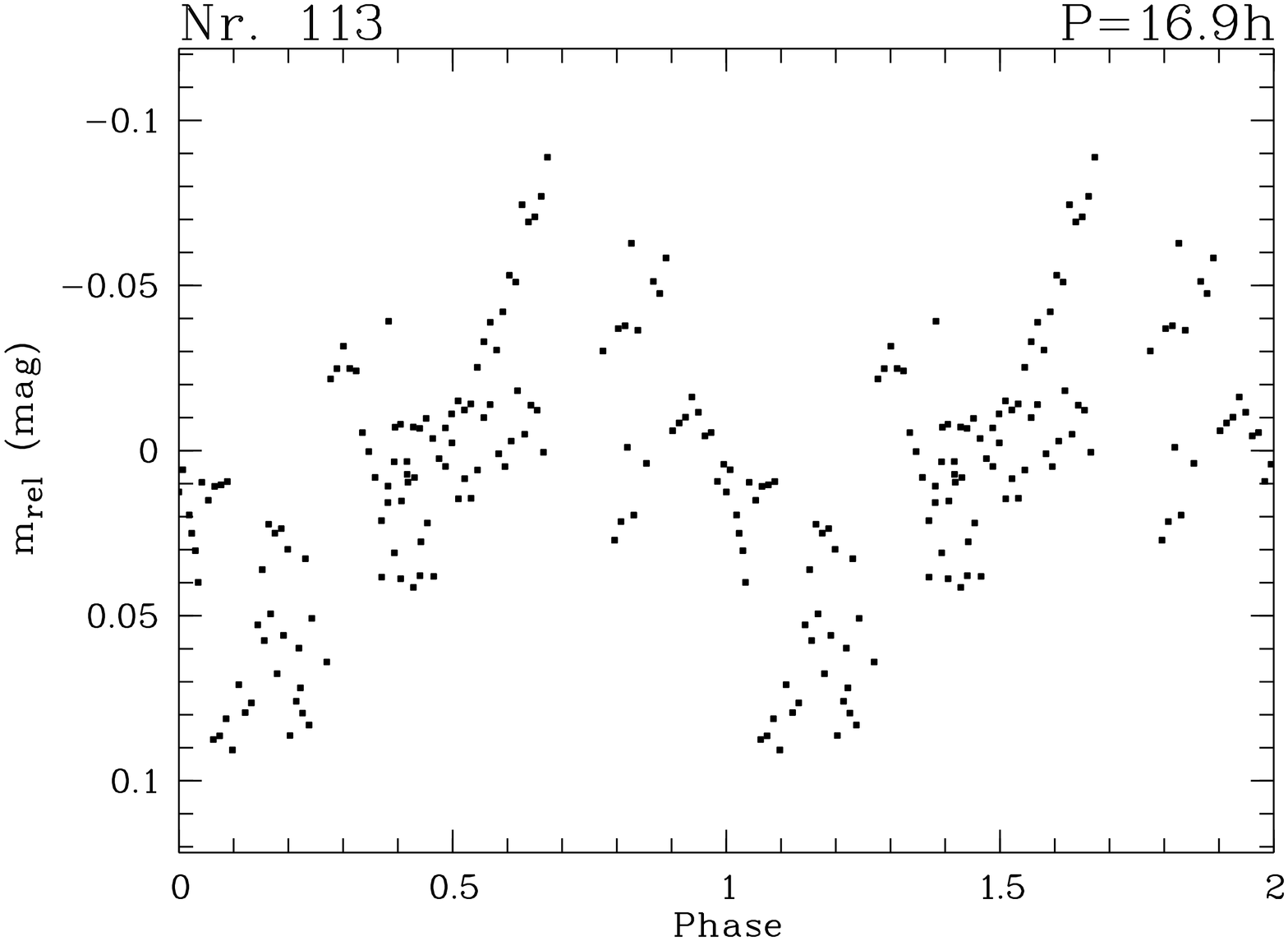}} \hfill %pp12287.ps
\resizebox{5.9cm}{!}{\includegraphics[angle=-90]{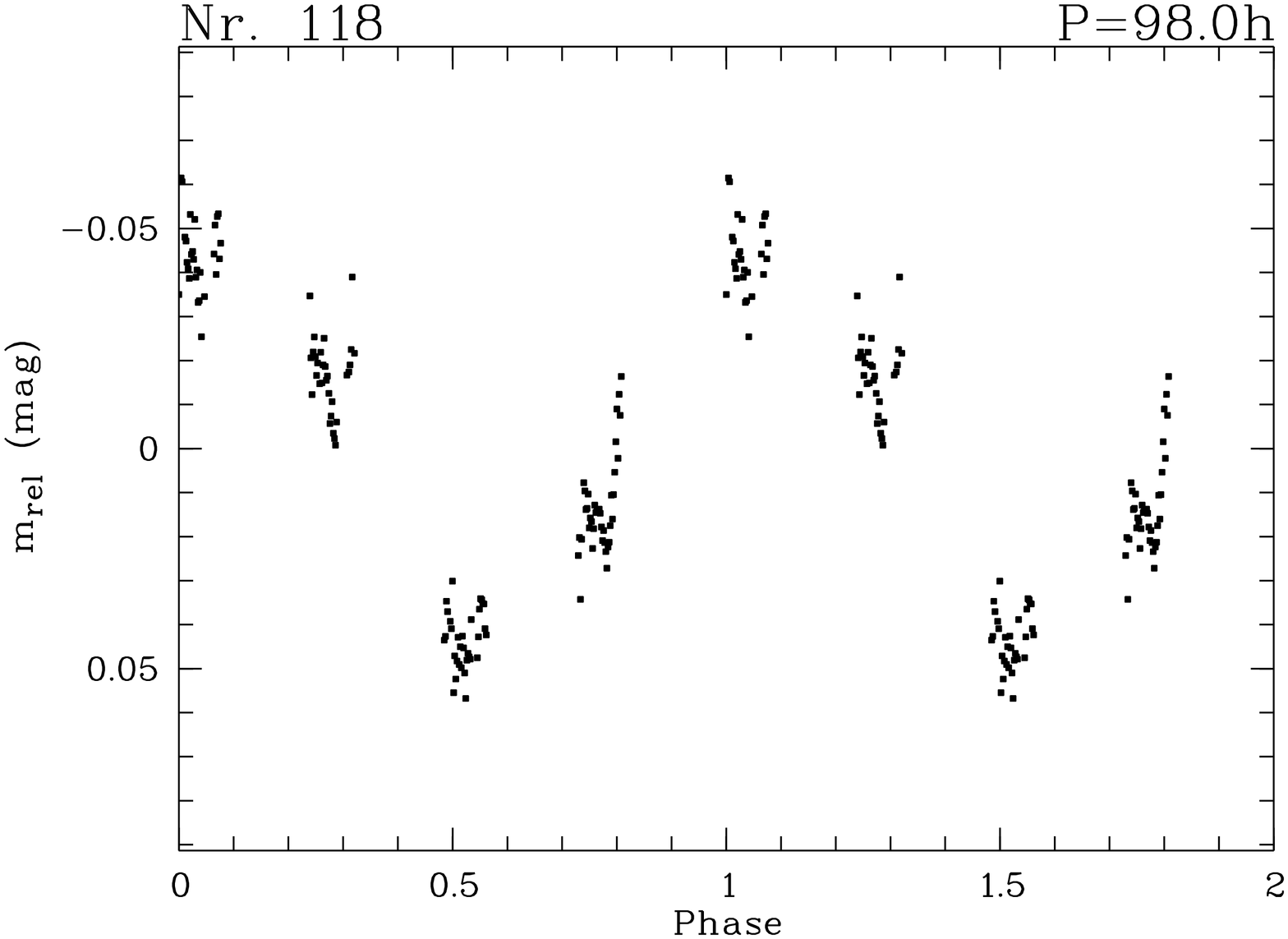}} \hfill %pp12068.ps
\resizebox{5.9cm}{!}{\includegraphics[angle=-90]{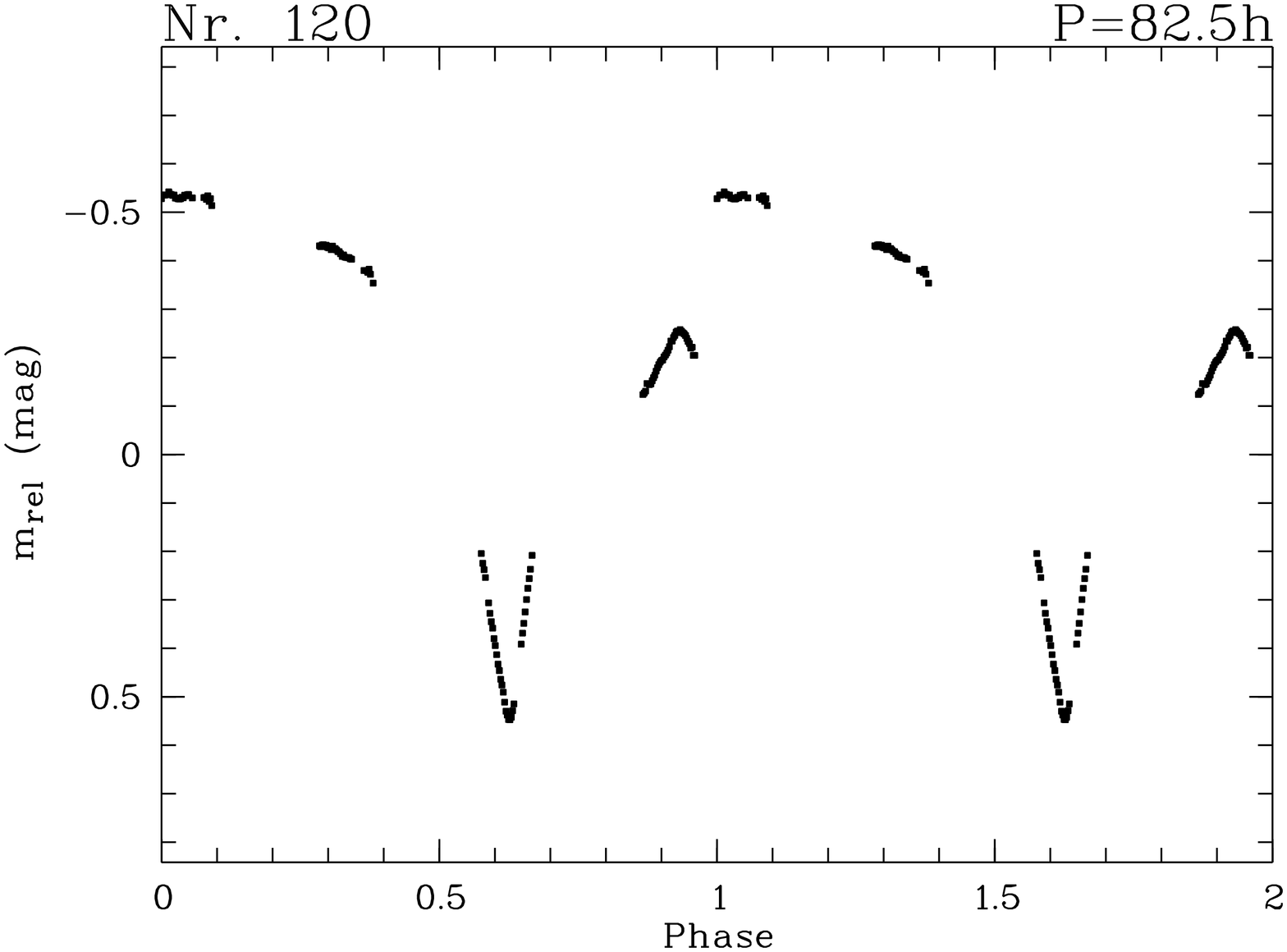}} \\     %pp10860.ps
\resizebox{5.9cm}{!}{\includegraphics[angle=-90]{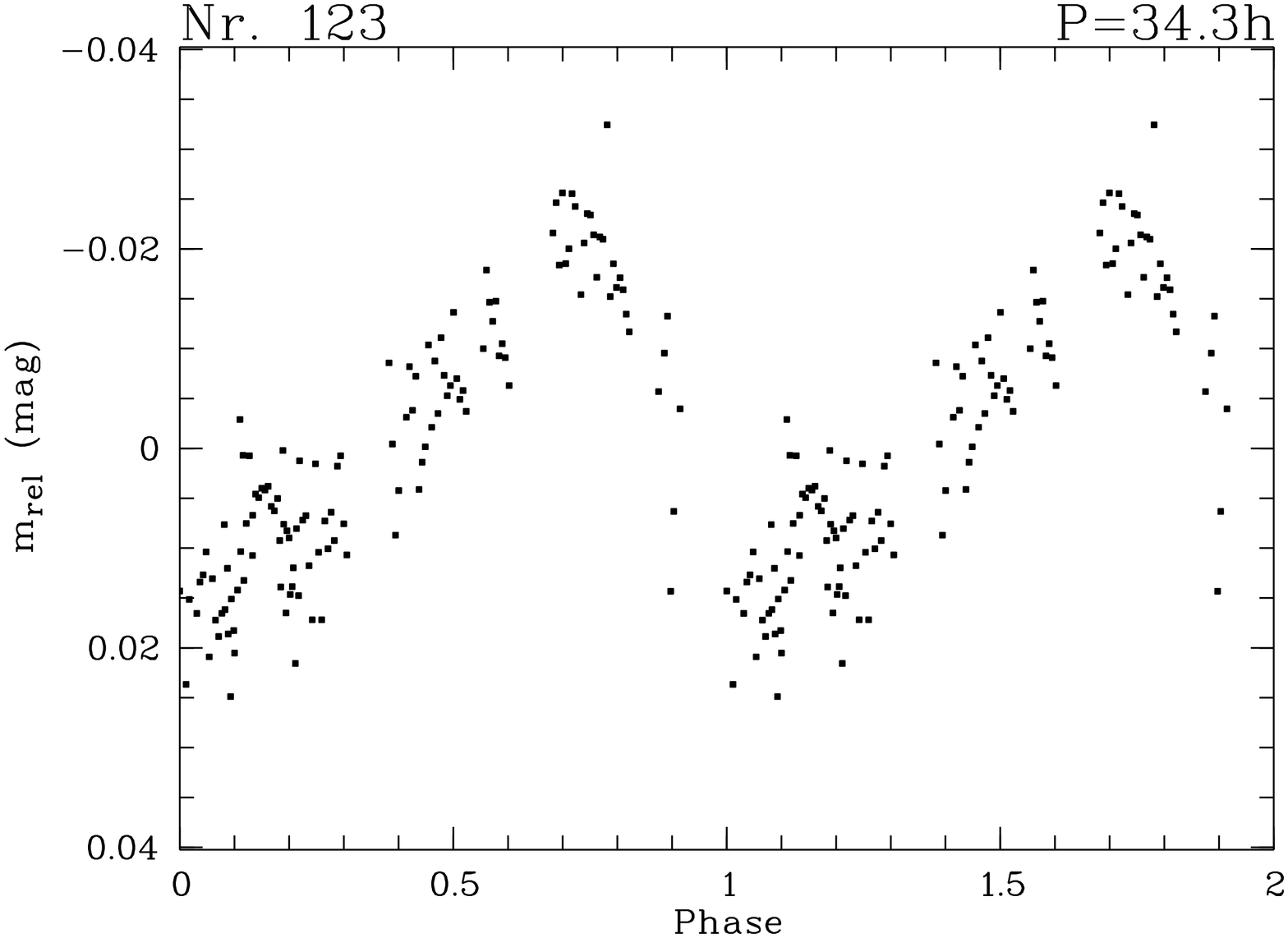}} \hfill   %pp845.ps
\resizebox{5.9cm}{!}{\includegraphics[angle=-90]{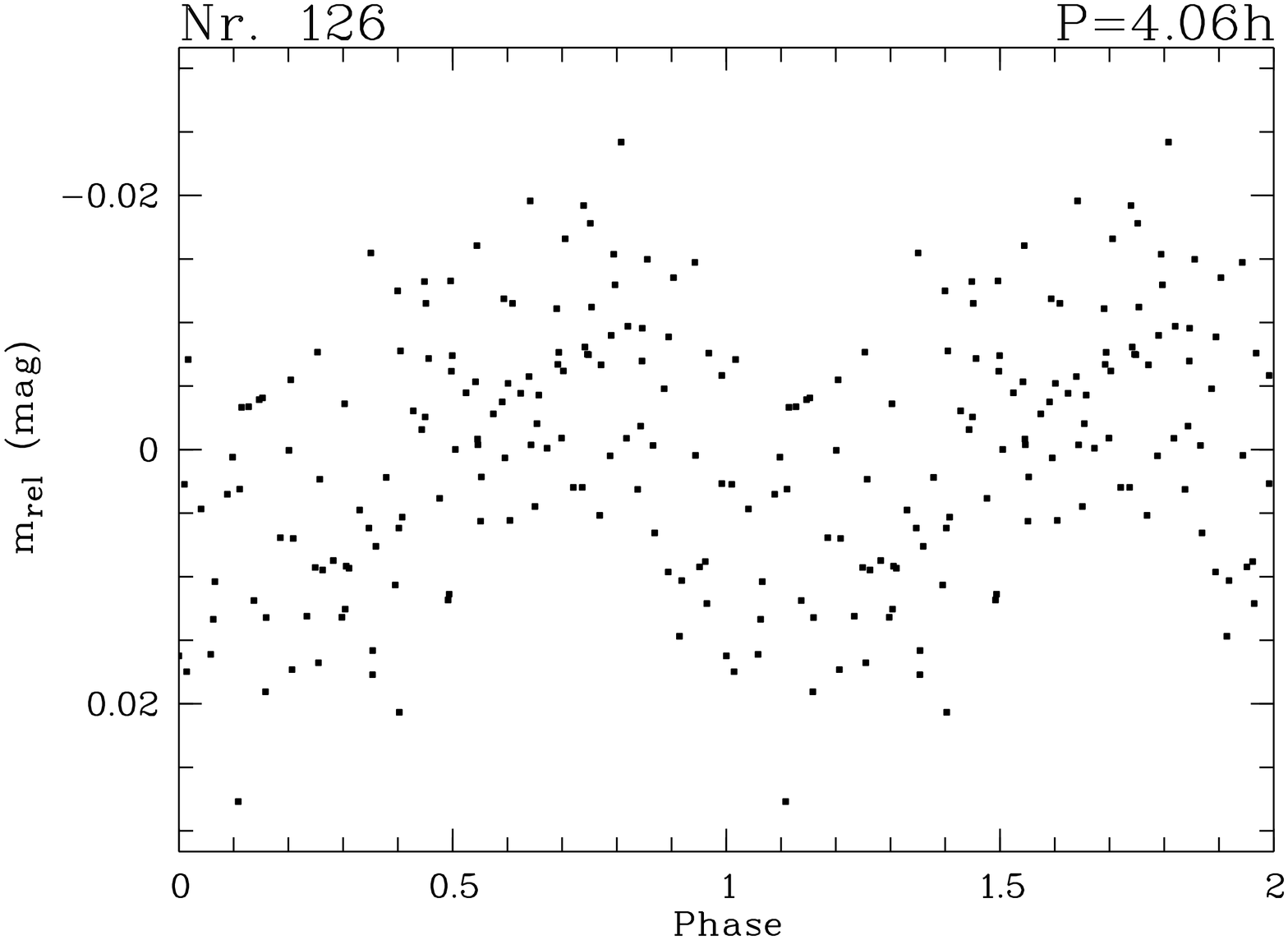}} \hfill  %pp1212.ps
\resizebox{5.9cm}{!}{\includegraphics[angle=-90]{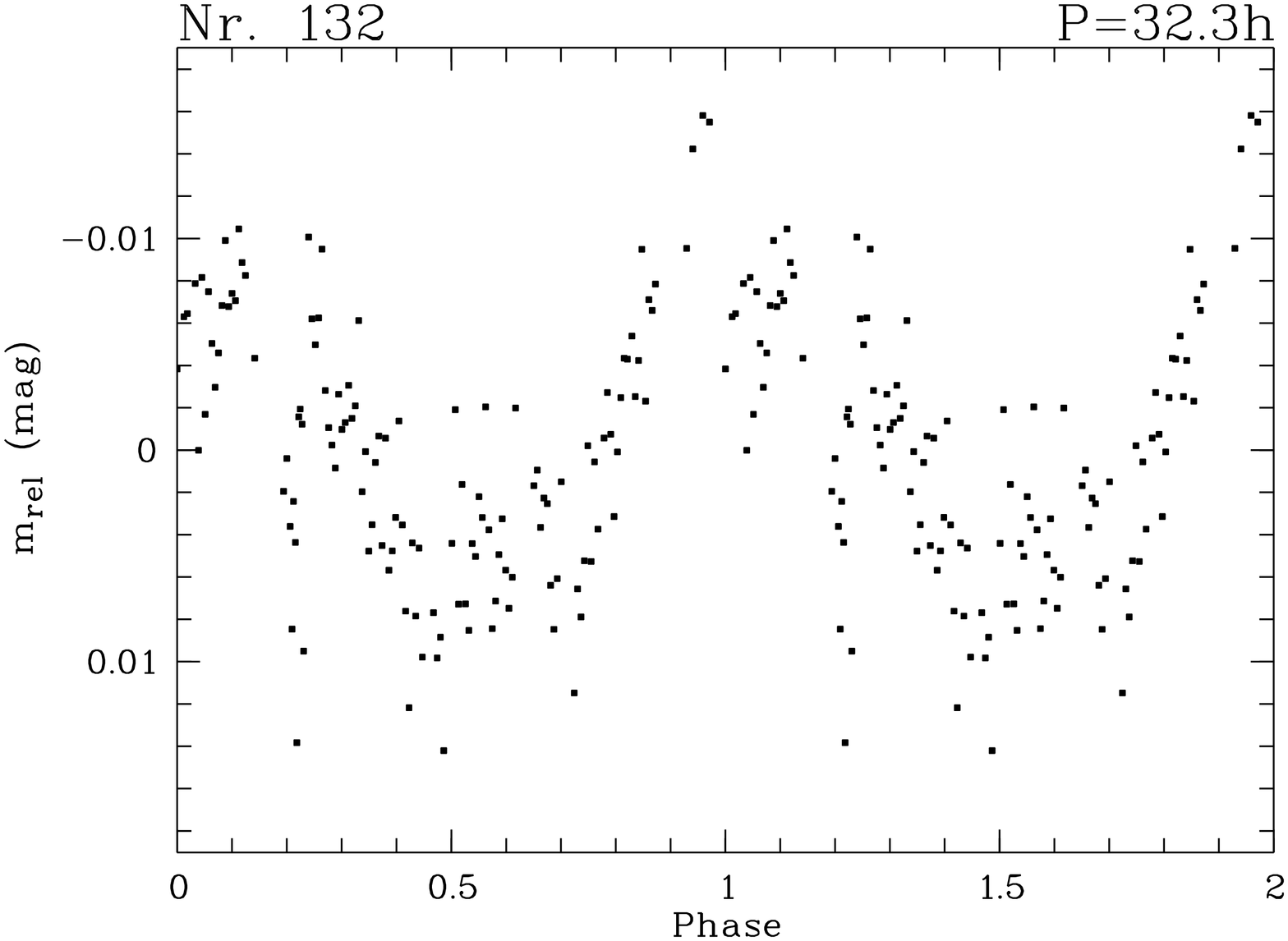}} \\     %pp11111.ps
\caption{Phased light curves (part 2).}
\label{phase2}
\end{figure*}

As the result of our period search we found 30 significant periodicities,
whose phased light curves are shown in Fig. \ref{phase1} and \ref{phase2}. 
All relevant data for the objects with periodic variability are given in 
Table \ref{periods}. For most of these periods the phase plots look
convincing, and the period can be approximated with a sinusoid. In the
light curves of the faintest targets the noise level is relatively
high, which does not mean that the period detection is unreliable
(see Sect. \ref{sens}). The exact period values in Table \ref{periods}
are determined by fitting the corresponding peak in the CLEANed periodogram
with a Gaussian and measuring its exact maximum. The error values
for the periods were estimated following the description given in
Horne \& Baliunas (\cite{hb86}). Typical errors are 2\% for $P=10$\,h and
10\% for $P=100$\,h. The amplitudes of the periodicities were determined 
by binning the phased light curves to ten data points (each bin corresponds 
to a 0.1 interval in phase space), and measuring the peak-to-peak value in 
these binned light curves. This approach guarantees that the amplitudes are 
not dominated by the noise level in the light curves.

The five highly variable objects whose light curves are shown in Fig.
\ref{strange} require special attention. From visual inspection, these
light curves could contain a periodic component, although the period search
is hampered by high-level irregular variability. The periodogram often
shows several highly significant peaks, so that an unambiguous period
detection is difficult. Nevertheless, for three of these objects we found 
a convincing period which satisfies our period search criteria. The 
phased light curves of these objects (no. 51, 63, 120) show the
period and superimposed short-term fluctuations. Because of these 
additional irregular variations, the exact period values for these two
objects have to be treated with caution. For the two remaining 
objects from Fig. \ref{strange} (no. 87, 104), a reliable period determination 
was impossible. Possible origins of the photometric behaviour for all these 
objects will be discussed in Sect. \ref{ori}.

\subsection{Sensitivity}
\label{sens}

As in our previous two variability studies (SE1, SE2),
we investigated the sensitivity of our period search using the following procedure:
We selected non-variable objects with low photometric noise, and added 
a sine shaped periodicity to their light curves. Then we calculated the Scargle
periodogram for these synthetical periodicities, varying the period length and
the signal-to-noise ratio (i.e. the amplitude of the co-added sine wave). For each
simulation, we recorded the frequency of the highest peak in the periodogram and
compared it with the real period. This gives us an estimate of the reliability
of our period search for a given signal-to-noise and a given period.

The first important result is that we are extremely sensitive even at very
low signal-to-noise levels. We can reliably detect periods with amplitude-to-noise
ratios (defined as ratio of amplitude of the periodicity and noise in the
original light curve) down to 0.75. For lower amplitude-to-noise,
the deviations between detected period and real period can exceed 5\%.
On the other hand, the lowest signal-to-noise of the periods, which we 
detected in the $\epsilon$\,Ori light curves is 1.8. Thus, all periods 
are reliable, even when the phased light curves look noisy. This is mainly 
due to the high number of data points in our time series, delivering a dense 
sampling of the period. In general, periodogram techniques are particularly 
well-suited for detecting low-amplitude periods, since the white noise in the 
light curves is equally distributed over all frequencies and thus the noise 
level in the periodogram is very low.

\begin{figure}[t]
\centering
\resizebox{\hsize}{!}{\includegraphics[angle=-90,width=6.5cm]{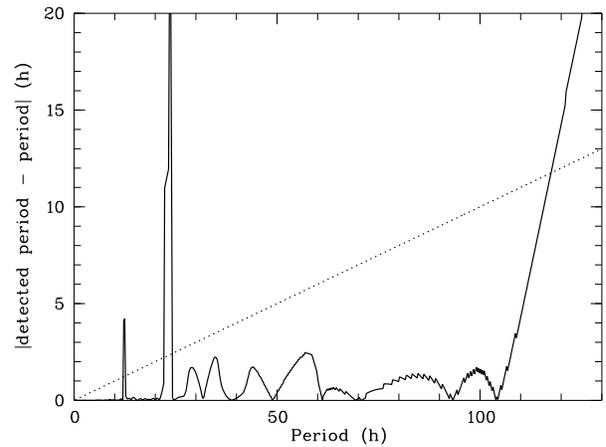}}
\caption{Deviation between real and detected period for a sample of periods
with amplitude-to-noise ratio of 2.5, which is typical for our detected periods.
The dotted line corresponds to a period error of 10\%. Our period search
is highly reliable up to periods of 110\,h.}
\label{sensfig}
\end{figure}

Fig. \ref{sensfig} shows the sensitivity of our period search as a function
of the period for an amplitude-to-noise level of 2.5, which is typical for
the detected $\epsilon$\,Ori periods. The plot demonstrates clearly that
we are able to reliably detect periods up to 110\,h. Above this limit,
the uncertainty of the period determination exceeds 10\%. There are a
few small windows where our period search is less reliable, namely around
12\,h and 24\,h, i.e. periods around these values should be treated with
caution. We note, however, that this test is only based on the Scargle
periodogram, whereas our period search procedure also includes several
plausibility checks, as outlined above. We should be able
to exclude most unreliable period detections by means of the CLEAN
algorithm and comparison with nearby reference stars. Thus we do not
expect many spurious period detections, but we may miss some periods
which lie in the narrow windows of uncertainty. 

\section{Origin of the variability - activity and accretion}
\label{ori}

It is apparent from our light curves that we observe three different kinds
of variability. Most variable targets have low-amplitude brightness modulations,
and their light curves show regular periodic behaviour. In many cases, the
shape of the light curve is well-approximated by a sine wave. We interpret
this behaviour as a consequence of cool spots co-rotating with the targets,
and discuss these objects in Sect. \ref{low}. On the other hand, five of our 
targets exhibit high-amplitude variations with at least partly irregular behaviour. 
The light curves for these objects are shown in Fig. \ref{strange}. For some of
these objects, a periodicity was found, but there are clearly superimposed 
irregular variations on timescales from one hour to one day. This behaviour
is most likely caused by a combination of accretion and rotation, and will 
be discussed in detail in Sect. \ref{high}. We note that this classification 
of the light curves is in agreement with the description of Herbst et al. 
(\cite{hhg94}), who distinguish between type I (periodic variations with 
low amplitude caused by cool spots) and type II variability (high-amplitude, 
partly irregular variations caused by hot spots) for T Tauri stars. Additionally, 
we observed one strong flare event, and we therefore include a discussion of 
the flare behaviour of VLM objects in Sect. \ref{flare}.

\subsection{Low-amplitude variables}
\label{low}

Low-amplitude variations with regular periodicities are well-known from a
large number of variability studies on solar-mass stars in open clusters 
(e.g. Patten \& Simon \cite{ps96}, Krishnamurthi et al. \cite{ktp98}, 
Herbst et al. \cite{hbm02}). The usual explanation for these periods is
the existence of magnetically induced cool spots co-rotating with the star
and thus modulating its brightness. This could also be the origin of
the behaviour of our low-amplitude targets.

It is well-established that magnetic activity, expressed as H$\alpha$ or
X-ray emission, persists down to late M spectral types (e.g., Mohanty \& Basri 
\cite{mb03}, Mokler \& Stelzer \cite{ms02}). From Doppler imaging studies,
we know that at least down to spectral type M2 the objects show
a strong spot pattern on the surface (Barnes \& Collier Cameron \cite{bc01}). All
these results suggest the existence of magnetically induced spots on
VLM objects.

An alternative explanation for the observed low-amplitude periodic light curves
is the existence of inhomogeneously distributed dust clouds in the atmospheres,
as reported for ultra-cool dwarfs in the field (e.g., Bailer-Jones \& Mundt
\cite{bm01}, Clarke et al. \cite{ctc02}). Such clouds, however, are believed 
to form at temperatures below $2600$\,K, whereas our targets in $\epsilon$\,Ori 
have effective temperatures $>2600$\,K. At these temperatures, dust condensation 
processes are improbable (Allard et al. \cite{aha01}). Thus it is unlikely 
that dust clouds are the source of the periodic variations on our targets.
Therefore the most plausible explanation for these low-amplitude variations
is, as indicated above, the existence of asymmetrically distributed magnetically
induced spots.

The amplitudes of the light curves are determined by the properties of the
surface features and the inclination of the rotational axis to the line of sight.
On the basis of the given interpretation of the low-level variability, 
the amplitudes can therefore be used to obtain information 
about the properties of the spots. If we exclude the objects with
high-amplitude, partly irregular light curves (see Sect. \ref{high}), the amplitudes
scatter between 0.016 and 0.13\,mag, with an average of 0.038\,mag. Compared
with similar studies for solar-mass stars, these values are rather low. For example,
the amplitudes of the solar-mass periodic variables in the Pleiades range between 
0.02 and 0.2\,mag, with a mean value of 0.08\,mag (e.g., Krishnamurthi et al. 
\cite{ktp98}). Thus, our $\epsilon$\,Ori data confirm the result from SE2: The 
photometric amplitudes are significantly reduced in the VLM regime. 

This has also been observed by Lamm (\cite{l03}) in the young cluster NGC2264 ,
where the light curve amplitudes show a sharp decrease at colours of $R-I=1.6$,
roughly corresponding to $T_\mathrm{eff}=3500$\,K or a mass of about $0.4\,M_{\odot}$
(Baraffe et al. \cite{bca98}). Below this limit, the average amplitude is 
0.037\,mag, whereas higher mass objects have mean amplitudes of 0.089\,mag.
All our periodic targets are cooler than 3500\,K, thus our results are in good 
agreement with those of Lamm (\cite{l03}). 

There are several possible explanations for this effect, which we will discuss in
the following. It could be that spots on VLM objects are more or less concentrated
in polar regions, leading to a reduced photometric amplitude. However, the Doppler
imaging results indicate otherwise: Whereas G- and K-type stars can show strong 
polar spots, the pattern for M-type stars concentrates at low latitudes (Barnes \&
Collier Cameron \cite{bc01}). Thus, this interpretation seems unlikely.

A second explanation for the low amplitudes would be a change in the spot
distribution in the sense that spots on VLM objects are distributed more
symmetrically than on more massive stars. This would require a change in the
magnetic field geometry. For the Pleiades objects, such an interpretation
seems to be reasonable, because the change of the amplitudes occurs just
at the mass limit, where the objects are expected to become fully 
convective. This could indeed induce a change in the magnetic field
topology, e.g. from a large-scale $\alpha\Omega$ type dynamo to a small-scale
turbulent field (e.g., Durney et al. \cite{ddr93}). For very young objects,
however, this approach is not useful, because the change to fully convective
objects occurs at higher masses, in $\epsilon$\,Ori at 0.7 and in NGC2264 at
$1.3\,M_{\odot}$ (D'Antona \& Mazzitelli \cite{dm94}). Thus, objects above
and below the mass limit where the amplitudes change are fully convective,
and there is no reason to believe that their magnetic field structure changes
at this point. Thus, at least for very young objects, this scenario cannot
explain the low amplitudes.

The third interpretation would be a decrease of the relative spotted area in the
VLM regime, either because the spots are very few or very small. This could
be a result of the decrease in effective temperature, leading to increased
resistivities and thus less coupling between gas and magnetic field, as proposed
by Mundt (\cite{mu04}, see Lamm \cite{l03}). This would, however, be surprising, since 
VLM objects show strong H$\alpha$ and X-ray activity down to spectral type M9 (e.g., 
Mohanty \& Basri \cite{mb03}, Mokler \& Stelzer \cite{ms02}), corresponding to
$T_\mathrm{eff}\approx 2500$\,K, whereas the change of the amplitudes occurs
at significantly higher temperatures. It is difficult to understand how these
objects are able to sustain high chromospheric activity levels, if the magnetic 
field-gas coupling is not sufficient to produce photospheric spots. Thus, this 
scenario clearly needs further investigation.

There are only very few observations available which constrain
the spot filling factor in the VLM regime. Terndrup et al. (\cite{tkp99}) derive
a filling factor of 13\% for a Pleiades member with $M=0.39\,M_{\odot}$, a value
very similar to solar-mass stars. The Doppler images for M-type stars also
show that a large fraction of the surface is covered with spots (Barnes \&
Collier Cameron \cite{bc01}). However, both studies are based on targets which
are just at the mass limit where the photometric amplitudes decrease. Thus,
they cannot be used to demonstrate that cooler objects exhibit many spots.
We conclude that a decrease of the relative spotted area could be an 
explanation for the low photometric amplitudes. This result should, however, 
definitely be verified with future investigations of spot properties for
objects with masses significantly lower than $0.4\,M_{\odot}$.

\subsection{High-amplitude variables}
\label{high}

Five of our targets exhibit variability with high peak-to-peak amplitudes of from 0.15
to 1.0\,mag. Their light curves are shown in Fig. \ref{strange}. Clearly, the
brightness variations are at least partly irregular. Nevertheless, for three of
these objects, we detected a significant periodicity, but these periods show
obvious deviations from the sine shape, caused by superimposed non-periodic
variability. In the other two cases (no. 51 and 87), the photometric behaviour is totally
irregular. 

It is not possible to explain this conspicuous photometric behaviour with the
existence of cool spots alone. Herbst et al. (\cite{hbm02}) and Carpenter et al.
(\cite{chs01}) argue that the maximum amplitude from cool spots is 0.4-0.5\,mag.
To produce higher amplitudes with only cool spots would require unphysical 
values for filling factor or temperature contrast. On the other hand, magnetically 
induced spots are typically stable over at least several days (Hussain \cite{h02}, 
SE2). One should expect regular periodicities, and no irregular light curve 
components. Thus, the origin of the high-amplitude variations is clearly 
different from that of the low-amplitude variables.

The light curves of objects no. 51, 63, 120 could be affected by an eclipse event.
The typical duration of the eclipse would be about one night, periods less
than one week, and the eclipse depth would range from 0.15 to 1.0\,mag. With 
these constraints, the eclipsing body should be an object with a size
similar to that of our targets. If this companion would have significant luminosity, 
we should expect that these three objects appear displaced in the colour-magnitude
diagram. This is not the case, all three targets lie exactly on the empirical
isochrone for this cluster. Thus, if the photometric behaviour were due to eclipses,
the companion should be a very cool, but similar sized object. This criterion could
be fulfilled by a giant planet on a close orbit, a so-called 'hot Jupiter'. To 
estimate the probability of finding an eclipsing object, we used the results from
Carpenter et al. (\cite{chs01}), who find 22 candidate eclipsing systems among
1235 variable stars in the ONC. Scaling to the number of variable objects in
our study, we should expect to find one eclipsing system among our targets.
The main argument against an eclipse scenario is, however, that the three light curves
which we consider here are, as outlined above, not strictly periodic, contrary to what 
we should expect if they were produced by eclipses. Thus eclipses might contribute 
to the strange behaviour, but are obviously not the best explanation for these 
light curves. 

We interpret the high-amplitude variations as the consequence of hot spots formed
by matter flow from an accretion disk onto the central object. According to 
Carpenter et al. (\cite{chs01}), cool and hot spots can be distinguished from 
the light curves: First, hot spots with temperatures several thousand Kelvin 
higher than the photosphere can easily produce amplitudes as high as 1\,mag,
even with moderate filling factors. Second, there is a clear difference in the
timescales over which the photometric behaviour changes. Whereas cool spots 
produce stable periodic light curves, variability from hot spots is often 
irregular as a result of accretion rate variations, unstable accretion flow,
or misalignment of rotation and magnetic dipole axes. Since our highly variable
objects show not only high amplitudes, but also partly irregular variations,
in contrast to the low-amplitude variables, the most plausible explanation for
their behaviour is co-rotating hot spots formed by accretion, comparable
to the usual variability interpretation for solar-mass classical T Tauri stars.
Indeed, our high-amplitude time series are very similar to the light curves of 
classical T Tauri stars (Fern\'andez \& Eiroa \cite{fe96}, Herbst et al. 
\cite{hmw00}).

\begin{figure}[t]
\centering
\resizebox{\hsize}{!}{\includegraphics[angle=-90,width=6.5cm]{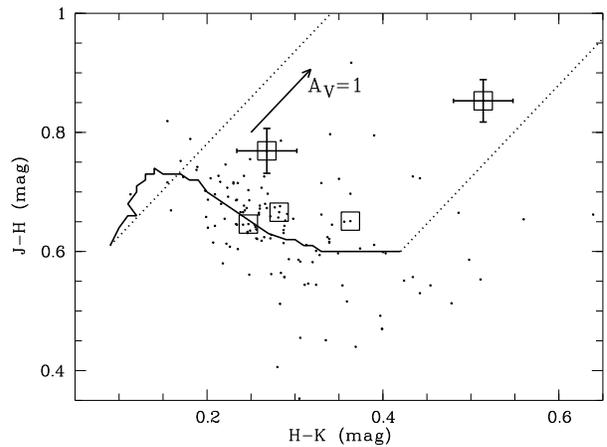}}
\caption{(H-K, J-H) colour-colour diagram for the $\epsilon$\,Ori targets constructed
from 2MASS data: The solid line is the 5\,Myr isochrone of Baraffe et al. \cite{bca98}. 
The dotted lines mark the extinction path, the arrow indicates a visual extinction of 
1\,mag. The five highly variable objects are marked with a square. For two of them, 
which clearly appear to be reddened, we over-plot their photometry errors. }
\label{cc}
\end{figure}

The existence of accretion disks around VLM stars and brown dwarfs has been established
by numerous recent studies. Comparable to T Tauri stars, young VLM objects often show
infrared excess emission (e.g., Natta \& Testi \cite{nt01}, Jayawardhana et al. 
\cite{jas03}) and strong H$\alpha$ emission in optical spectra (Mohanty et al. \cite{mjb03},
Barrado y Navascu\'es et al. \cite{bmj04}), clear signs for the existence of a 
disk and for ongoing accretion. The disk lifetime in the VLM regime seems to be not vastly 
different that of solar-mass stars, i.e. in the range of a few Myr, thus we can expect 
that at least a small subsample of our $\epsilon$\,Ori targets still possesses an accretion 
disk. It has also been shown that the typical high-amplitude, partly irregular photometric 
variations of classical T Tauri stars are also present in VLM objects (SE1).

To evaluate whether the highly variable objects are surrounded by accretion disks, we
investigated their near-infrared colours. In SE1 we demonstrated for the $\sigma$\,Ori 
cluster that high-amplitude variations are indeed correlated with near-infrared colour 
excess and accretion signature in optical spectra. Therefore we can expect a similar 
behaviour in $\epsilon$\,Ori. In Fig. \ref{cc}
the (H-K, J-H) colour-colour diagram is shown, together with an unreddened 5\,Myr 
isochrone (Baraffe et al. \cite{bca98}, solid line) and the reddening path (dotted lines).
The highly variable objects are marked with squares. The majority of our targets 
is grouped around the isochrone, confirming that this cluster does not suffer 
significant interstellar extinction. The objects with the largest photometric variations, 
no. 104 and 120, clearly are offset from the isochrone, and thus probably affected by 
intrinsic reddening. We cannot expect that all highly variable objects show up in the
reddening path, since with near-infrared colours it is certainly not possible to detect
all disks (see Natta \& Testi \cite{nt01}). The fact that two of our objects 
with high variations show a near-infrared colour excess, indicative of the existence of 
a circumstellar disk, confirms our interpretation of their photometric behaviour.
We conclude that the high-amplitude variations observed for our $\epsilon$\,Ori 
targets are most probably caused by ongoing accretion. Thus, these five highly variable 
objects can be identified as VLM analogues of classical T Tauri stars.

Only five of our 143 targets show a T Tauri like photometric behaviour. Taking account of a
contamination of 16\%, this corresponds to a fraction of 4\%. Thus, strong accretors
are probably very rare in the $\epsilon$\,Ori cluster. This frequency of objects with
high-amplitude variations is somewhat lower than in the $\sigma$\,Ori cluster 
(5-7\%, SE1). This could indicate that the targets in $\epsilon$\,Ori are, on average, 
older than those in the $\sigma$\,Ori cluster, confirming the age estimate given in Sect. 
\ref{intro}. Thus, most VLM objects lose their accretion disk within a few Myr, similar 
to solar-mass stars, as already found for example by Jayawardhana et al. (\cite{jas03}) 
and Barrado y Navascu\'es \& Mart\'{\i}n (\cite{bm03}). 

\subsection{Flares on VLM objects}
\label{flare}

As part of the time series analysis, we searched for flares in the light curves and find only
one event with an I-band amplitude of 0.3\,mag (Sect. \ref{tsa}). This result can be translated 
into a flare rate. Our observations cover about 7.5\,hours per night, in all four nights 
30\,hours in total. We analysed the light curves of 143 candidate objects, i.e. the data cover
4290 object hours. This translates to a flare rate of $2.3 \cdot 10^{-4}\,h^{-1}$ in the I-band.

It is difficult to compare these results with studies for solar mass stars, since we 
do not have flare statistics for them in the I-band. Also, we have little information 
about the spectral energy distribution of flares on VLM objects. Nevertheless,
some first tentative conclusions can be drawn with the available data.
Guenther \& Ball (\cite{gb99}) used a spectroscopic time series with a similar time resolution 
as in our campaign to determine a flare rate of $6 \cdot 10^{-2}\,h^{-1}$ for T Tauri stars with 
ages roughly comparable to the $\epsilon$\,Ori cluster. With this flare rate, we would expect to 
detect 257 flares in our light curves, but we found only one. The observations of Guenther \& Ball 
(\cite{gb99}), however, are based on spectra with a wavelength coverage from 360 to 610\,nm, whereas 
the I-band covers $\lambda>800$\,nm. Multi-wavelength studies by de Jager et al. (\cite{jha86}) and 
Stepanov et al. (\cite{sfk95}) show that the intensity of the flare flux substantially decreases 
towards redder wavelengths. Therefore, with our current knowledge, it is not possible to definitely 
decide if VLM objects are really flare-inactive compared with solar mass stars.

We investigated the photometric behaviour of a flare event observed by Liebert et al. 
(\cite{lkr99}). They caught a M9.5 field dwarf spectroscopically in a huge flare event.
This object has a mass near the substellar limit, i.e. comparable to our $\epsilon$\,Ori flare object, 
but since it is older, its effective temperature is significantly lower. We folded the spectrum in 
flare state (from 7th Dec 1997, 4:52 UT) and in quiet state (from 23th January 1998) with the I-filter 
transmission function and the sensitivity function for one of the WFI CCDs. The quotient between the 
flare spectrum and the quiet spectrum gives the net flare flux in the I-band. We obtained an integrated 
flux increase of 0.33\,mag. Thus, this flare would have produced an I-band eruption similar to the one 
flare observed in our $\epsilon$\,Ori light curves. Even flares which are only a tenth as strong would 
result in an I-band flux increase of 0.03\,mag and would have been easily detectable for nearly all 
our candidates.

Judging from the flare studies for solar-mass stars, we would expect to see only very energetic flares in
the I-band. According to the observations of Stepanov et al. (\cite{sfk95}), even a flare with U-band 
amplitude larger than 1\,mag does not show any clear I-band eruption. Assuming a similar energy 
distribution for flares on VLM objects, our 0.3\,mag flare in the I-band should have been extraordinarily 
strong in the blue wavelength range. Such huge events are rare for solar-mass stars, at least less 
frequent as less energetic flares. Thus, it seems surprising that we see only one very energetic flare, 
which is also comparable in its amplitude with the event observed by Liebert et al. (\cite{lkr99}), 
but no weaker events, in spite of our extended time coverage. Thus, VLM 
objects possibly show only very few flares, but if they flare, the event is very strong. However,
the statistics is rather poor, therefore we leave a more detailed discussion of the flare
behaviour of VLM objects for future investigations.

\section{Rotation periods}
\label{rot}

Our period sample for objects in the $\epsilon$\,Ori cluster significantly increases the
number of rotation periods for young VLM objects. In particular, it contains periods for
nine brown dwarf candidates, which is the largest sample of BD periods so far. 
In the following section we will analyse these periods and compare them with
literature data.

\subsection{Mass-period relationship}
\label{massper}

As found by recent studies, the rotation periods of VLM objects are clearly mass dependent,
in the sense that the periods decrease with decreasing mass. This positive correlation between
period and mass has recently been found for VLM stars in the Pleiades (SE2), 
but it has also been detected at very young ages for the Trapezium cluster (Herbst et al. \cite{hbm01}),
NGC2264 (Lamm \cite{l03}, Lamm et al. \cite{lbm04}, \cite{lmb04}), and the $\sigma$\,Ori cluster 
(SE1). For these very young objects, the mass-period relation can only be recovered from the median of the 
periods. At this stage of evolution the scatter of the periods is considerable, probably because 
their rotational behaviour is still significantly influenced by accretion, strong activity, and/or 
initial conditions. With our 30 new periods for $\epsilon$\,Ori VLM objects, it is now again possible 
to probe the mass-period relationship.

\begin{figure}[htbd]
\centering
\resizebox{\hsize}{!}{\includegraphics[angle=-90,width=6.5cm]{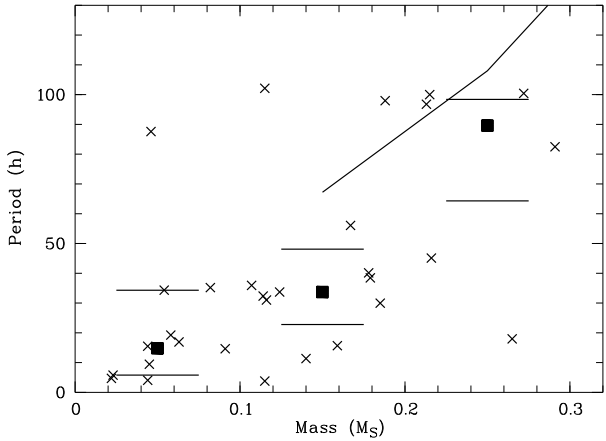}}
\caption{Period-mass relationship for the $\epsilon$\,Ori cluster. The crosses are
the periods measured in this study. Filled squares mark the median period for the mass
bins $0<M\le 0.1\,M_{\odot}$, $0.1<M \le 0.2\,M_{\odot}$, and $0.2<M\le 0.3\,M_{\odot}$.
The horizontal lines are the quartiles for these values. The solid line shows the 
period-mass-relationship for the ONC, as derived by Herbst et al. (\cite{hbm01}).}
\label{permass}
\end{figure}

In Fig. \ref{permass} we plot the $\epsilon$\,Ori periods vs. mass. The filled squares mark the
period median for three mass bins ($M<0.1\,M_{\odot}$, $0.1<M<0.2\,M_{\odot}$, $0.2<M<0.3\,M_{\odot}$),
the horizontal lines show the quartiles for these values, which can be considered as a measure for
the scattering. Clearly, the median period increases with increasing mass, confirming
the results from previous studies. In Fig. \ref{permass} we also plot the mass-period relationship
for the Trapezium cluster derived by Herbst et al. (\cite{hbm01}). Obviously, the slope is steeper
for the 1\,Myr young Trapezium objects. To investigate this in more detail, both period-mass
relationships were fitted linearly. The ratio between the slope for the Trapezium data and the
slope for the $\epsilon$\,Ori periods is 1.3, i.e. on average the periods in $\epsilon$\,Ori are 
shorter by this factor. 
%lower limit of 0.9 and an upper limit of 2.1. 

This behaviour is expected, since $\epsilon$\,Ori is considerably older, and the objects are expected 
to contract in the course of their pre-main sequence evolution. As a consequence, their rotation
should accelerate. The age of the Trapezium cluster is about 1\,Myr (Hillenbrand \cite{h97}), where the age of the 
$\epsilon$\,Ori objects should be between 2 and 10\,Myr. Using the radii from the models of the Lyon 
group (Chabrier \& Baraffe \cite{cb97}, Chabrier et al. \cite{cba00}) and assuming angular momentum
conservation, we should expect that for a $0.1\,M_{\odot}$ object the period should decrease by a 
factor of 1.7 and 7.3, where the exact value depends on the age of $\epsilon$\,Ori. For a $0.2\,M_{\odot}$ 
object, the factors are somewhat smaller, and lie between 1.3 and 5.6. Whereas the lower limit of these 
values is in good agreement with the observed period decrease, the upper limit is clearly too high. 

%the radius of a $0.1\,M_{\odot}$ object should decreases by a factor of about 1.3 (2.7) between ages 
%of 1 and 2 (10)\,Myr. With angular momentum conservation, the period evolves proportional to the square 
%of the radius.  

This could simply indicate that the $\epsilon$\,Ori objects are mostly very young, with ages
below 5\,Myr. The alternative explanation is that our assumption of angular momentum conservation 
is not valid. For some objects the rotation might be significantly braked by magnetic interaction 
between star and disk (so-called 'disk-locking'). The study of Lamm et al. (\cite{lmb04}) in NGC2264 
showed that the impact of disk-locking on the rotational regulation is dependent on mass, in the 
sense that VLM objects are probably less affected by disk-locking than solar-mass stars (so-called 
'imperfect' disk-locking). Nevertheless, even in the VLM regime the disk has a certain influence on 
the rotation of young objects. For VLM objects in the $\sigma$\,Ori cluster we noted that stars which
show evidence for the existence of the accretion disk rotate on average slower than
stars without disk (SE1). This is consistent with the scenario where a few objects are 
still braked by interacting with their disks. This result has been confirmed by a recent 
angular velocity study of Mohanty (\cite{m04}), where the accreting objects are mostly 
slow rotators. Thus, disk-locking might be less efficient than in solar-mass stars, but it 
still works in the VLM regime. This could be the explanation for the moderate decrease of the 
average period in $\epsilon$\,Ori compared with the periods in the Trapezium cluster.

\subsection{Rotation of brown dwarfs}

In contrast to the variability studies in the Trapezium cluster and NGC2264, our analyses
for $\sigma$\,Ori (SE1) and $\epsilon$\,Ori (this paper) extend well down into the substellar 
regime, down to $0.03\,M_{\odot}$. Rotation periods were measured for nine BDs for each of 
these two clusters, enabling us to give a first detailed description of the rotational
behaviour of substellar objects.

One of the main conclusions of Fig. \ref{permass} is that the period-mass relationship
extends down into the substellar regime. Thus, in comparison with stars, BDs rotate very fast.
For $\sigma$\,Ori the median period is 43.4\,h for VLM stars and 14.7\,h for BDs. For $\epsilon$\,Ori 
the values are very similar, 35.9\,h for VLM stars and 15.5\,h for BDs. Only in a very few cases 
do BDs show periods longer than two days in these two clusters. The previously published periods
for substellar objects are in good agreement with this result: Bailer-Jones \& Mundt (\cite{bm01})
and Zapatero Osorio et al. (\cite{zcb03}) obtained periods shorter than 10\,h for three BDs in 
the $\sigma$\,Ori cluster. Additionally, the field brown dwarf Kelu-1 shows a period of only
1.8\,h (Clarke et al. \cite{ctc02}), and Koen (\cite{k04}) found tentative periods between 2 and 
7\,h for three ultracool field dwarfs. Rotational velocity studies also indicate that 
brown dwarfs in general are rapid rotators with a lower $v\sin i$ limit of about 10\,kms$^{-1}$
(Mohanty \& Basri \cite{mb03}, Bailer-Jones \cite{b04}). Slowly rotating BDs with periods
of a few days were only detected in the Chamaeleon I star-forming region (Joergens et al. 
\cite{jfc03}). However, since these objects are only about 1\,Myr year old, they could
still be subject to significant rotational braking by interaction with the circumsubstellar
disk, as argued by Joergens et al. (\cite{jfc03}). Thus, based on our current knowledge, 
BDs are very rapid rotators with average periods shorter than one day, as soon as they are
no longer influenced by the disk.

It is of particular interest to have a closer look at the lower limit of the periods,
since this value is constrained by the breakup period, i.e. the period where the 
centrifugal forces exceed the gravitational forces. For the $\sigma$\,Ori cluster
the shortest period measured so far is about 3\,h (Zapatero Osorio et al. \cite{zcb03}).
In our own study in $\sigma$\,Ori, we found a lower limit of 6\,h (SE1). In $\epsilon$\,Ori,
which is probably slightly older than $\sigma$\,Ori, we obtain a lower boundary of 4\,h,
whereas our period search was sensitive down to periods below one hour. In our variability
study in the Pleiades we found periods of 3 and 4\,h for two VLM stars with masses
very close to the substellar boundary. Finally, the above mentioned period of 2\,h measured
for Kelu-1 defines the lower period limit for field brown dwarfs. This is confirmed by
the available rotational velocities for field objects, which show an upper limit between
40 and 60\,kms$^{-1}$, corresponding to a period of 2-3\,h. Thus, the lower period limit 
of 2-4\,h seems to be independent of age, implying that a fraction of BDs evolve with 
nearly constant period over one Gyr, in clear contrast to the period evolution of stars. 
This is particularly surprising because all these objects should undergo 
significant rotational acceleration as a consequence of the contraction during the first 
Gyr of their evolution. Assuming angular momentum conservation, their period should decrease 
by a factor of about ten as they evolve from 3\,Myr to 1\,Gyr. Thus, these very fast rotating 
objects should experience strong angular momentum loss.

It seems unlikely that disk-locking can account for the required rotational braking,
since, as mentioned already in Sect. \ref{massper}, the impact of disk-locking decreases 
with decreasing object mass (Lamm et al. \cite{lmb04}). It is also implausible to assume 
excessive angular momentum loss through stellar winds in the BD regime. As BDs evolve 
they will rapidly cool down, and the resistivity in their atmospheres will increase. This
will very likely lead to a decline of the efficiency of angular momentum loss by stellar 
winds. It is thus unclear how the rotation of the fast rotating substellar objects is
regulated. 

\begin{figure}[t]
\centering
\resizebox{\hsize}{!}{\includegraphics[angle=-90,width=6.5cm]{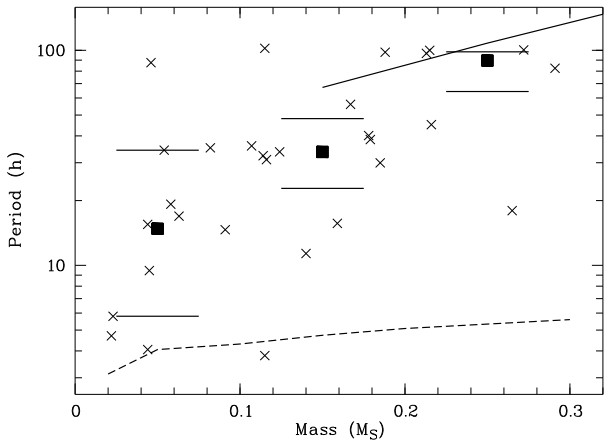}}
\caption{Period-mass relationship for the $\epsilon$\,Ori and the Trapezium cluster
in comparison with the breakup period (dashed line). The filled squares show the
median period from our $\epsilon$\,Ori sample for the mass bins $0<M\le 0.1\,M_{\odot}$, 
$0.1<M \le 0.2\,M_{\odot}$, and $0.2<M\le 0.3\,M_{\odot}$. Horizontal lines are the
quartiles for these bins. The solid line is the period-mass relationship for 
the Trapezium cluster (Herbst et al. \cite{hbm01}). }
\label{permasscrit}
\end{figure}

As mentioned above, there is a natural lower limit to the rotation period, namely
the so-called breakup period where the centrifugal exceeds the gravitational
force. No object can rotate faster than with its breakup period. According to 
Porter (\cite{p96}), the breakup period can be expressed as:
\begin{equation}
P_\mathrm{crit} = 0.143 \, \frac{(R/R_{\odot})^{3/2}}{(M/M_{\odot})^{1/2}}\,d 
\end{equation}
Thus, for evolved BDs with radii of about $0.1\,R_{\odot}$, the breakup period is
below 1\,h, i.e. well below the observed period limit. At very young ages, however,
the fastest rotating BDs approach the breakup limit. In Fig. \ref{permasscrit}
we plot the period-mass relationship for the Trapezium cluster (solid line)
and for our $\epsilon$\,Ori objects (filled squares and horizontal lines, symbols
as in Fig. \ref{permass}), together with the breakup period (dashed line), calculated
with the 5\,Myr radii from Chabrier \& Baraffe (\cite{cb97}) and Chabrier et al. 
(\cite{cba00}). Apparently, the period range of the substellar objects extends down to 
the critical period, which lies between 3 and 5\,h. Thus, BDs with periods of a few 
hours do indeed rotate extremely fast in terms of their breakup limit. As a consequence,
these objects are probably strongly oblate and they might lose material because
of the strong centrifugal forces at the equator. If and how this has influence
on their rotational evolution should be investigated in detail in the future.

\section{Conclusions}
\label{conc}

We present the results of an extended variability study of young VLM objects near
the bright star $\epsilon$\,Ori. This region harbours a rich population of young
stars and brown dwarfs (called $\epsilon$\,Ori cluster), which probably belongs 
to the Ori\,OB1b association. The ages of these objects are probably between 2 and 10\,Myr. 
In a field of 0.3 sq. deg. we identified 143 young VLM objects by means of colour-magnitude 
diagrams. For these objects, the photometry in five bands (RIJHK) is consistent with 
membership of the $\epsilon$\,Ori cluster. From simulated star counts for this Galactic 
direction, we estimate a contamination rate of 16\% for this sample. We cover a mass range 
from 0.02 to 0.66$\,M_{\odot}$, where most of our candidates have masses below 
$0.4\,M_{\odot}$. 

These objects were observed in a photometric time series with the ESO/MPG wide field imager
at the 2.2-m telescope on La Silla. We covered four complete consequent nights, and
obtained at least 29 I-band images per night for a total of 129 images. The instrumental 
magnitudes from these time series images were calibrated relative to non-variable reference 
stars in the field. We reach a mean precision of 5\,mmag for the brightest targets. A time 
series analysis procedure was carried out for all candidate light curves. The main focus was 
on the period search, which was based on periodogram techniques and included several
independent reliability checks. 

A large fraction of the candidates shows clear signs of variability. We identified 
three types of variability in our targets:

a) For one VLM star, a large flare event was detected, with a sudden I-band brightness 
increase of 0.3\,mag, and a subsequent exponential decline over several hours. From this 
finding, we estimate an I-band flare rate of VLM objects of about $2 \cdot 10^{-4}$ per 
hour. Since the flare intensity usually decreases drastically toward red wavelengths,
our single flare event should have been extraordinarily strong in the blue wavelength range.
Observations of flares on VLM objects are rare, but the very few detected flares, including
the event in $\epsilon$\,Ori, are quite strong. Whether this is a characteristic feature
of these objects or just a result of poor statistics has to be verified.

b) Five objects, including two brown dwarf candidates, show strong and partly irregular
variations, with amplitudes of up to 1\,mag. For three of these candidates, we nonetheless
detected a significant period (see c)), but clearly strong irregular brightness modulations
are superimposed on their periodicities. Because this photometric behaviour is comparable
to that of classical T Tauri stars, we interpret the variability of these highly variable targets 
as a consequence of ongoing accretion processes. Thus, this study again provides  
independent confirmation of the existence of accretion disks around young brown dwarfs and
VLM stars. Since the fraction of highly variable objects is only 4\%, we conclude that
most VLM objects lose their disks on timescales of a few Myr.

c) For 30 targets, including nine brown dwarf candidates, we detected highly significant 
periods, which we interpret as the rotation periods. The periods range from 4\,h up to 
100\,h, while our analysis is sensitive for periods from 0.2\,h up to 110\,h. With the
exception of the three highly variable objects discussed under item b), all these periodic
light curves have amplitudes below 0.15\,mag. The most reasonable explanation for their
variability is the existence of cool magnetically induced surface spots co-rotating with
the objects. The amplitudes are smaller than in similar studies for solar-mass stars.
This could be a consequence of a change in the dynamo in the VLM regime or a decrease
of the relative spotted area induced by the high resistivities in cool atmospheres.

We interpreted the 30 periods as rotation periods of our objects. This is the largest
sample of VLM objects with known rotation periods for ages $>2$\,Myr. The mass-period
relationship was analysed in comparison with the data sets in $\sigma$\,Ori (3\,Myr, SE1), 
NGC2264 (2\,Myr, Lamm et al. \cite{lmb04}), and the Trapezium cluster (1\,Myr, 
Herbst et al. \cite{hbm02}). In agreement with these previous studies, we find that 
the median period decreases with decreasing mass. In $\epsilon$\,Ori, 
the slope of this relationship is by a factor of 1.3 lower than in the younger 
Trapezium cluster. In case of angular momentum conservation, we would expect a rotational
acceleration by a factor between 1.3 and 7.3, where the exact value depends on the age
of the $\epsilon$\,Ori targets. Thus, either most of the $\epsilon$\,Ori objects have 
an age of about 2-3\,Myr, or there is still significant rotational braking, e.g. by magnetic 
interaction with the accretion disk ('disk-locking').

Our period sample in combination with literature data allows us to investigate
the rotational evolution of brown dwarfs. BDs with ages $>$1\,Myr are very fast rotators
with average rotation periods below one day. According to the available data, the lower
period limit lies between 2 and 4\,h and is more or less independent of age. Thus, at ages
between 3\,Myr and 1\,Gyr a fraction of the BDs rotate with very short but more or less 
constant rotation periods, in clear contrast to the period evolution of stars. Taking into 
account the hydrostatical contraction, this implies that these fast rotators experience 
strong rotational braking, which cannot be done only by disk-locking or stellar winds. 
For the young objects in $\epsilon$\,Ori the lower limit is very close to the breakup 
period, which might have an influence on their rotational evolution.

\begin{acknowledgements}
      An implementation of the CLEAN algorithm was kindly provided by 
      David H. Roberts. It is a pleasure to acknowledge the support of James 
      Liebert and Davy J. Kirkpatrick, who made their spectra of the flare 
      on the field dwarf available to us. We thank the referee, Dr. Luisa
      Rebull, for her rapid response and helpful comments.
      This work was supported by the German 
      \emph{Deut\-sche For\-schungs\-ge\-mein\-schaft, DFG\/} grants Ei~409/11--1 
      and Ei~409/11--2. 
\end{acknowledgements}

\end{document}